\documentclass[
  aps,
  prd,superscriptaddress,
  nofootinbib,twocolumn,
  longbibliography
]{revtex4-2}

\usepackage[T1]{fontenc}
\usepackage[utf8]{inputenc}
\usepackage{amsmath,amssymb,amsfonts}
\usepackage{bm}
\usepackage{graphicx}
\usepackage[caption=false]{subfig}
\usepackage{booktabs}
\usepackage{hyperref}
\usepackage{xcolor}
\usepackage{placeins}

\newcommand{\dd}{\mathrm{d}}

\begin{document}

\title{Dynamically Stable Magnetic Fields in Relativistic Neutron Stars: A Diverse Landscape of GRMHD Equilibria}

\author{Mai Li}
\affiliation{Department of Physics, University of Illinois Urbana-Champaign, Urbana, IL 61801, USA}
\author{Vasileios Paschalidis}
\affiliation{Department of Astronomy, University of Arizona, Tucson, AZ 85719, USA}
\affiliation{Department of Physics, University of Arizona, Tucson, AZ 85719, USA}
\author{Antonios Tsokaros}
\affiliation{Department of Physics, University of Illinois Urbana-Champaign, Urbana, IL 61801, USA}
\affiliation{National Center for Supercomputing Applications, University of Illinois Urbana-Champaign, Urbana, IL 61801, USA}
\affiliation{Research Center for Astronomy and Applied Mathematics, Academy of Athens, Athens 11527, Greece}
\author{Maria Mutz}
\affiliation{Department of Physics, University of Arizona, Tucson, AZ 85719, USA}
\author{Koji Uryu}
\affiliation{Department of Physics, University of the Ryukyus,
Senbaru 1, Nishihara, Okinawa 903-0213, Japan}

\date{\today}

\begin{abstract}
Neutron stars endowed with purely poloidal or purely toroidal magnetic fields are known to be unstable, and settle into mixed poloidal-toroidal configurations. The end state of these instabilities is still poorly understood. It is unclear whether this state is unique or whether the dynamically stable configuration has memory of the initial conditions. In this work, we perform long-term, fully general relativistic  magnetohydrodynamic simulations of equilibrium configurations of slowly rotating neutron stars endowed with self-consistent mixed poloidal and toroidal magnetic fields. Our initial configurations differ in their initial magnetization and magnetic field geometry. We evolve the initial equilibria for tens of Alfv\'en timescales until well after any instabilities have saturated. Employing an array of diagnostic tools from visualizations to the structure of the plasma four-current and a vector spherical harmonic decomposition of the settled magnetic fields, we characterize the structure of the dynamically stable, settled magnetic equilibria. We find that the settled configurations are not unique. Instead, they exhibit diverse multipolar structures, with higher-order modes contributing substantially. Our findings demonstrate that there is memory of the magnetic-field initial conditions, and that there is no unique dynamically stable magnetic-field geometry for neutron stars. Nevertheless, we find that the angle-averaged radial profile of the magnetic field toroidal to poloidal amplitude in the bulk of the settled configurations exhibits some degree of universality with  typical values of order \(20\)--\(40\%\).

\end{abstract}

\maketitle

\section{Introduction}

The detailed geometric structure of the magnetic field of neutron stars (NSs) plays a fundamental role in dictating a wide array of high-energy astrophysical phenomena associated with these compact objects. Within the stellar interior and crust, strong magnetic fields govern the elastic properties and magnetothermal evolution of magnetar crusts (see, e.g.,~\cite{Ciolfi:2009bv,Chamel:2012uz,Thompson:2016dkd,Gourgouliatos:2016fnl,Pons:2019zyc}), driving catastrophic crustal ruptures that power observed non-thermal magnetar flares and non-thermal emission~\cite{Kaspi:2017fwg}. Beyond the surface, the precise configuration of the global magnetic field lines anchors the pulsar magnetosphere, steering the coherent radio and high-energy particle acceleration processes that power pulsar emission (see e.g.~\cite{BeskinPulsarReview2018,PhilippovAAReview}) and potentially sourcing the engine behind fast radio bursts (FRBs) (see ~\cite{Zhang:2022uzl} for a recent review).

Crucially, reverse-engineering the exact magnetic field structure at and near the neutron star surface is a major target of high-cadence X-ray observations, particularly with the Neutron Star Interior Composition Explorer (NICER). NICER utilizes pulse-profile modeling of thermal X-ray emissions from surface "hot spots" to place joint constraints on the masses and radii of millisecond pulsars, which are subsequently mapped onto the nuclear equation of state (EOS). The adopted ray-tracing frameworks used for the interpretation of NICER observations rely heavily on modeling the shape and size of these X-ray emitting hot spots, which are believed to arise from the topology of the surface magnetic field, which traces where accelerating magnetospheric particles impact the stellar atmosphere~\cite{Baubock:2013gna,Gralla:2016fix,Gralla:2017nbw,Lockhart:2019nch,Riley:2019yda,Miller:2021qha,Salmi:2024bss,Miller:2025qfq,Mauviard:2025dmd,Kini:2026rjx}. Therefore, knowing the precise neutron star surface magnetic-field structure is important for ab-initio models used to interpret NICER observations.

Historically, theoretical models of pulsar magnetospheres and surface emission have routinely adopted idealized, pure magnetic dipole topologies, or linear combinations of lower-order multipoles, see e.g.~\cite{Gralla:2016fix,Gralla:2017nbw} and references therein). However, empirical results from phenomenological X-ray light-curve modeling by the NICER collaboration favor hot spot configurations on real millisecond pulsars that are explicitly non-axisymmetric and non-antipodal~\cite{Bilous:2019knh}. Such multi-component, asymmetric surface-emission geometries indicate that the magnetic and magnetospheric structure of real neutron stars may depart substantially from a simple centered dipole~\cite{Chen:2020rud,Cao:2026xnp,Kundu:2026guq}. Higher-order and nonaxisymmetric magnetic multipoles provide a natural mechanism for producing such complexity, although the mapping between an inferred thermal-emission geometry and the underlying stellar magnetic field is not unique. Bridging the gap between theoretical pulsar-magnetosphere models and these observations therefore requires moving beyond idealized dipolar configurations toward self-consistent, dynamically stable magnetic topologies.

Furthermore, standard configurations involving purely poloidal or purely toroidal  magnetic fields are physically unviable due to the so-called "kink", "Tayler" and "pinch" instabilities~\cite{Tayler_1957,10.1093/mnras/161.4.365,1973MNRAS.163...77M,10.1093/mnras/162.4.339,1977ApJ...215..302F}. Analytical and numerical work has firmly established that purely poloidal, purely toroidal fields and even certain combinations of poloidal and toroidal magnetic fields are universally unstable on dynamical timescales~\cite{2007A&A...469..275B,Ciolfi:2011xa,Lasky:2011un,2011MNRAS.412.1730L,Lasky:2012ju,Ciolfi:2012en,Ciolfi:2013dta,Sur:2020hwn,Tsokaros:2021pkh,Sur:2021awe,Pinas:2025bpq}. These instabilities operate on the order of a few Alfvén timescales, causing the field to rapidly rearrange or decay, giving rise to mixed poloidal--toroidal configurations, including twisted-torus-like geometries, as candidates for long-lived magnetic states. On the other hand, the simulations of \cite{Tsokaros:2021pkh} found that the stability of mixed poloidal-toroidal configurations  depends 
strongly on the mixture of the toroidal and the poloidal components, with some models being highly unstable while others being 
stable over many Alfv\'en timescales. Rapid rotation can also have a stabilizing effect~\cite{Pinas:2025bpq,Joshi:2026Interior}, but the vast majority of pulsars are not rapidly rotating. 

Despite the acknowledged instability of the simplest magnetic field configurations, comparatively little work has been done to systematically extract the multipolar structure of stable, fully general relativistic, mixed-field neutron stars following the saturation of these dynamical instabilities and subsequent relaxation. Moreover, a fundamental open question about neutron star magnetic fields is whether the ultimate dynamically stable magnetic field geometry is unique or not. To properly address the aforementioned questions, one must model the magnetized neutron stars under full general relativity, tracking the non-linear magnetohydrodynamic (MHD) evolution until a settled, quasi-equilibrium state is achieved, while treating significantly different initial magnetic field geometries. Prior work suggests that a twisted torus configuration appears to be a universal outcome. However, prior studies have considered magnetic fields whose geometry can be approximated by the field generated by a loop current on the equatorial plane and/or a line current perpedicular to that plane running throught the center of mass of the star. However, this geometry is very specific, and it is unlear if more complex geometries such as superposing multiple current loops and off the equatorial plane give rise to a twisted torus configuration.

In this work, we address the aforementioned challenges by conducting long-term, fully dynamical GRMHD simulations of slowly rotating neutron star equilibria endowed with self-consistent mixed poloidal and toroidal fields, that have varying complex structure consistent with multiple current loops. Utilizing initial models generated with the {\tt COCAL} code~\cite{Uryu:2014tda,Uryu:2019ckz} and evolving them for tens of Alfv\'en timescales with the Einstein Toolkit infrastructure~\cite{Loffler:2011ay,Zilhao:2013hia}, we track the nonlinear development of the instability and dynamical stabilization of the magnetic field until a quasistationary state is reached. We then employ a vector spherical harmonic (VSH) decomposition pipeline paired with a radial discontinuous Galerkin (DG) representation to quantitatively analyze the settled fields. 

In this work, dynamical stability means that, following saturation of the initial magnetic-field
rearrangement, the configuration exhibits no further rapidly growing ideal-GRMHD instability or
global disruption over many Alfv\'en times. This statement does not address resistive or other
secular timescales.

Our main findings demonstrate that the saturated, dynamically stable magnetic fields occupy distinct mixed poloidal--toroidal and multipolar states rather than converging toward a unique magnetic topology. Higher-order multipoles make substantial contributions, while the relative importance of the toroidal and poloidal components depends strongly on direction, radius, and model. The four-current distributions further corroborate these conclusions.
 
To enable broader community engagement, Python tools that reconstruct these dynamically stable fields are available upon request.

The remainder of this paper is organized as follows. In Sec.~\ref{sec:methods}, we describe the initial data we adopt, GRMHD and spacetime evolution framework, as well as our diagnostic tools. In Sec.~\ref{sec:results}, we present the multipolar structure of the settled magnetic fields together with multidimensional magnetic field and four-current. In Sec.~\ref{sec:discussion}, we discuss the implications of our findings and summarize our main conclusions. Throughout the paper, we adopt geometrized units in which \(G=c=1\), unless otherwise stated.

\section{Methods}\label{sec:methods}

\subsection{Initial data}

We construct stationary, axisymmetric equilibria of uniformly rotating magnetized neutron stars with the {\tt COCAL} code~\cite{Uryu:2014tda,Uryu:2019ckz}. The construction solves the coupled Einstein, Maxwell, and ideal-magnetohydrodynamic equations self-consistently under the assumption of perfect conductivity. The electromagnetic stress-energy and Lorentz force are included in determining the spacetime and fluid equilibrium, so both poloidal and toroidal magnetic field components are incorporated into the equilibrium construction. We adopt the polytropic equation of state $P=K\rho_b^\Gamma$ with $\Gamma=2$, where $\rho_b$ denotes the fluid rest-mass density throughout this work.. 

In the COCAL formulation, the gravitational and electromagnetic potentials satisfy coupled elliptic equations, while first integrals of the fluid equations determine the matter variables. These equations are solved iteratively to obtain a consistent magnetized equilibrium. Freely specified functions entering the integrability conditions control the magnetic field distribution; the formulation and numerical solution procedure are described in Refs.~\cite{Uryu:2014tda,Uryu:2019ckz}.

We consider three initial models whose properties are summarized in Table~\ref{tab:initial_data}. We label the three models {\tt A}, {\tt B}, and {\tt C}. These slowly rotating models, which are also weakly magnetized (in terms of their ratio of total magnetic to gravitational potential energy), differ in their initial magnetization and magnetic field geometry, as is likely to be the case in different proto-neutron stars in nature. Throughout the text and figures, $t_A$ denotes the central Alfv\'en timescale $t_{A,c}$ of the corresponding initial equilibrium, as defined in Table~\ref{tab:initial_data}. We  express the evolution coordinate time in units of $t_A$.

\begin{table*}[t]
\centering
\caption{
Initial equilibrium \(\Gamma=2\) polytropic models of uniformly rotating neutron stars evolved in this work. The dimensionless quantity \(\Omega R_e\) is reported in the second column. \(\rho_{b,c}\) is the central rest-mass density, \(M\) is the ADM mass, \(P\) is the rotational period, and \(R_p/R_e\) is the polar-to-equatorial coordinate-radius ratio. The ratio \(T/|W|\) uses the kinetic and gravitational energies entering the COCAL virial relation. The quantities \(\mathcal{M}/|W|\), \(\mathcal{M}_{\rm tor}/|W|\), and \(\mathcal{M}_{\rm pol}/|W|\) are the total electromagnetic, toroidal magnetic, and poloidal magnetic energies, respectively, normalized by the same \(|W|\). These energies are defined in Appendix~\ref{App:initial_energies}; the total \(\mathcal{M}\) includes the electric contribution. The last four columns give the dynamical time \(t_d\equiv\rho_{b,c}^{-1/2}\), the central Alfv\'en time \(t_{A,c}\equiv R_e\sqrt{4\pi\rho_{b,c}}/B_c\), the surface Alfv\'en time \(t_{A,\rm surf}\equiv R_e\sqrt{4\pi\rho_{b,c}}/B_{\rm pole}\), and the total simulated duration in units of the central Alfv\'en time. Here \(B_c\) and \(B_{\rm pole}\) are the magnetic field strengths at the stellar center and pole, respectively. Throughout the text, \(t_A\) denotes \(t_{A,c}\). The times \(t_d\), \(t_{A,c}\), and \(t_{A,\rm surf}\) are reported in units of \(GM_\odot/c^3\simeq4.93\,\mu{\rm s}\). The physical scaling adopts the polytropic equation of state \(P=K\rho_b^2\) with \(K=70.97\) in units where \(G=c=M_\odot=1\).
}
\label{tab:initial_data}
\begin{ruledtabular}
\begin{tabular}{l*{13}{c}}
Model
& \(\Omega R_e\)
& ${\rho_{b,c}}$
& $M$
& $P$
& $R_p/R_e$
& $T/|W|$
& $\mathcal{M}/|W|$
& $\mathcal{M}_{\rm tor}/|W|$
& $\mathcal{M}_{\rm pol}/|W|$
& $t_d$
& $t_{A,c}$
& $t_{A,\rm surf}$
& $t_{\rm evol}/t_{A,c}$
\\
& 
& $(10^{15}{\rm g\,cm^{-3}})$
& $(M_\odot)$
& 
& 
& $(10^{-2})$
& $(10^{-3})$
& $(10^{-5})$
& $(10^{-3})$
& 
& 
& 
& 
\\
\hline
{\tt A} & 0.0300 & 1.07 & 1.18 & \(1.47\times10^{3}\) & 0.956 & 0.101 & {10.2} & {41.6} & {9.82} & {24.0} & {46.9} & {\(3.70\times10^{2}\)} & {42.7} \\
{\tt B} & 0.0457 & {1.51} & {1.29} & {856} & {0.988} & {0.195} & {0.128} & {0.0547} & {0.126} & {20.2} & {114} & {\(2.41\times10^{3}\)} & {26.1} \\
{\tt C} & 0.0300 & {1.76} & {1.33} & {\(1.23\times10^{3}\)} & {0.994} & {0.0661} & {0.218} & {0.0381} & {0.217} & {18.7} & {73.0} & {\(1.06\times10^{3}\)} & {27.4} \\
\end{tabular}
\end{ruledtabular}
\end{table*}

\begin{figure*}[!tp]
\centering
\includegraphics[width=0.99\textwidth]{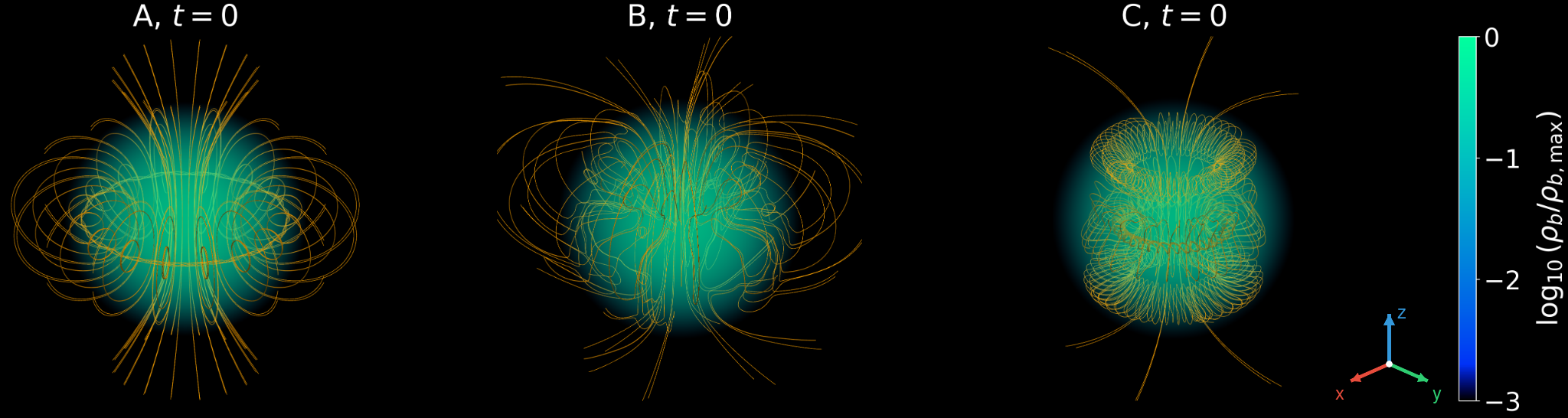}
\caption{Three-dimensional magnetic field structure of the initial equilibrium configurations. Selected magnetic field lines are shown together with a volume rendering of the rest-mass density. The field lines display the principal topological families identified in the corresponding two-dimensional slices in Fig.~\ref{fig:Bfield_2d_t0}. Their number and spacing do not indicate magnetic-field strength. The orientation triad identifies the Cartesian coordinate directions.}
\label{fig:Bfield_3d_t0}
\end{figure*}

\begin{figure*}[!tp]
\centering
\includegraphics[width=0.99\textwidth]{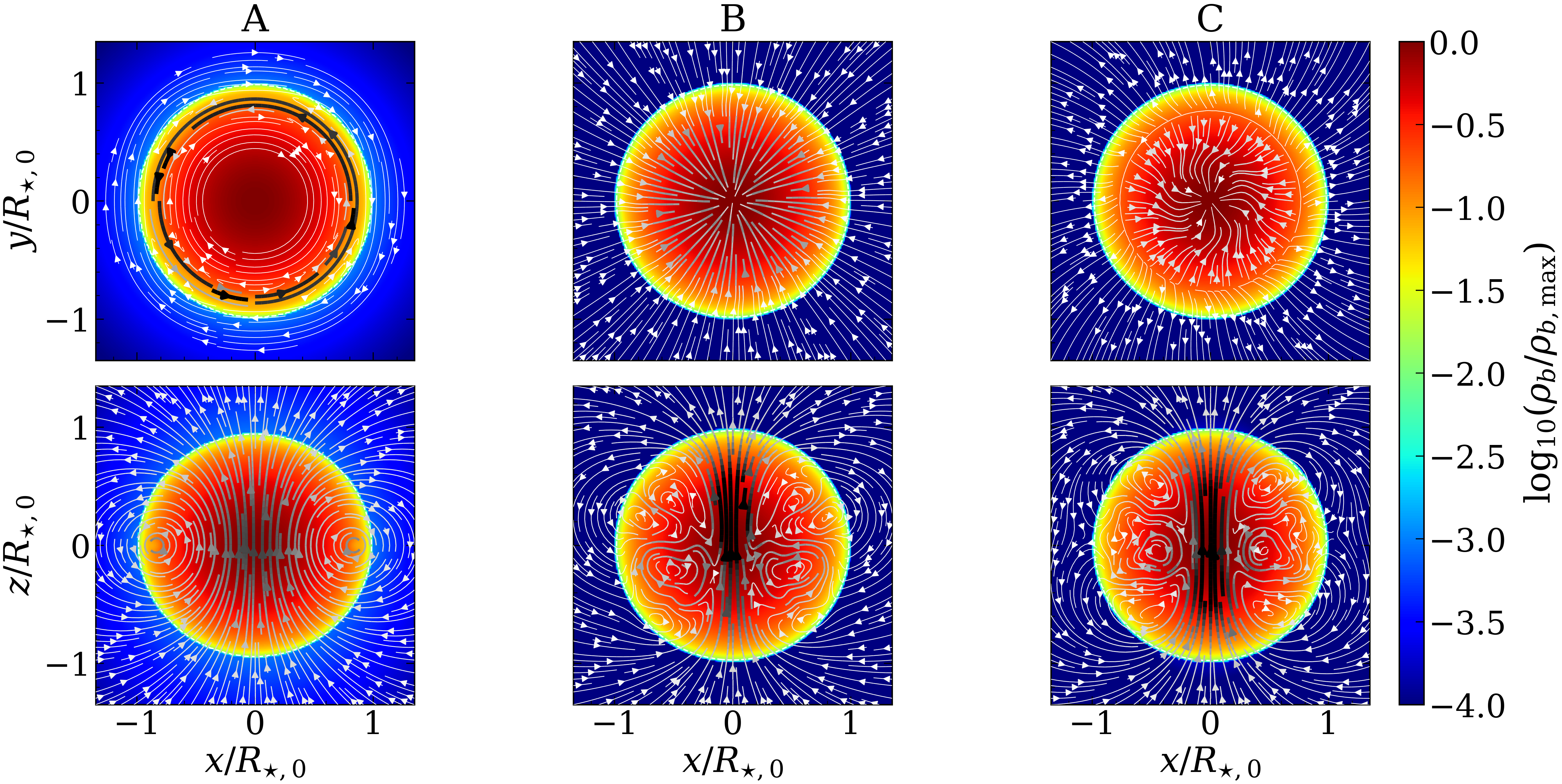}
\caption{Two-dimensional magnetic field lines of the initial equilibrium configurations. For each model, the upper panel shows the equatorial \(xy\) plane and the lower panel shows the meridional \(xz\) plane. Coordinates are normalized by the initial stellar radius \(R_{\star,0}\). }
\label{fig:Bfield_2d_t0}
\end{figure*}

The initial equilibria possess distinct magnetic field geometries to allow us to study their impact on the final configuration. Figure~\ref{fig:Bfield_3d_t0} shows the three-dimensional magnetic field structures of each model, while Fig.~\ref{fig:Bfield_2d_t0} shows the corresponding equatorial \(xy\) and meridional \(xz\) slices. Together, these figures establish the distinct large-scale magnetic geometries of the three initial equilibria. However, the equatorial and meridional slices are not exhaustive, because a localized azimuthal component away from the equatorial plane need not appear in the in-plane field lines displayed in Fig.~\ref{fig:Bfield_2d_t0}.

Model {\tt A} contains an axial poloidal backbone along with a spiral structure that carries the toroidal component, as shown by the concentric streamlines in its \(xy\) slice in Fig.~\ref{fig:Bfield_2d_t0}. 
The figure shows no clearly visible toroidal field-line structure in the equatorial slice for Model {\tt B}. However, off-equatorial slices (not shown) reveal coherent, azimuthal field components, which are seen as field line loops in the meridional slice of  Model {\tt B}. Model {\tt C} contains poloidal field lines, as well as spiral equatorial and off-equatorial magnetic field lines, which carry the toroidal magnetic field component.

The meridional two-dimensional slices reveal the corresponding poloidal structures more clearly. Model {\tt A} possesses a relatively simple large-scale axial poloidal backbone. Models {\tt B} and {\tt C} contain multiple organized poloidal cells surrounding a central axial field bundle. Higher-order initial multipolar structure is evident for models {\tt B} and {\tt C}.

To establish the multipolar content already present in the initial equilibria, we perform a VSH decomposition and mode-ranking procedure described in Sec.~\ref{sec:Diagnostics}. Figure~\ref{fig:vsh_poloidal_top4_t0_ABC} shows the four strongest initial poloidal modes for each model and provides the baseline needed to distinguish multipoles inherited from the initial data from those generated or amplified during the evolution.

The initial poloidal spectra are strongly model dependent and already contain resolved axisymmetric structure beyond $\ell=4$. In model {\tt A}, the four strongest modes are $(1,0)$, $(3,0)$, $(5,0)$, and $(7,0)$, forming a decreasing sequence of odd-$\ell$ axisymmetric contributions dominated by the dipole. Model {\tt B} has the ordering $(1,0)$, $(2,0)$, $(3,0)$, and $(5,0)$, while model {\tt C} has the ordering $(3,0)$, $(1,0)$, $(4,0)$, and $(5,0)$. Thus, the leading octupolar component of model {\tt C}, as well as several higher-order axisymmetric components in all three models, are inherited from the initial equilibria. We note that although for clarity we show the dominant modes for models {\tt A}, {\tt B}, and {\tt C} in Fig.~\ref{fig:vsh_poloidal_top4_t0_ABC} there are initial non-zero contributions from higher axisymmetric modes as well.

\subsubsection{Initial atmosphere}
Our evolution methods require a tenuous atmosphere in the stellar exterior.  We initialize this exterior atmosphere using the constant-plasma beta prescription of Ref.~\cite{Paschalidis:2014qra} to mimic a magnetosphere in the sense that it is magnetic-pressure dominated. After \(B^i\) has been initialized, we define
\begin{equation}
\beta\equiv\frac{P}{P_{\mathrm{mag}}}=\frac{2P}{b^2},
\qquad
P_{\mathrm{mag}}\equiv\frac{b^2}{2},
\end{equation}
where \(b^2\equiv g_{\mu\nu}b^\mu b^\nu\), and \(b^\mu/\sqrt{4\pi}\) is the magnetic field measured by an observer comoving with the fluid. We start by assigning a preliminary atmosphere with rest-mass density  \(\rho_{b,\mathrm{atm}}=10^{-14}\) in code units. The magnetized exterior initially satisfies \(\beta<\beta_0\) and is selected by the reset routine, whereas points with initial \(\beta\geq\beta_0\), including the stellar interior, are left unchanged. Here $\beta_0$ is the target plasma parameter $\beta$ of the atmosphere.

For \(\beta<\beta_0\), we first set the coordinate 3-velocity of the fluid \(v^i\equiv u^i/u^0\) to corotate with the star
\begin{equation}
v^x=-\Omega_0 y,
\qquad
v^y=\Omega_0 x,
\end{equation}
using the angular velocity \(\Omega_0\) of the corresponding initial stellar configuration, and \(v^z=0\). Past the light cylinder the fluid exceeds the speed of light, we therefore impose an upper limit on the Lorentz factor measured by a normal observer of \(W\leq2.29\), which corresponds to fluid velocity measure by a normal-observer of approximately $0.9c$. Next, holding \(B^i\) fixed, we  recompute \(b^2\) using the new atmospheric velocity and then reset the atmospheric pressure and density such that the atmospheric plasma beta matches $\beta_0$ as follows~\cite{Paschalidis:2014qra}
\begin{equation}
\begin{aligned}
\rho_b
&=
\max\!\left[
\rho_{b,\mathrm{atm}},
\left(
\frac{\beta_0b^2}{2\kappa_{\mathrm{eff}}}
\right)^{1/2}
\right],
\\
P
&=
\kappa_{\mathrm{eff}}\rho_b^2,
\qquad
\epsilon
=
\frac{P}{\rho_b},
\end{aligned}
\label{eq:atmosphere}
\end{equation}
where
\begin{equation}
\kappa_{\mathrm{eff}}
\equiv
K_{\mathrm{ext}}K,
\end{equation}
where $K$ is the polytropic constant of the equilibrium star. The parameter \(K_{\mathrm{ext}}\) controls the density assigned for a given magnetic pressure but does not set beta. In this work we set it to \(K_{\mathrm{ext}}=60\). Wherever the hard density floor is inactive, substitution into Eq.~\eqref{eq:atmosphere} gives
\begin{equation}
P=\frac{\beta_0b^2}{2},
\qquad
\beta=\beta_0=0.05,
\qquad
\frac{b^2}{2P}=20.
\end{equation}
Thus, $\kappa_{\rm eff}$ allows us to control the exterior density such that we do not add too much rest mass to the system.
This reset is applied only during initialization. During the subsequent ideal-GRMHD evolution we do not reimpose constant beta, but impose the floor rest-mass density $\rho_{b,\mathrm{atm}}$. The value of $\beta_0$ and $\kappa_{\rm eff}$ are chosen such that the atmosphere is magnetic pressure dominated, while the total rest mass changes by less  than \(1\%\) in each model. Thus, the construction provides an initially constant-beta (unless very far from the star), magnetic-pressure dominated atmosphere. We point out that this is not a force-free magnetosphere, because the magnetic field does not dominate over the inertia of the fluid. The way we set up the atmosphere allows for a more reliable evolution of the exterior magnetic fields. It has been shown that standard numerical GRMHD schemes cannot be reliable when the magnetization exceeds values $\mathcal{O}(10)$~\cite{Duez:2005sf}. The atmosphere in our simulations is initialized with a magnetization $b^2/{\rho_b}\lesssim 0.1$. In our evolutions the magnetization remains of that order and safely below unity without needing to impose any ceilings on $b^2/{\rho_b}$ as in~\cite{Capobianco:2026ots}. This does not imply that the atmosphere in our simulations is a realistic magnetosphere, only that our atmosphere is less affected by failures due to extremely high magnetization.

\subsection{Evolution}

We evolve the magnetized neutron-star initial data in full general relativity using the Einstein Toolkit~\cite{Loffler:2011ay,Zilhao:2013hia}. The spacetime is evolved with the {\tt McLachlan} thorn in the Baumgarte--Shapiro--Shibata--Nakamura formulation~\cite{Shibata:1995we,Baumgarte:1998te,Brown:2008sb}. The ideal-GRMHD equations are evolved in conservative form with {\tt IllinoisGRMHD}~\cite{Etienne:2015cea}. This is the same Einstein Toolkit/{\tt McLachlan}/{\tt IllinoisGRMHD} framework used in ~\cite{Etienne:2015cea}. 

We use fourth-order finite differencing for the spacetime variables and fourth-order Runge--Kutta time integration through the {\tt MoL} thorn. The computational domain is covered by nine nested Cartesian refinement levels provided by {\tt Carpet}, with fifth-order spatial prolongation and second-order time prolongation between levels. The finest-grid spacing adopted for models {\tt A}, {\tt B}, and {\tt C} is \(\Delta x_{\min}=0.10\), \(0.08\), and \(0.08\), respectively. These correspond to approximately \(140\), \(156\), and \(147\) zones across the initial equatorial diameter. The Outer boundary is located at \(\approx 400R_{\star,0}\) for all models. To assess resolution dependence, we additionally evolve model {\tt C} with \(\Delta x_{\min}=0.10\), corresponding to approximately \(118\) finest-grid spacings across the initial stellar equatorial diameter.

We adopt a non-advective moving-puncture-type gauge. The lapse obeys
\begin{equation}
  \partial_t \alpha = -\alpha K ,
  \label{eq:lapse_condition}
\end{equation}
where $\alpha$ is the lapse and $K$ is the trace of the extrinsic curvature. The shift is evolved using the non-advective Gamma-driver condition in {\tt McLachlan}, with damping parameter
$
  \eta = 0.75/M_\odot$.

The lapse and shift choices belong to the same gauge family as the singularity-avoiding lapse and Gamma-driver shift conditions commonly used in moving-puncture compact-object evolutions~\cite{Alcubierre:2002kk,Campanelli:2005dd,Baker:2005vv,vanMeter:2006vi}.

We adopt a $\Gamma$-law equation of state for the evolution, thereby not enforcing the initial cold, barotropic equation of state throughout the evolution, which could preclude the existence of stable magnetic equilibria~\cite{2015MNRAS.447.1213M,2022MNRAS.517..560B}. The magnetic field is evolved through a vector potential formulation in the generalized Lorenz gauge~\cite{Farris:2012ux,Etienne:2012te}, which damps electromagnetic gauge modes near mesh-refinement boundaries. The generalized Lorenz-gauge damping parameter is set to
  $\xi = 9.0/M_\odot$.

\subsection{Diagnostics}\label{sec:Diagnostics}

We monitor the Hamiltonian and momentum constraints, the maximum rest-mass density, the baryonic rest mass, the ADM mass, and the ADM angular momentum. The constraint norms remain bounded during the evolutions. For the representative model {\tt B}, the ADM mass changes by less than \(0.1\%\) over the full run, while the ADM angular momentum is conserved at approximately \(2\%\). Similar behavior is found for the other models. For models {\tt A}, {\tt B}, and {\tt C}, respectively, the maximum value of dimensionless Hamiltonian constraint late in the evolution (at \(20t_A\)) \(\max_{\mathbf{x}}|H(t,\mathbf{x})|\,M_{\rm ADM}^{2}(t)\)
are \(7.25\times10^{-3}\), \(2.37\times10^{-2}\), and
\(1.51\times10^{-2}\) for models A, B, C respectively. These diagnostics indicate that the spacetime and matter errors in our long-term simulations are sufficiently well controlled.


\subsubsection{Vector spherical harmonics}

To characterize the angular structure of the magnetic field, we perform a VSH decomposition of the spatial vector potential \(A_i\) and to the densitized magnetic field \(\tilde B^i=\sqrt{\gamma}B^i\), where \(B^i\) is the magnetic field measured by a normal observer and \(\gamma\) is the determinant of the spatial metric. The reason for using the densitized magnetic field, is that there is a straightforward test for checking that the VSH decomposition preserves its divergence-free character, unlike in the case of $B^i$ (see App.~\ref{App:VSH} for more details).

All magnetic field amplitudes reported below are obtained from the decomposition of \(\tilde B^i\).

For every selected snapshot, we define the instantaneous "Newtonian" coordinate center of mass \(\mathbf{x}_{\rm c}(t)\) as
\begin{widetext}
\begin{equation}
\mathbf{x}_{\rm c}(t)
=
\frac{
\displaystyle
\int_{\mathcal{D}(t)}
\rho_bW\sqrt{\gamma}\,\mathbf{x}\,d^3x
}{
\displaystyle
\int_{\mathcal{D}(t)}
\rho_bW\sqrt{\gamma}\,d^3x
},
\qquad
\mathcal{D}(t)
=
\left\{
\mathbf{x}:
\rho_b\geq10^{-3}\rho_{b,\max}
\right\}.
\label{eq:vsh_coordinate_com}
\end{equation}
\end{widetext}
All VSH extraction spheres in our analysis are centered on \(\mathbf{x}_{\rm c}(t)\) by introducing translated coordinates
\begin{equation}
\mathbf{x}'
=
\mathbf{x}-\mathbf{x}_{\rm c}(t),
\qquad
r
=
|\mathbf{x}'|,
\qquad
\hat{\mathbf r}
=
\frac{\mathbf{x}'}{r}.
\label{eq:vsh_centered_coordinates}
\end{equation}
The \(z\) direction remains aligned with the initial rotation axis. The centered radial coordinate \(r\) is normalized by the fixed initial equatorial coordinate radius \(R_{\star,0}\).

Given the scalar spherical harmonics $Y_{\ell m}$, we use the vector spherical harmonic basis
\begin{align}
\mathbf{Y}_{\ell m} &= Y_{\ell m}\hat{\mathbf r}, \\
\mathbf{\Psi}_{\ell m} &= \nabla_{\Omega}Y_{\ell m}, \\
\mathbf{\Phi}_{\ell m} &= \hat{\mathbf r}\times\nabla_{\Omega}Y_{\ell m},
\end{align}
where $\nabla_{\Omega}$ is the angular derivative on the unit sphere. The densitized magnetic field is expanded as
\begin{widetext}
\begin{equation}
\tilde{\mathbf B}(r,\theta,\phi)=\sum_{\ell=1}^{\ell_{\max}}\sum_{m=-\ell}^{\ell}\left[a_{\ell m}(r)\mathbf{Y}_{\ell m}+b_{\ell m}(r)\mathbf{\Psi}_{\ell m}+c_{\ell m}(r)\mathbf{\Phi}_{\ell m}\right].
\end{equation}
\end{widetext}
The poloidal and toroidal parts of the densitized magnetic field are therefore reconstructed as
\begin{align}
\tilde{\mathbf B}_{\rm pol} &= \sum_{\ell=1}^{\ell_{\max}}\sum_{m=-\ell}^{\ell}\left[a_{\ell m}\mathbf{Y}_{\ell m}+b_{\ell m}\mathbf{\Psi}_{\ell m}\right], \\
\tilde{\mathbf B}_{\rm tor} &= \sum_{\ell=1}^{\ell_{\max}}\sum_{m=-\ell}^{\ell}c_{\ell m}\mathbf{\Phi}_{\ell m}.
\label{eq:vsh_poloidal_toroidal_fields}
\end{align}
Thus, $a_{\ell m}$ and $b_{\ell m}$ determine the poloidal contribution, whereas $c_{\ell m}$ determines the toroidal contribution. We note that this decomposition is not gauge invariant.

For each extraction sphere of radius \(r\) on which the vector spherical harmonic decomposition is performed, the Cartesian components of the vector-potential, magnetic field, and spatial-metric data are interpolated via trilinear interpolation using the finest mesh available at the interpolation point. The angular grid  on the sphere has \(N_\theta=64\) Gauss--Legendre points, \(\mu_j=\cos\theta_j\), and \(N_\phi=128\) uniformly spaced points in \(\phi\).  The radial domain \(0.08\leq r/R_{\star,0}\leq1.2\) is divided into 16 elements, and the radial dependence of each VSH coefficient is represented by a Legendre discontinuous-Galerkin expansion of order seven. All VSH diagnostics retain modes through \(\ell_{\max}=63\). In in App.~\ref{App:VSH_Lmax_accuracy} we discuss the error budget from the magnetic field reconstruction using the VSH and radial DG decomposition.

For \(q\in\{{\rm pol},{\rm tor}\}\), let
\(\boldsymbol{\mathcal{B}}_{\ell m}^{(q)}\) denote the generally complex contribution of the single VSH index \((\ell,m)\) to the magnetic field \(B^i\), which is real. To obtain a real vector the \(+m\) and \(-m\) contributions must be added up. We therefore define one real contribution for each nonnegative value of \(m\) by
\begin{equation}
\mathbf{B}_{\ell m}^{(q)}
=
\begin{cases}
\boldsymbol{\mathcal{B}}_{\ell 0}^{(q)},
& m=0, \\[1mm]
\boldsymbol{\mathcal{B}}_{\ell m}^{(q)}
+
\boldsymbol{\mathcal{B}}_{\ell,-m}^{(q)},
& 1\leq m\leq\ell .
\end{cases}
\label{eq:real_vsh_sector}
\end{equation}
The nonnegative index \(m\) in Eq.~\eqref{eq:real_vsh_sector} is the \(|m|\) label displayed in the figures. Thus, a displayed contribution with \(m>0\) contains both the \(+m\) and \(-m\) complex VSH contributions.

For $1\leq\ell\leq\ell_{\max}$, the total number of signed $(\ell,m)$ VSH contributions is
\begin{equation}
N(\ell_{\max})=\sum_{\ell=1}^{\ell_{\max}}(2\ell+1)=\ell_{\max}(\ell_{\max}+2).
\end{equation}

According to Eq.~\eqref{eq:real_vsh_sector}, each pair of contributions with $+m$ and $-m$, for $m>0$, is combined into one real contribution. Including the $m=0$ contribution therefore gives $\ell+1$ displayed contributions for each $\ell$, and hence
\begin{equation}
N_{\rm displayed}(\ell_{\max})=\sum_{\ell=1}^{\ell_{\max}}(\ell+1)=\frac{\ell_{\max}(\ell_{\max}+3)}{2}.
\end{equation}
For $\ell_{\max}=63$, the decomposition retains all $N(63)=4095$ total contributions, which are combined into $N_{\rm displayed}(63)=2079$ real contributions labeled by nonnegative $m$.

We characterize the amplitude of each real contribution using the solid-angle root-mean-square
\begin{equation}
\begin{aligned}
B^{(q)}_{\ell m}(r)
&=
\left[
\frac{1}{4\pi}
\int_{S^2}
\gamma_{ij}
B^{i,(q)}_{\ell m}(r,\theta,\phi)
B^{j,(q)}_{\ell m}(r,\theta,\phi)
\,d\Omega
\right]^{1/2}.
\end{aligned}
\label{eq:vsh_sector_rms}
\end{equation}

The angular integral is evaluated with the same quadrature used for the VSH projection. The total poloidal or toroidal amplitude is obtained by summing all retained contributions of the corresponding field before applying the same solid-angle root-mean-square operation. The VSH basis definitions, projection formulae, analytic tests, consistency checks, and direct accuracy test of the \(\ell_{\max}=63\) reconstruction are presented in Appendix~\ref{App:VSH}.

\subsubsection{Four-current}

In ideal magnetohydrodynamics, the electric field is determined by the magnetic field and the fluid four-velocity, so the inhomogeneous Maxwell equation can be used as a diagnostic to compute the four-current. In Heaviside--Lorentz units, the four-current is given by
\begin{equation} 
J^\mu=\nabla_\nu F^{\mu\nu}, 
\end{equation} 
where \(J^\mu\) is the four-current, \(F^{\mu\nu}=-F^{\nu\mu}\) is the Faraday tensor, and \(\nabla_\nu\) is the covariant derivative associated with the spacetime metric \(g_{\mu\nu}\). Antisymmetry of \(F^{\mu\nu}\) gives 
\begin{equation}
\label{eq:4curr} 
J^\mu=\frac{1}{\sqrt{-g}}\left[\partial_t\!\left(\sqrt{-g}F^{\mu t}\right)+\partial_i\!\left(\sqrt{-g}F^{\mu i}\right)\right], 
\end{equation} 
where \(g\equiv\det(g_{\mu\nu})\) is the determinant of the spacetime metric and \(\partial_t\) and \(\partial_i\) denote partial derivatives. In the \(3+1\) decomposition the metric is given by
\begin{equation} 
ds^2=-\alpha^2dt^2+\gamma_{ij}(dx^i+\beta^i dt)(dx^j+\beta^jdt), 
\end{equation} 
where \(\alpha\) is the lapse, \(\beta^i\) is the shift, and \(\gamma_{ij}\) is the spatial metric. Defining \(\gamma\equiv\det(\gamma_{ij})\), one has \(\sqrt{-g}=\alpha\sqrt{\gamma}\), and the future-directed unit normal to each spatial slice is \(n^\mu=\alpha^{-1}(1,-\beta^i)\). The electric and magnetic fields \(E^\mu\) and \(B^\mu\) measured by the normal observer satisfy \(n_\mu E^\mu=n_\mu B^\mu=0\). The coordinate three-velocity is \(v^i=u^i/u^t\), where \(u^\mu\) is the fluid four-velocity, is converted to the three-velocity measured by a normal-observer  as \(v_\perp^i=(v^i+\beta^i)/\alpha\). Then under the conventions of the {\tt IllinoisGRMHD} code, the ideal-MHD condition gives 
\begin{equation} 
E_i=\epsilon_{ijk}v_\perp^jB^k,\qquad \epsilon_{ijk}=\sqrt{\gamma}\,[ijk], 
\end{equation} 
where \(\epsilon_{ijk}\) is the spatial Levi-Civita tensor and \([ijk]\) is the fully antisymmetric symbol with \([123]=+1\). The Faraday tensor is reconstructed as \begin{equation} 
F^{\mu\nu}=n^\mu E^\nu-n^\nu E^\mu-\epsilon^{\mu\nu\rho\sigma}B_\rho n_\sigma, \end{equation} 
where \(\epsilon_{\mu\nu\rho\sigma}\) is the spacetime Levi-Civita tensor. Spacetime tensor indices are raised and lowered with \(g_{\mu\nu}\), while purely spatial tensor indices are raised and lowered with \(\gamma_{ij}\). We then form \(\sqrt{-g}F^{\mu\nu}\). The time derivative in Eq.~\eqref{eq:4curr} is evaluated with a first-order backward difference, while the spatial derivatives use second-order centered differences. Appendix~\ref{App:4curr} presents an analytic test showing that our implementation recovers the correct four-current and converges at the expected order.

\begin{figure}[!t]
\centering
\includegraphics[width=\linewidth]{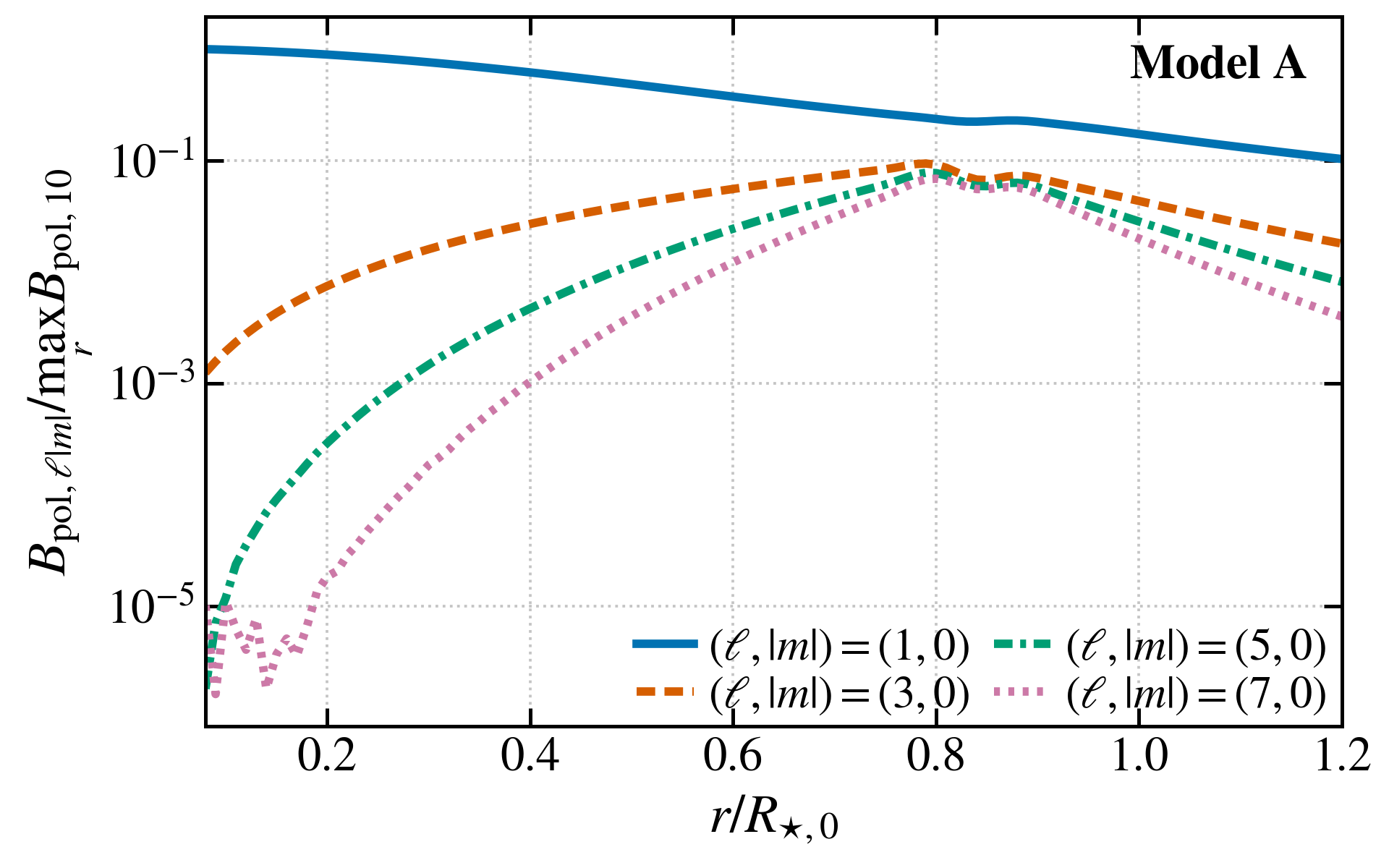}
\includegraphics[width=\linewidth]{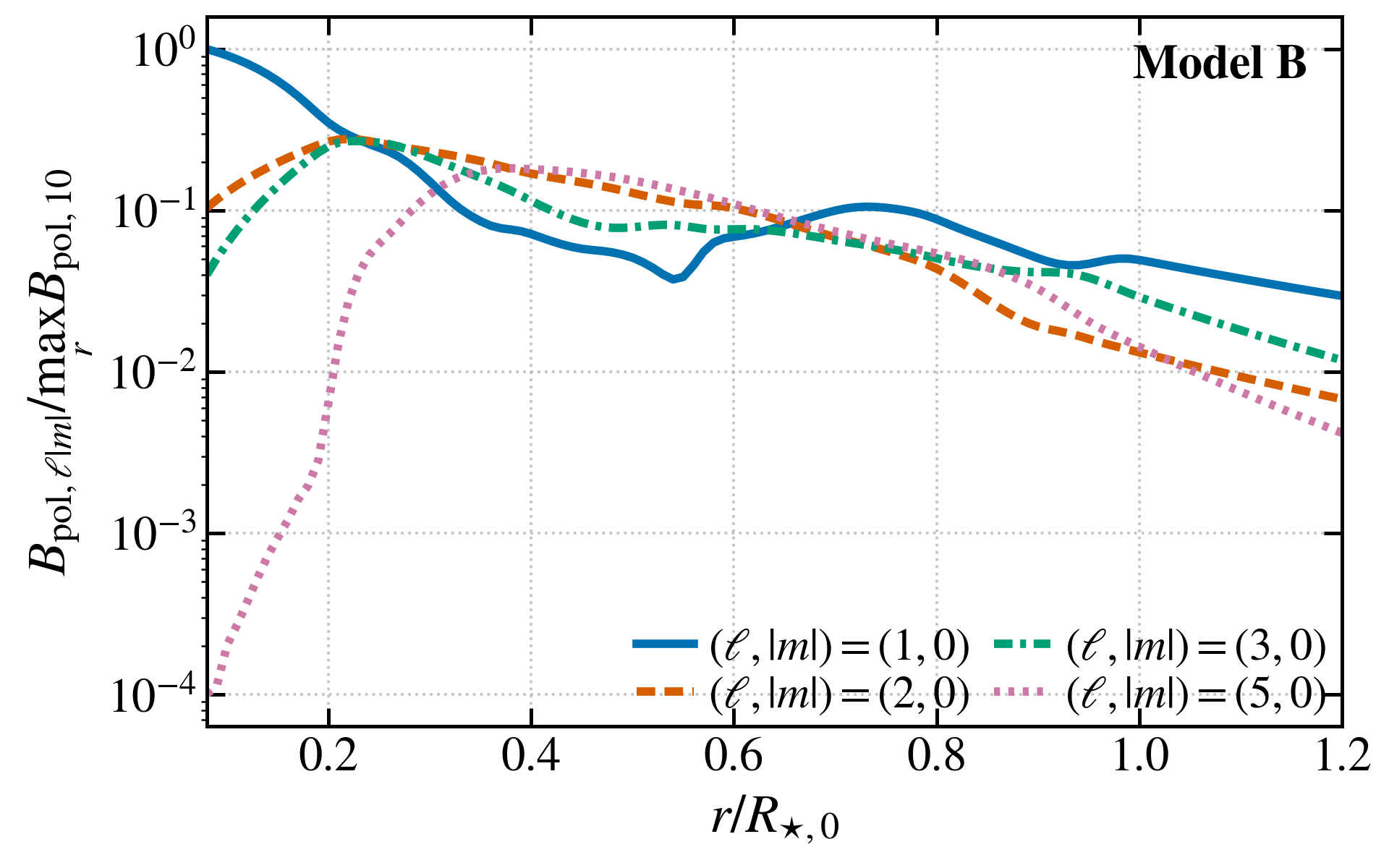}
\includegraphics[width=\linewidth]{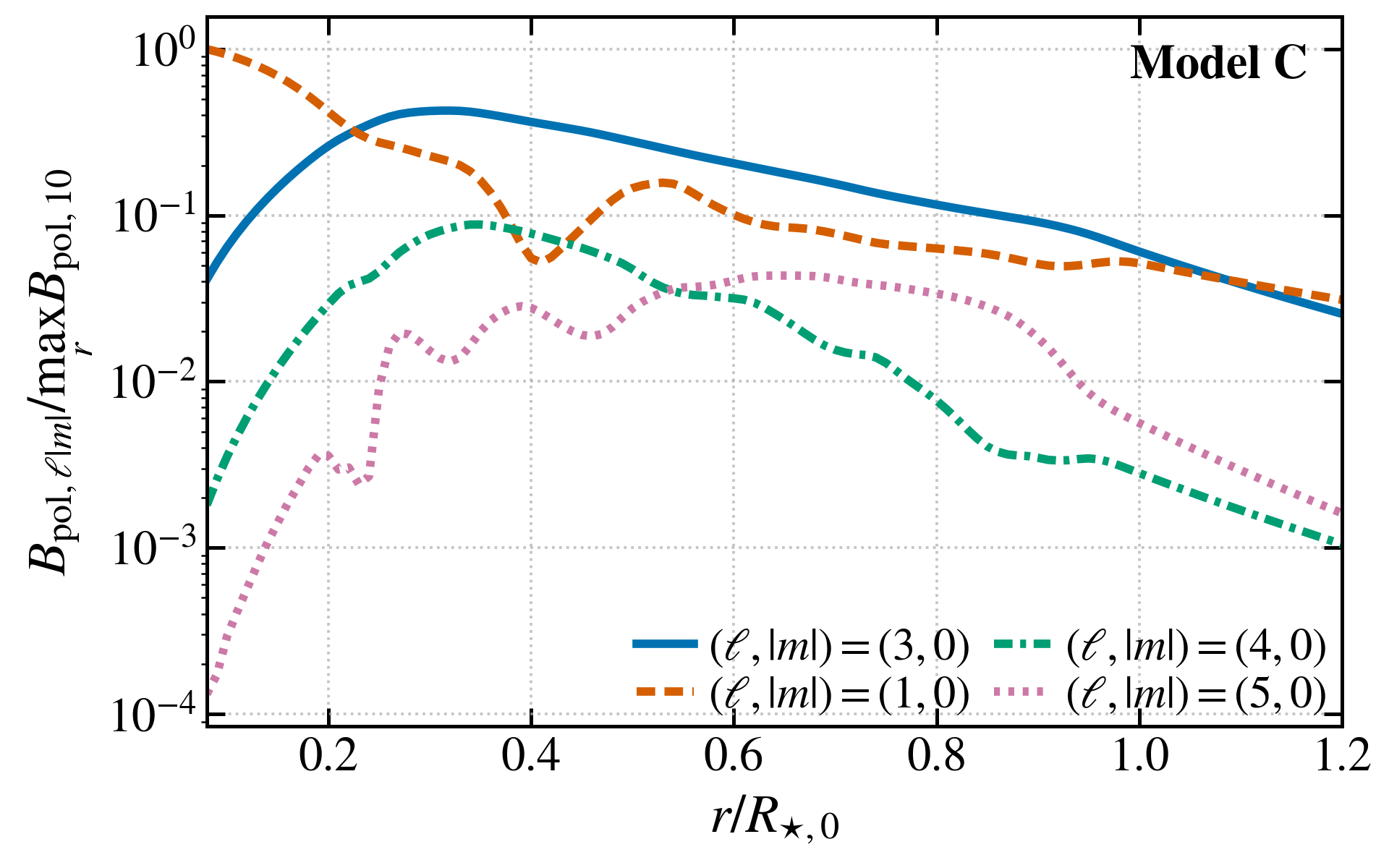}
\caption{
Initial poloidal VSH amplitudes at \(t=0\), computed on coordinate spheres centered on the center of mass. Each curve is normalized by \(\max_r B^{({\rm pol})}_{10}(r)\) for the same model.
The panels from top to bottom show models {\tt A}, {\tt B}, and {\tt C}. }
\label{fig:vsh_poloidal_top4_t0_ABC}
\end{figure}

\section{Results}\label{sec:results}

\begin{figure*}[!t]
\centering
\includegraphics[width=0.46\textwidth]{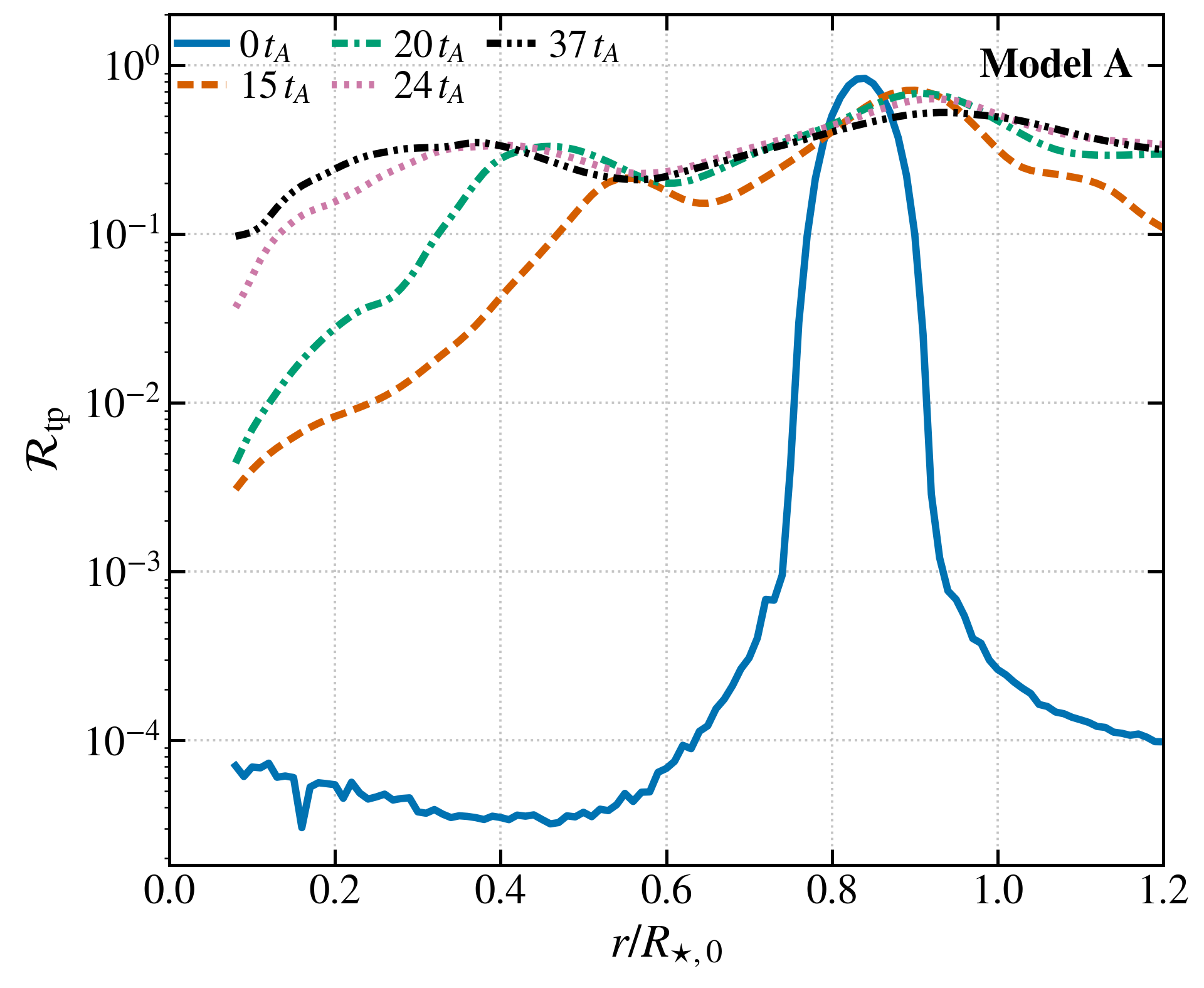}\hfill
\includegraphics[width=0.46\textwidth]{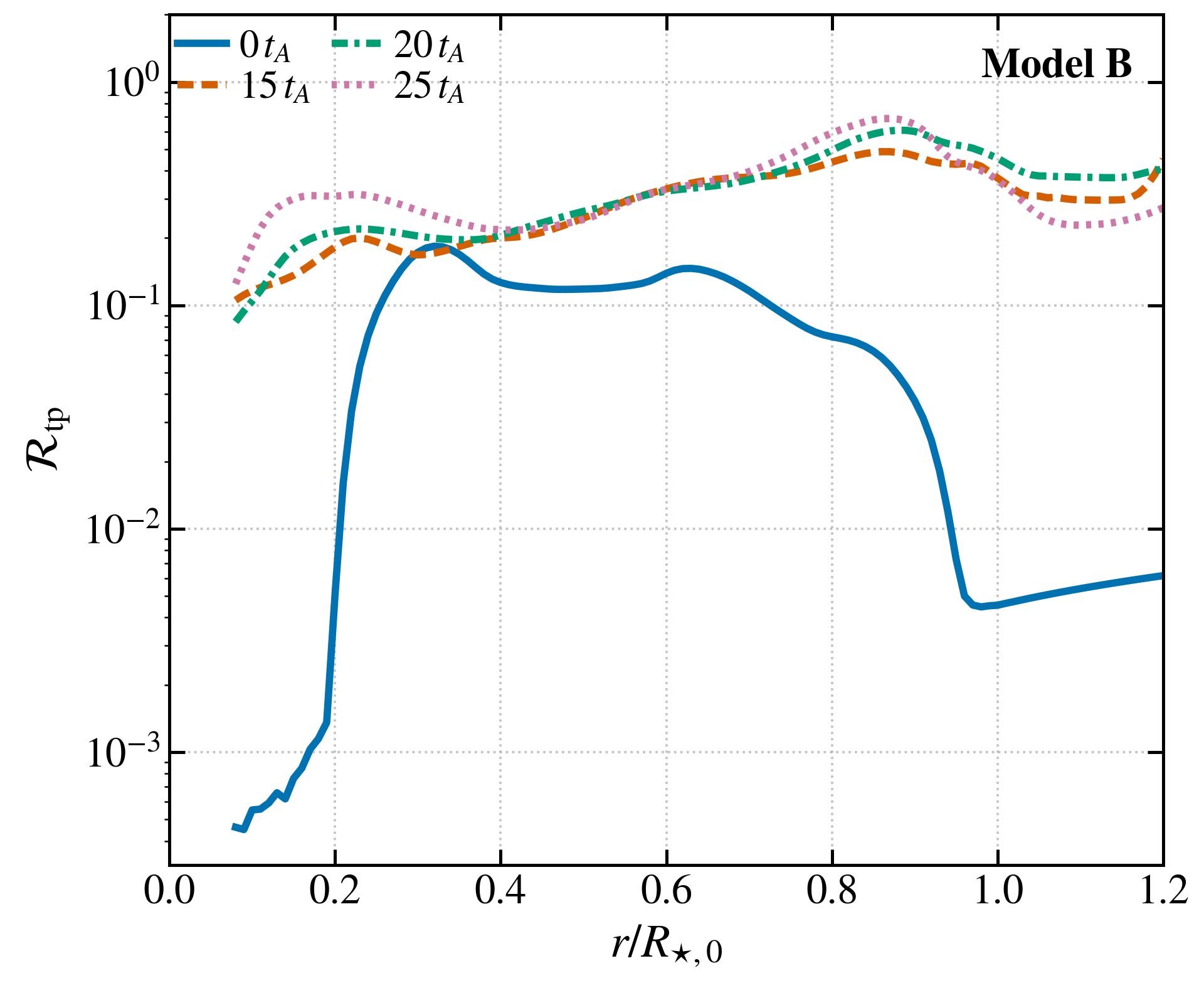}
\par\medskip
\includegraphics[width=0.46\textwidth]{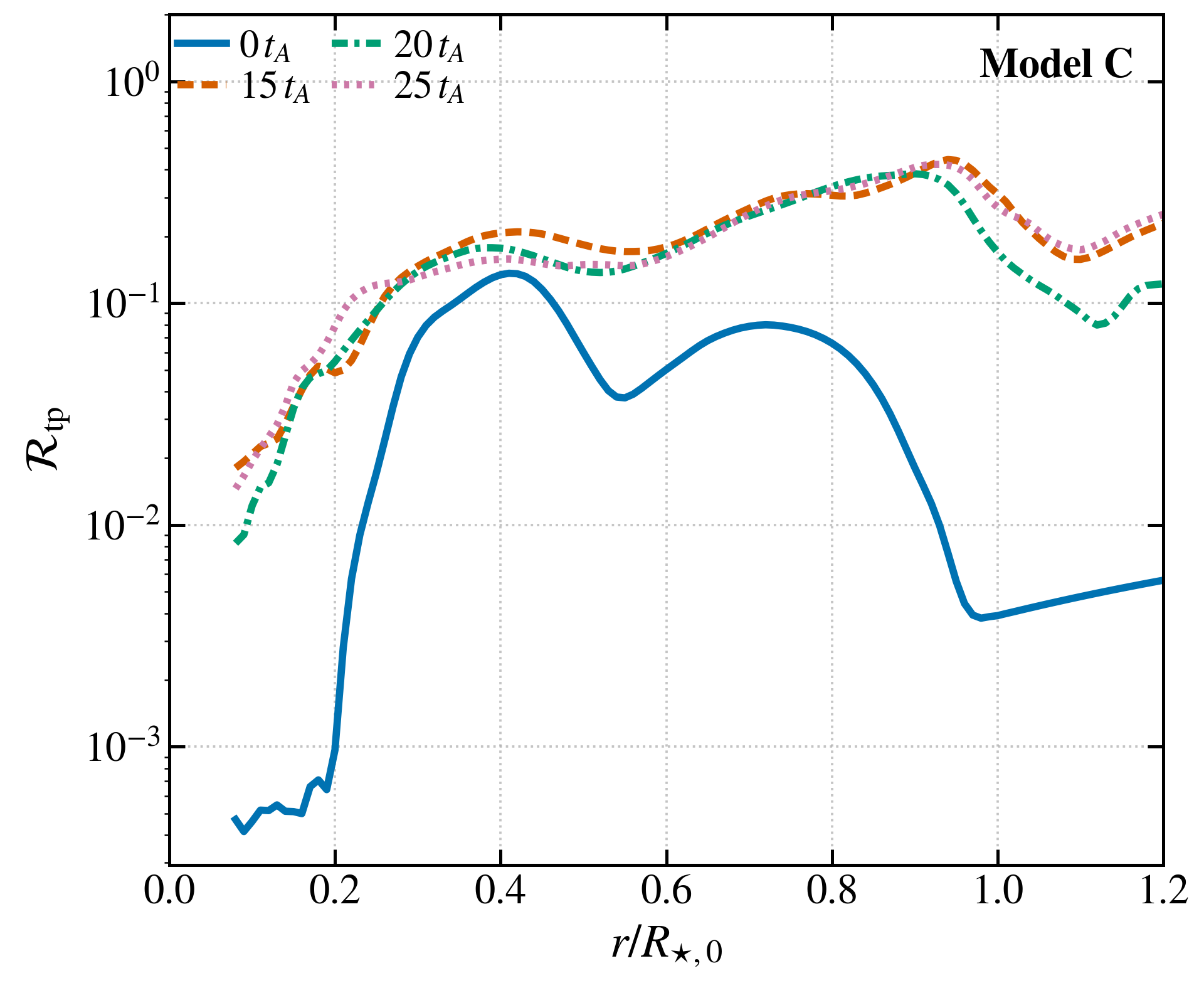}\hfill
\includegraphics[width=0.46\textwidth]{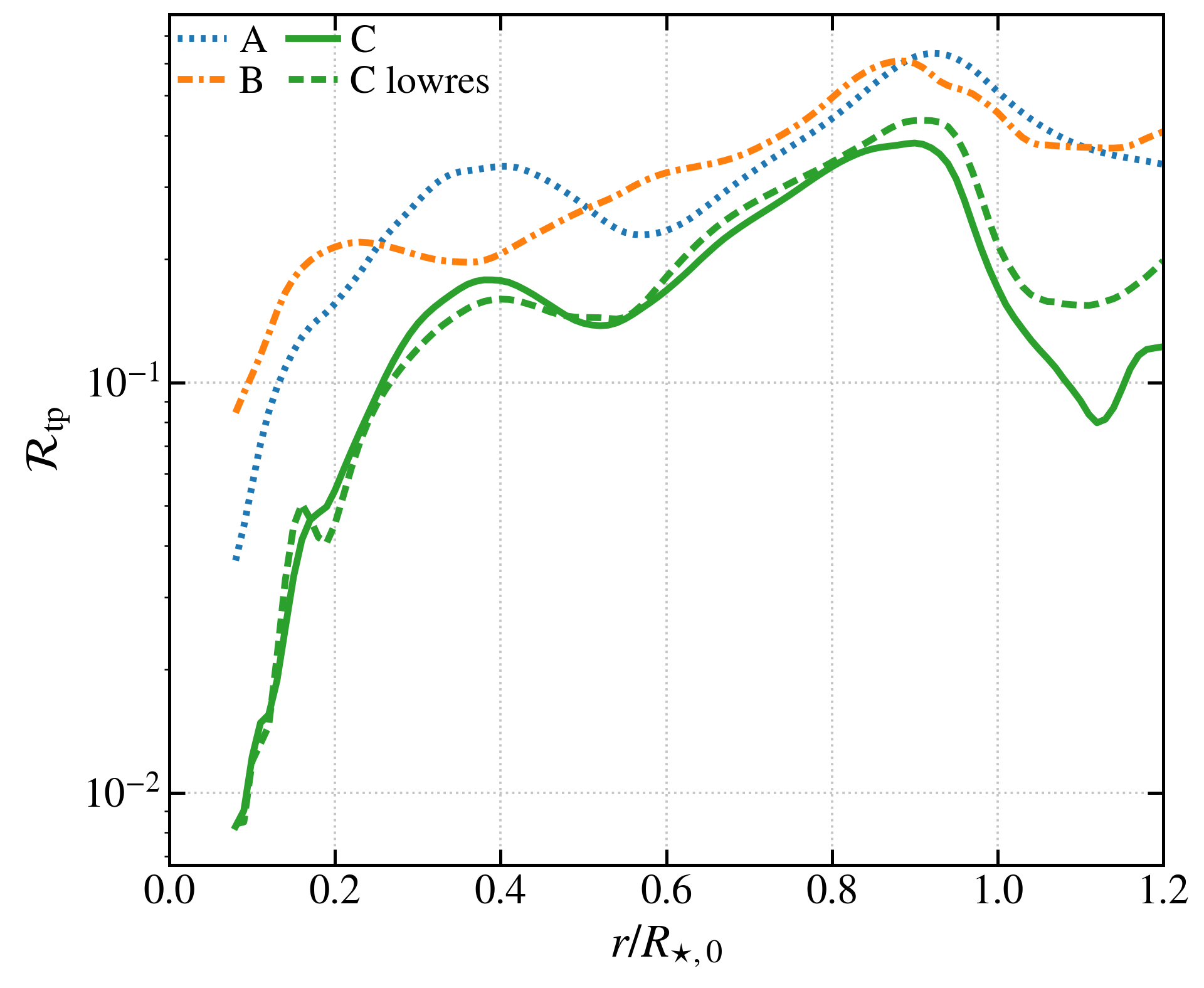}
\caption{
Evolution and cross-model comparison of the toroidal-to-poloidal magnetic field amplitude ratio
\(\mathcal{R}_{\rm tp}=B_{\rm tor}^{\rm rms}/B_{\rm pol}^{\rm rms}\), defined in Eq.~\eqref{eq:vsh_total_tor_pol_ratio}. The total poloidal and toroidal fields are reconstructed from all VSH contributions with \(1\leq\ell\leq63\).
The horizontal axis is \(r/R_{\star,0}\), where \(r=|\mathbf{x}-\mathbf{x}_{\rm c}(t)|\), \(\mathbf{x}_{\rm c}(t)\) is the center of mass defined in Eq.~\eqref{eq:vsh_coordinate_com}, and \(R_{\star,0}\) is the initial equatorial coordinate radius.
The upper-left panel shows model {\tt A} at \(0t_A\), \(15t_A\), \(20t_A\), \(24t_A\), and \(37t_A\). The upper-right and lower-left panels show models {\tt B} and {\tt C}, respectively, near \(0t_A\), \(15t_A\), \(20t_A\), and \(25t_A\). The lower-right panel compares model {\tt A} near \(24t_A\) with models {\tt B} and {\tt C} near \(20t_A\); the lower-resolution realization of model {\tt C} is included only as a resolution comparison. The lower-right panel indicates a limited degree of universality.
}
\label{fig:settled_vsh_profiles}
\end{figure*}

\subsection{Quasistationarity and quasi-universality}
\label{sec:settled_criterion}

The total duration of each evolution is listed in the last column of Table~\ref{tab:initial_data} in units of the central Alfv\'en time \(t_{A,c}\). We evolve each model through its initial magnetic field rearrangement, and continue the evolution until the dominant large-scale magnetic-field geometry becomes approximately quasistationary. Diagnostic quantities for each configuration are reported within the resulting settled portion of each model.

We use the late-time stabilization of the radial toroidal-to-poloidal amplitude ratio as an operational diagnostic of magnetic field quasistationarity. We regard a configuration as approximately quasistationary when the broad radial form of this profile changes only weakly between snapshots separated by several Alfv\'en times. At every selected time, the extraction spheres are centered independently on the center of mass \(\mathbf{x}_{\rm c}(t)\) defined in Eq.~\eqref{eq:vsh_coordinate_com}. We reconstruct the total poloidal and toroidal magnetic fields from all VSH modes with \(1\leq\ell\leq63\). For \(q\in\{{\rm pol},{\rm tor}\}\), we then compute the solid-angle root-mean-square amplitudes and form
\begin{align}
B_{(q)}^{\rm rms}(r,t)
&=
\left[
\frac{1}{4\pi}
\int_{S^2}
\gamma_{ij}B_{(q)}^i(r,\theta,\phi,t)
B_{(q)}^j(r,\theta,\phi,t)
\,d\Omega
\right]^{1/2},
\\
\mathcal{R}_{\rm tp}(r,t)
&=
\frac{B_{\rm tor}^{\rm rms}(r,t)}
{B_{\rm pol}^{\rm rms}(r,t)}.
\label{eq:vsh_total_tor_pol_ratio}
\end{align}
with \(\mathcal{R}_{\rm tp}\) the ratio of toroidal to poloidal amplitudes.

Figure~\ref{fig:settled_vsh_profiles} shows that the initial toroidal-to-poloidal profile differs substantially from the profiles obtained after magnetic rearrangement for all models. For models {\tt B} and {\tt C}, the profiles near \(20t_A\) and \(25t_A\) have similar broad radial behavior over most of the stellar interior. Model {\tt A} exhibits more noticeable late-time evolution, particularly in the inner region, and therefore requires a longer evolution to relax. Its profile near \(24t_A\) is appreciably closer to the profile near \(37t_A\). We therefore use \(24t_A\) for model {\tt A} in the subsequent late-time comparisons and describe the configurations as approximately quasistationary.

The lower-right panel of Fig.~\ref{fig:settled_vsh_profiles} compares the relaxed models. The lower-resolution realization of model {\tt C} is also shown, which reproduces well the broad radial behavior of the higher-resolution case, supporting the robustness of our results with resolution.

Despite the different initial magnetization and magnetic-field geometries, all three models are poloidally dominated as demonstrated by their \(\mathcal{R}_{\rm tp}\) profiles. The radial profile of \(\mathcal{R}_{\rm tp}\) exhibits broad similarities among the different models with values of  order \(20\)--\(40\%\) in the bulk of the star. These values correspond to toroidal-to-poloidal magnetic field energy ratio is approximately of \(4\)--\(16\%\), because the energy scales with the square of the magnetic field amplitude. The above suggest some limited degree of universality in the coarse radial poloidal--toroidal balance.

This limited universality concerns the radial locations and overall sequence of the principal features, rather than their precise profile and values of \(\mathcal{R}_{\rm tp}\). Moreover, \(\mathcal{R}_{\rm tp}\) compresses the complete three-dimensional magnetic field into a single scalar at each radius and therefore contains no information about the individual \((\ell,m)\) contributions or the field-line topology. Configurations with similar double-peaked profiles can consequently possess different multipolar structures, as demonstrated below by the VSH decomposition and magnetic field-line visualizations.

\begin{figure*}[!t]
\centering
\includegraphics[width=0.99\textwidth]{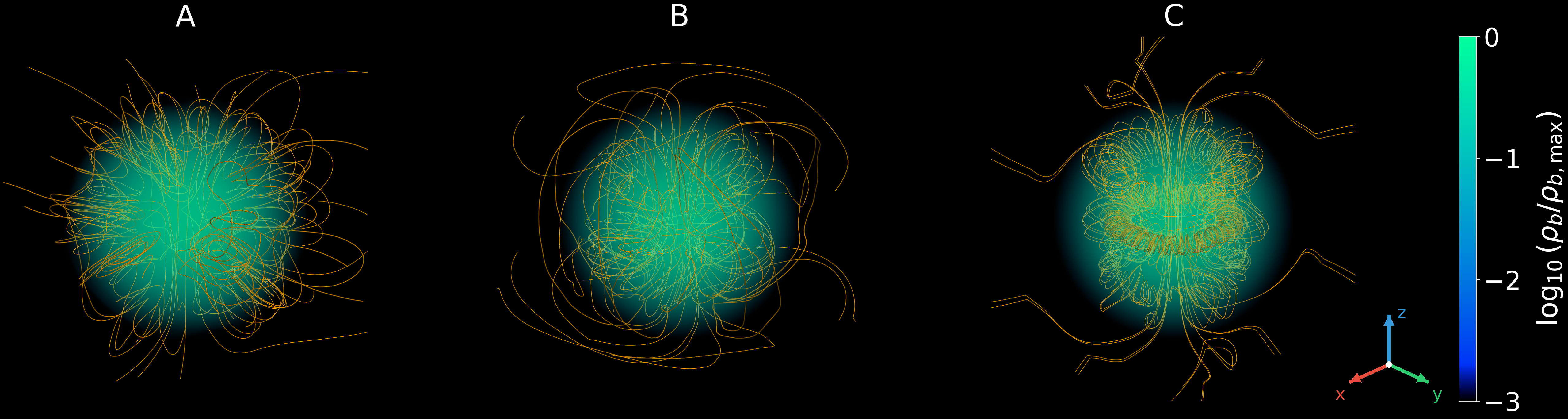}
\caption{Three-dimensional magnetic field lines of the settled configurations along with a volume rendering of the rest-mass density colored by $\log_{10}(\rho_b/\rho_{b,\max})$. Model {\tt A} is shown near $24t_A$, while models {\tt B} and {\tt C} are shown near $20t_A$. The orientation triad identifies the Cartesian coordinate directions. The selected field lines reproduce the principal field-line families seen in the corresponding two-dimensional slices in Fig.~\ref{fig:Bfield_2d_evolved}.}
\label{fig:Bfield_3d_evolved}
\end{figure*}

\begin{figure*}[!t]
\centering
\includegraphics[width=0.99\textwidth]{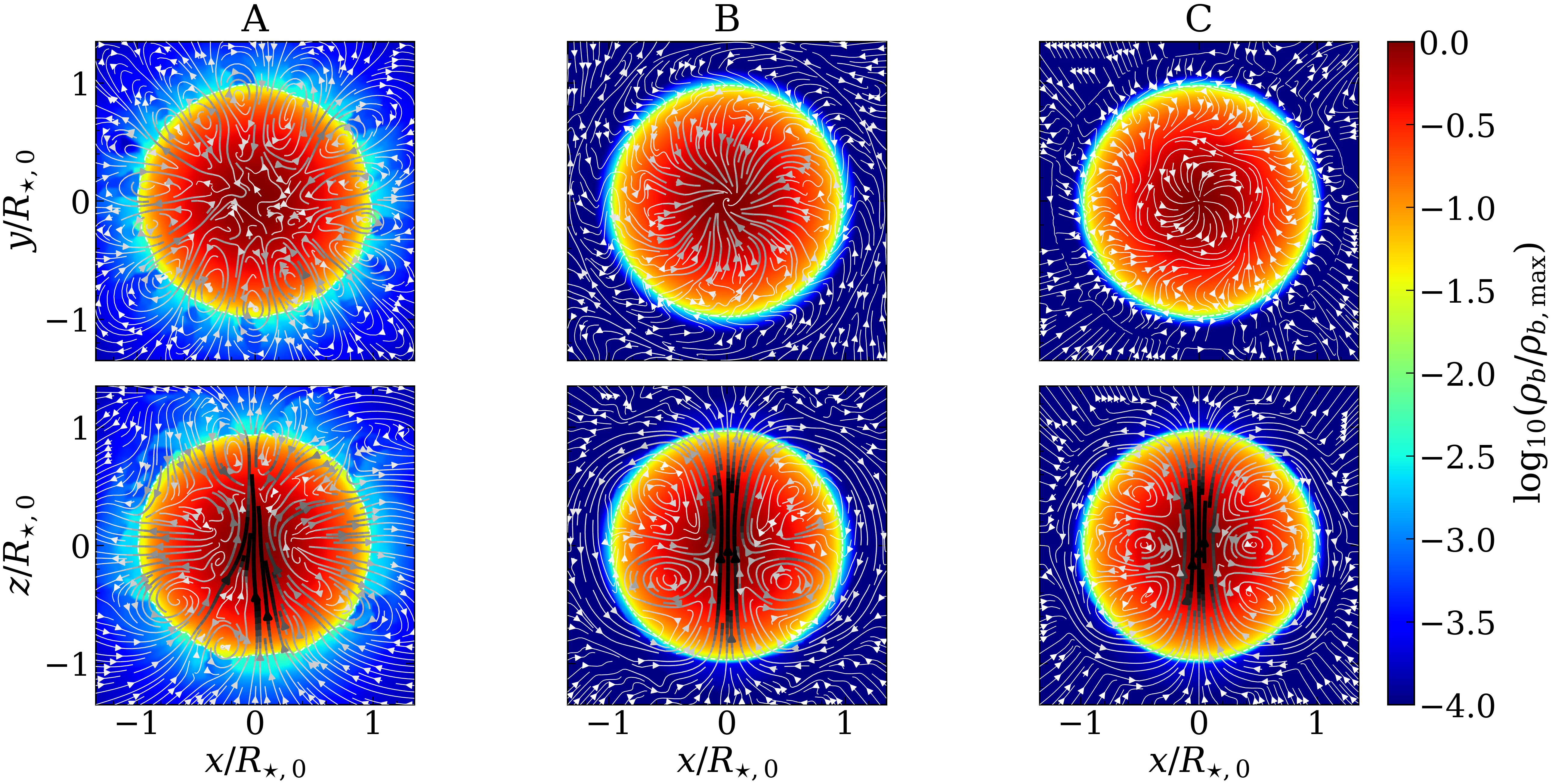}
\caption{Two-dimensional magnetic field structure of the settled configurations. Model {\tt A} is shown near \(24t_A\), while models {\tt B} and {\tt C} are shown near \(20t_A\). The upper row shows the equatorial \(xy\) plane, while the lower row shows the meridional \(xz\) plane. The rest-mass density is plotted as a color map of \(\log_{10}(\rho_b/\rho_{b,\max})\). The color of the field lines indicates the relative in-plane magnetic field amplitude.}

\label{fig:Bfield_2d_evolved}
\end{figure*}

\subsection{Two- and Three-dimensional settled magnetic fields}
\label{sec:Bfield_3D}

In this section, we present representative two- and three-dimensional visualizations of the magnetic field together with a volume rendering of the rest-mass density. 

Figure~\ref{fig:Bfield_3d_evolved} shows the three-dimensional counterparts of the magnetic structures identified in the 2D slices of Fig.~\ref{fig:Bfield_2d_evolved}. Model {\tt A} retains a prominent axial poloidal backbone surrounded by multipolar interior and near-surface structure. Its settled magnetic-field geometry is the most visibly nonaxisymmetric of the three configurations, which is consistent with its broad nonaxisymmetric VSH spectrum shown in the next section.

The two-dimensional visualizations of model {\tt B} clearly separate the two components identified in its meridional \(xz\) slice. A strong central axial bundle and the surrounding system of four organized poloidal cells arranged into two approximately reflection-symmetric pairs are visible. In three dimensions, the axial field lines pass through the rotation axis, while the surrounding field lines showcase a multipolar structure. The coherent azimuthal winding visible in the \(xy\) slice deforms these trajectories into three-dimensional helical field lines. The enclosing magnetic-field line bundles can therefore be interpreted as parts of an organized mixed poloidal--toroidal geometry.

Model {\tt C} exhibits a still more compact and regular hierarchy of nested field-line families. In the 2D meridional slice the central axial magnetic field-line bundle is surrounded by the three approximately symmetric pairs of meridional closed field-line loops, while the coherent equatorial winding produces the corresponding helical structures in three dimensions. The resulting organization retains a recognizable connection to the higher-order poloidal structure associated with the three current loops used to construct its initial magnetic field.

The two- and three-dimensional diagnostics together demonstrate that models {\tt B} and {\tt C} retain organized, predominantly axisymmetric higher-order structures, whereas model {\tt A} develops the strongest nonaxisymmetric complexity. We quantify these differences in the following subsection using the vector spherical harmonic decomposition of the settled configurations.

\subsection{Multipolar structure of the steady-state magnetic fields}
\label{sec:multipolar_structure}

The two- and three-dimensional visualizations and two-dimensional slices presented above show that the settled configurations possess qualitatively distinct and visibly multipolar magnetic field structures. In this section we quantify their multipolar structure via a vector spherical harmonic decomposition of \(\tilde B^i=\sqrt{\gamma}B^i\).

For each model and snapshot, we quantify the contribution associated with each pair \((\ell,m)\), where \(m\geq0\), using the solid-angle root-mean-square amplitude \(B_{\ell m}^{(q)}(r)\) defined in Eq.~\eqref{eq:vsh_sector_rms}.


To compare the relative radial prominence of different modes, we define
\begin{equation}
\widehat B_{\ell m}^{(q)}(r)=\frac{B_{\ell m}^{(q)}(r)}{\displaystyle \max_{0.08\leq r'/R_{\star,0}\leq1.20}B_{10}^{(q)}(r')},\qquad q\in\{{\rm pol},{\rm tor}\}.
\label{eq:vsh_mode_normalization}
\end{equation}
The denominator is the maximum radial amplitude of the \((\ell,m)=(1,0)\) sector of the same poloidal or toroidal component, evaluated over the displayed radial interval for the same model and snapshot.

\begin{figure}[!tp]
\centering
\includegraphics[width=0.99\linewidth]{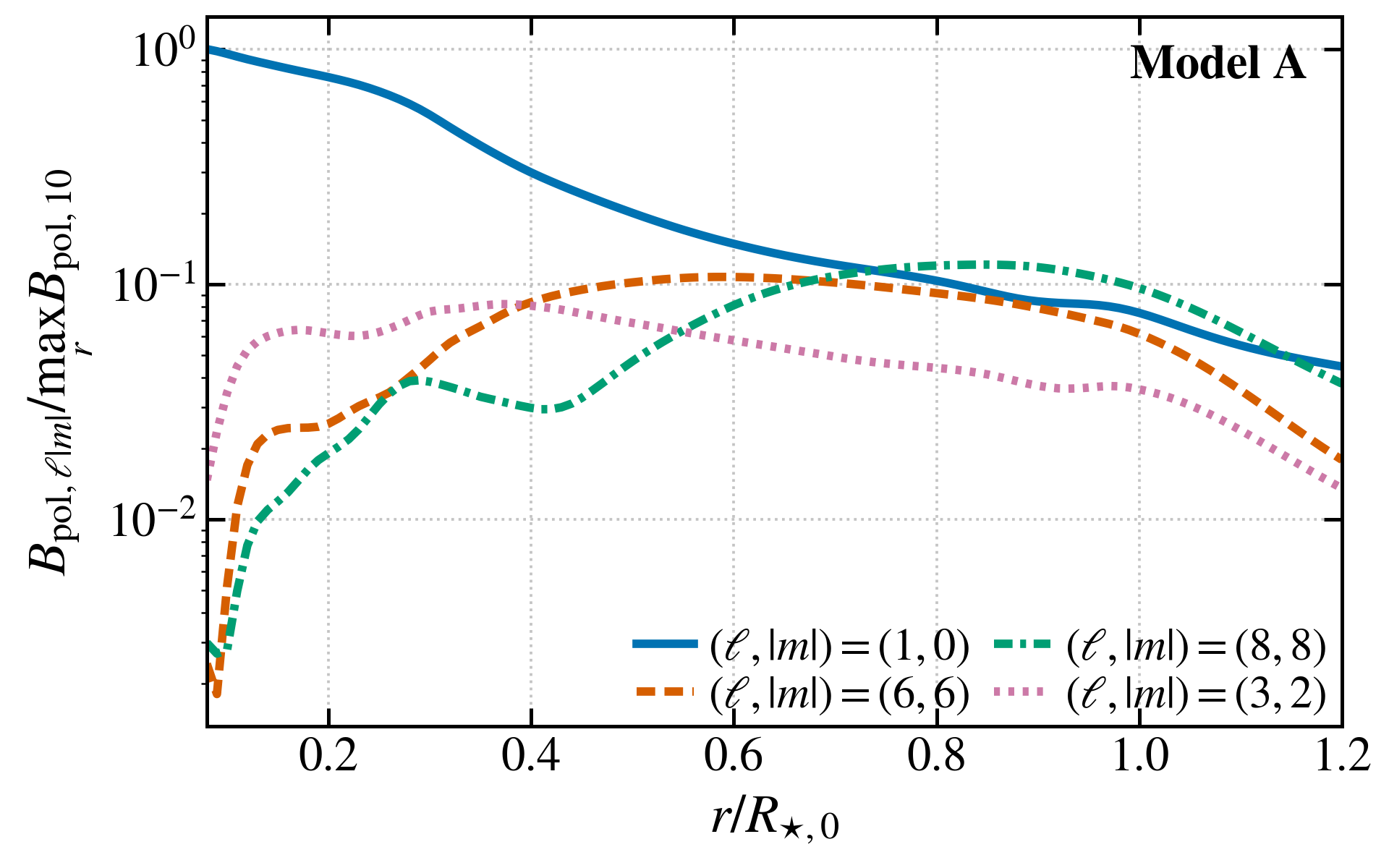}
\includegraphics[width=0.99\linewidth]{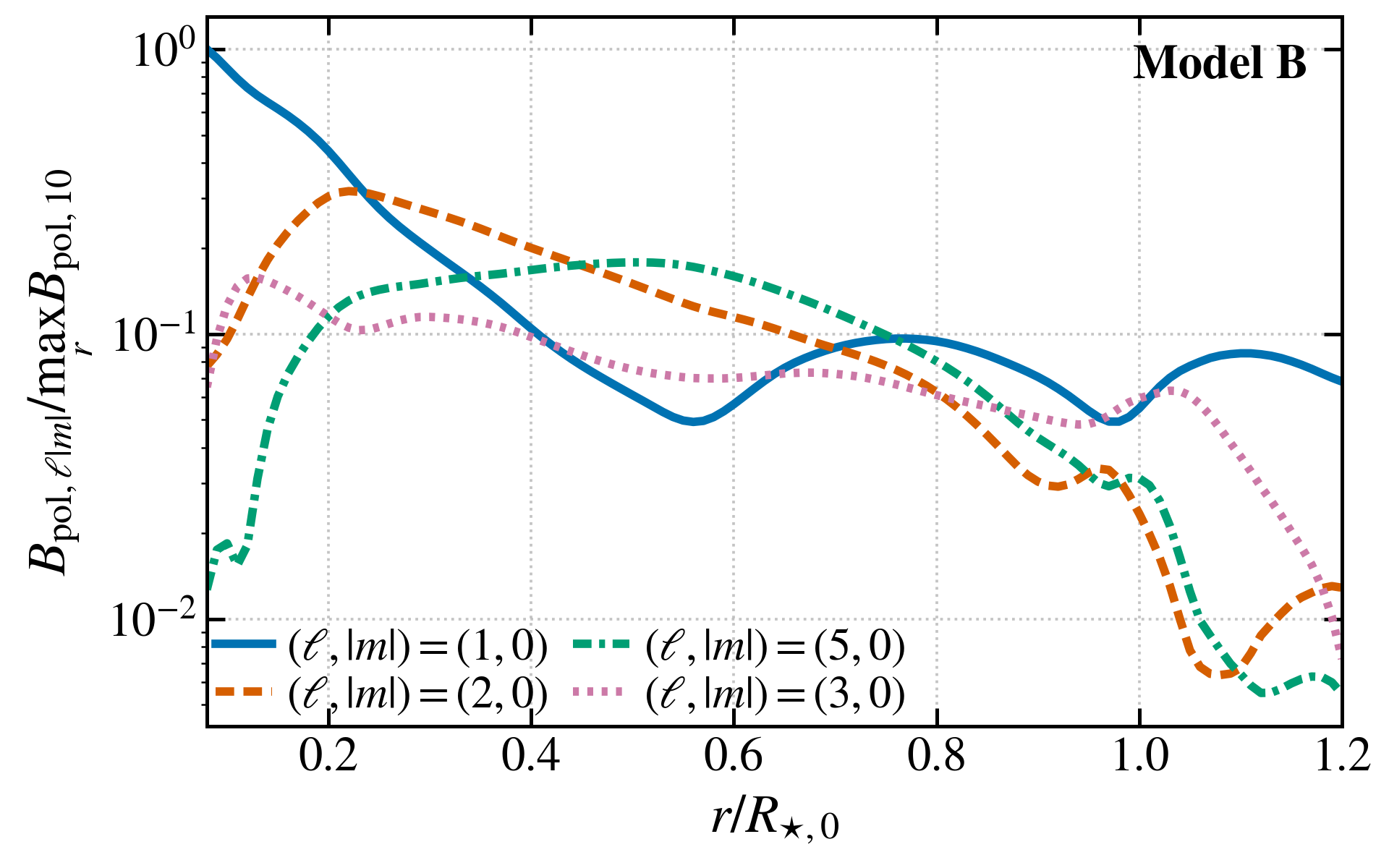}
\includegraphics[width=0.99\linewidth]{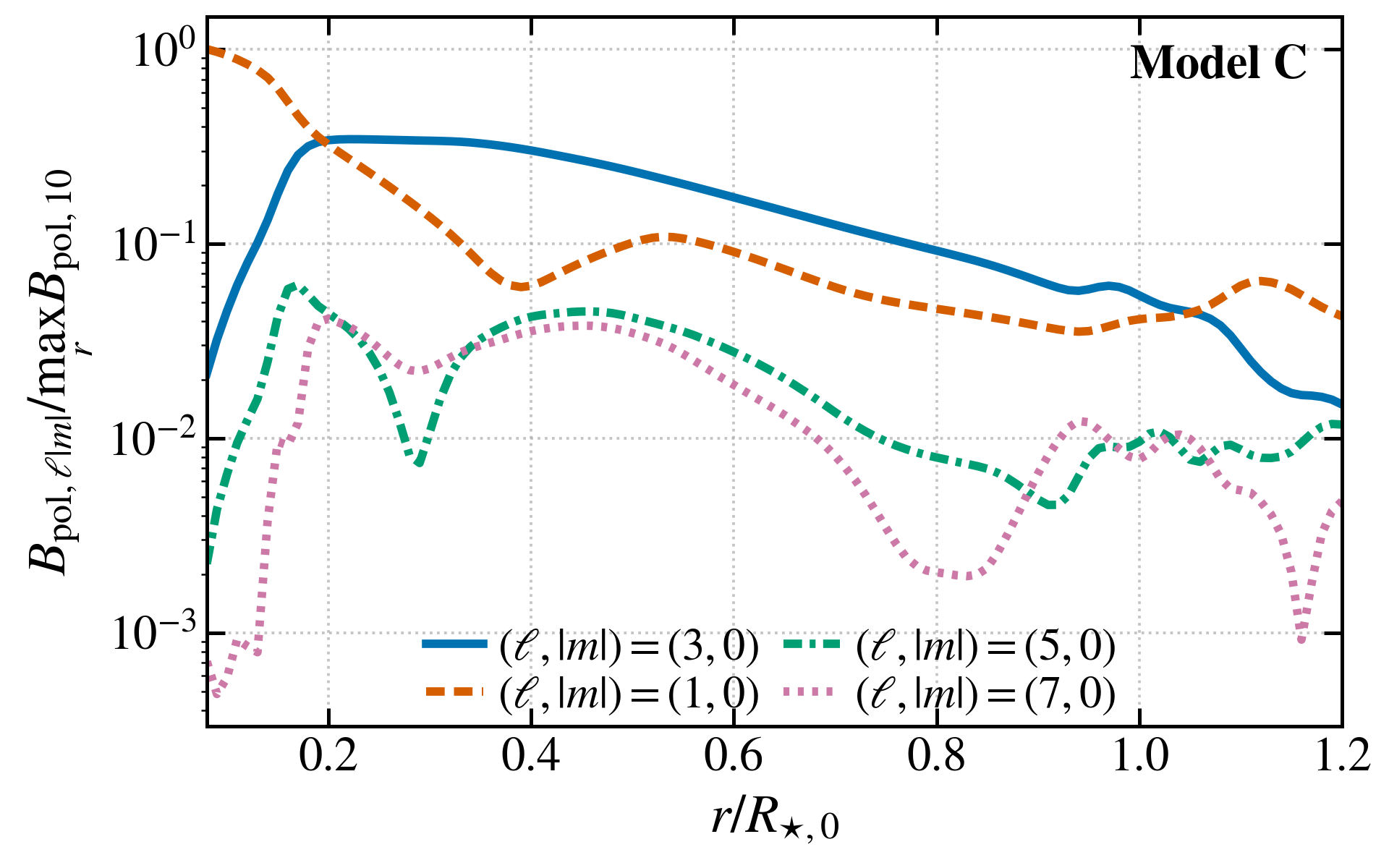}
\caption{
Strongest poloidal VSH contributions in the representative late-time configurations. Model {\tt A} is evaluated at \(t\approx 24t_A\), while models {\tt B} and {\tt C} are evaluated at \(t\approx 20t_A\). Each snapshot is decomposed about its center of mass. Each curve shows the normalized radial amplitude \(\widehat B_{\ell m}^{({\rm pol})}\) defined in Eq.~\eqref{eq:vsh_mode_normalization}. For each model, the four modes with the largest values of \({\cal A}_{\ell m}^{({\rm pol})}\) among all \(1\leq\ell\leq63\) and \(0\leq|m|\leq\ell\) are shown. The panels from top to bottom show models {\tt A}, {\tt B}, and {\tt C}.
}
\label{fig:vsh_poloidal_top4_ABC}
\end{figure}
%

In Fig.~\ref{fig:vsh_poloidal_top4_ABC} each panel displays the four most dominant modes for each model based on their radially averaged amplitudes
\begin{equation}
{\cal A}_{\ell m}^{(q)}=\frac{1}{\Delta \zeta}\int_{0.08}^{1.20}\widehat B_{\ell m}^{(q)}(\zeta)\,d\zeta,\qquad \zeta=\frac{r}{R_{\star,0}},
\label{eq:vsh_mode_ranking}
\end{equation}
where $\Delta \zeta=1.20-0.08=1.12$ is the dimensionless radial integration width and $q\in\{{\rm pol},{\rm tor}\}$. The lower integration limit avoids the spherical-coordinate singularity, while the upper limit includes the stellar interior and immediate near-surface region. The four displayed modes are selected for visual clarity only, as other modes can have average amplitudes comparable to that of the fourth-ranked mode.

Comparing the initial spectra in Fig.~\ref{fig:vsh_poloidal_top4_t0_ABC} with the evolved spectra in Fig.~\ref{fig:vsh_poloidal_top4_ABC} shows that the settled configurations retain partial memory of their initial poloidal multipolar content, although the degree of rearrangement differs among the models.

In model {\tt A}, the \((1,0)\) mode remains the strongest poloidal contribution near \(24t_A\), preserving the large-scale dipolar backbone already present initially. The next three contributions are the nonaxisymmetric \((6,6)\), \((8,8)\), and \((3,2)\) modes. The \((7,2)\) and initial axisymmetric \((3,0)\) contributions are ranked fifth and sixth, respectively. Model {\tt A} therefore retains its leading dipolar component while its detailed higher-order hierarchy is substantially rearranged into a broad nonaxisymmetric spectrum, consistently with its highly multipolar equatorial magnetic field structure in Figs.~\ref{fig:Bfield_3d_evolved} and~\ref{fig:Bfield_2d_evolved}.

Model {\tt B} retains the same set of four highest-ranked poloidal modes as its initial equilibrium, but their  ordering in the settled state is \((1,0)\), \((2,0)\), \((5,0)\), and \((3,0)\). Thus, the two leading modes are unchanged, while the \((5,0)\) and \((3,0)\) modes are swapped. The immediately following \((4,0)\) mode has approximately \(0.76\) of the average amplitude of the fourth-ranked \((3,0)\) contribution. The leading spectrum remains predominantly axisymmetric and preserves a clear, though not exact, memory of the initial magnetic field geometry.

In model {\tt C}, the octupolar \((3,0)\) mode remains the strongest one, while the dipolar \((1,0)\) mode remains second. The third- and fourth-ranked evolved modes are \((5,0)\) and \((7,0)\). The fifth-ranked \((9,0)\) mode has approximately \(0.71\) of the average amplitude of \((7,0)\), whereas the initially prominent \((4,0)\) mode is ranked eighth after the evolution. The leading \((3,0)\) and \((1,0)\) components and the strongly axisymmetric character of the spectrum persist, while the weaker higher-order hierarchy is rearranged.

The ratios of the fifth- to fourth-ranked averaged amplitudes are approximately \(0.99\), \(0.76\), and \(0.71\) for models {\tt A}, {\tt B}, and {\tt C}, respectively.

\begin{figure*}[!tp]
\centering
\includegraphics[width=0.49\textwidth]{multi_model_vsh_Lmax63/mode_diagnostics/C/t20_top4_Bpol_phys_ranked_by_area_over_max_Bpol10.png}
\includegraphics[width=0.49\textwidth]{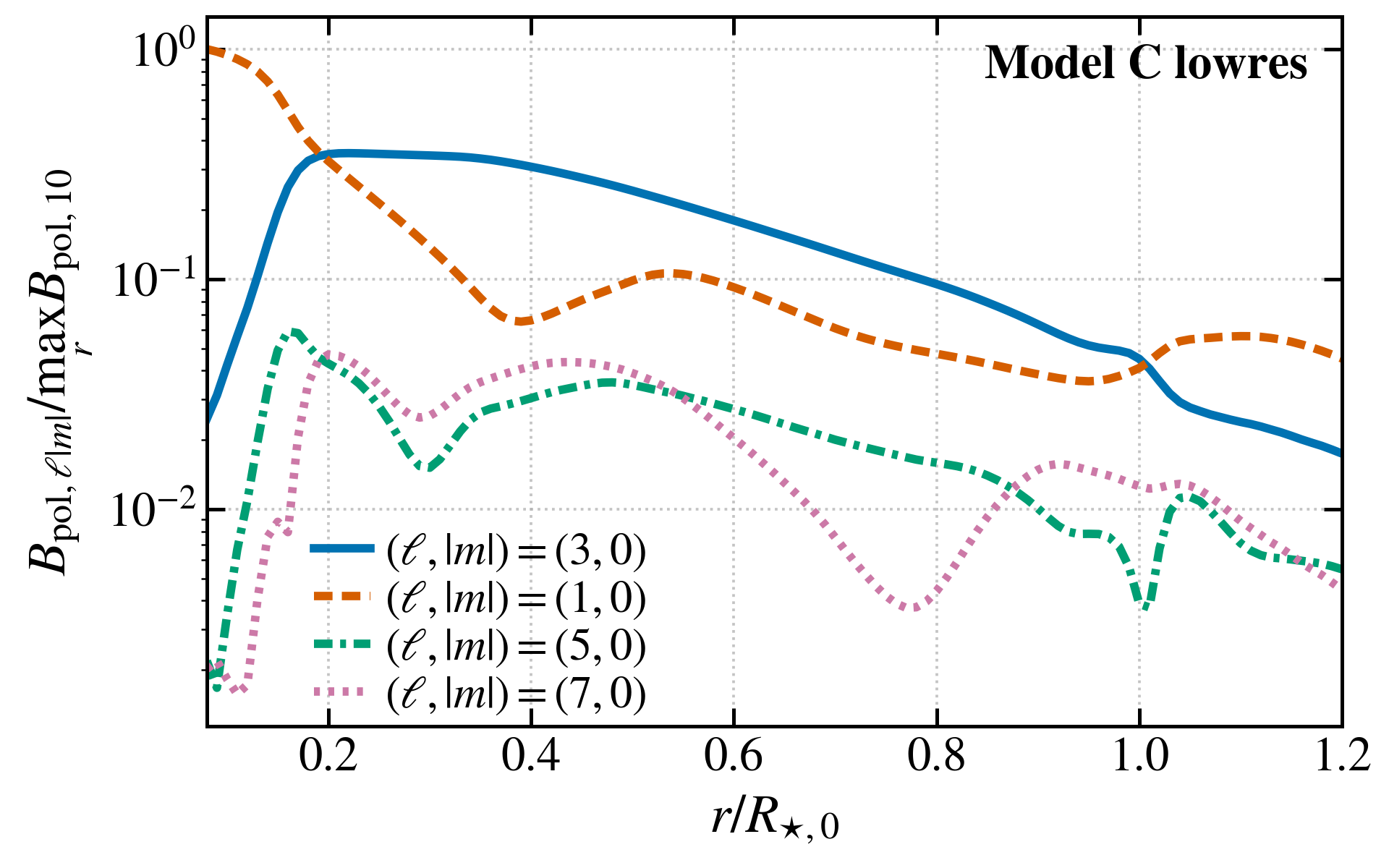}
\caption{
Resolution comparison of the settled poloidal VSH spectrum for model {\tt C} at \(t\approx 20t_A\). The left and right panels show the higher- and lower-resolution evolutions, respectively. 
Both resolutions give the same ordering of the four highest-ranked contributions: \((3,0)\), \((1,0)\), \((5,0)\), and \((7,0)\).}
\label{fig:vsh_C_resolution_poloidal}
\end{figure*}

Figure~\ref{fig:vsh_C_resolution_poloidal} shows that the qualitative poloidal hierarchy of model {\tt C} is unchanged by the resolution variation. The average amplitude \({\cal A}_{\ell m}^{({\rm pol})}\) of the \((3,0)\), \((1,0)\), and \((5,0)\) contributions differ by approximately \(1.68\%\), \(1.32\%\), and \(0.27\%\), respectively, between the two resolutions. The weaker \((7,0)\) contribution differs by approximately \(22.7\%\). The ordering of the modes immediately below the displayed four are resolution dependent. A resolution study would be required to determine how much of model {\tt A}'s detailed high-\(m\) ordering is intrinsic to the stellar magnetic field. However, such a resolution study goes beyond the scope of this paper.

\begin{figure}[!tp]
\centering
\includegraphics[width=0.99\linewidth]{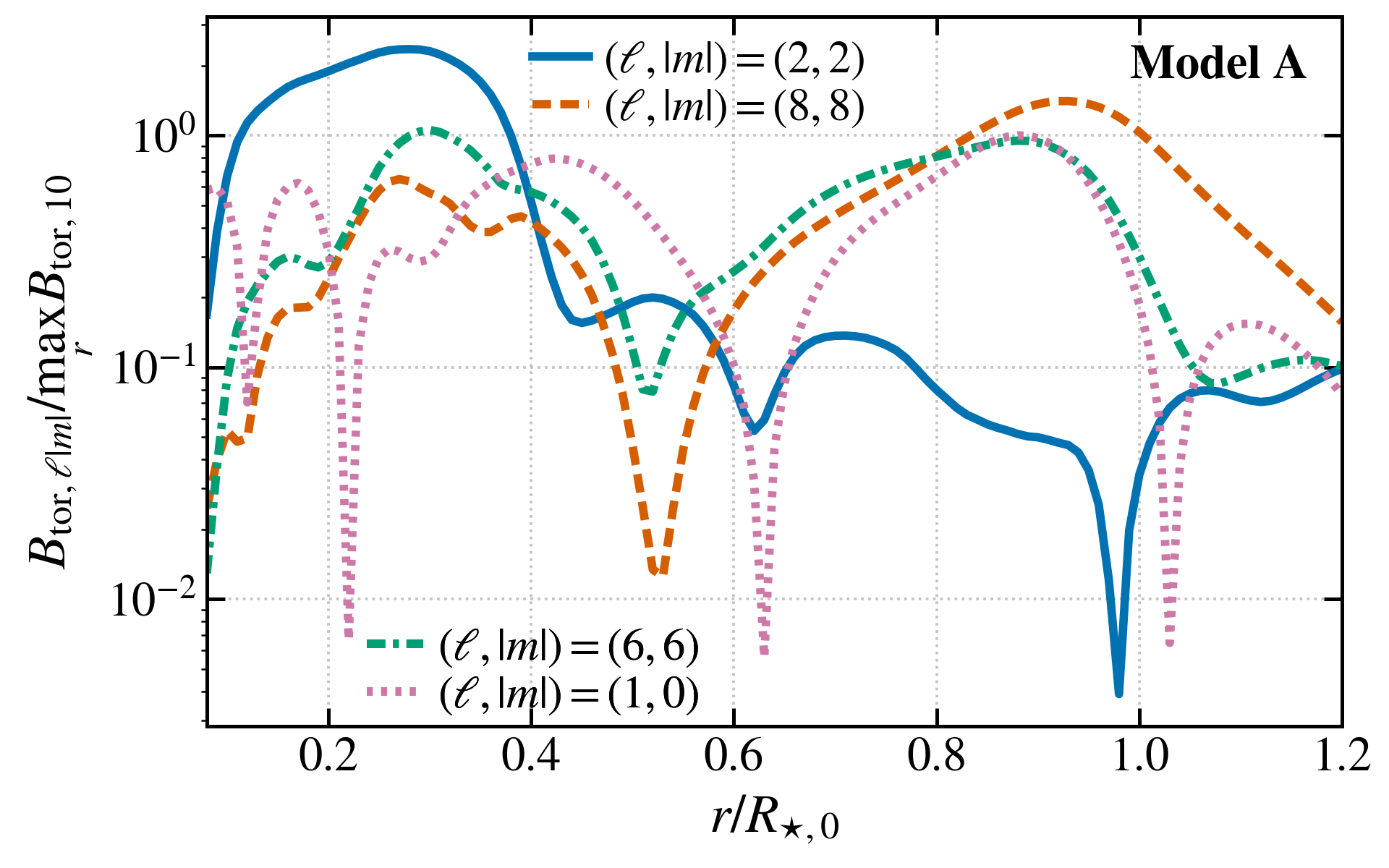}
\includegraphics[width=0.99\linewidth]{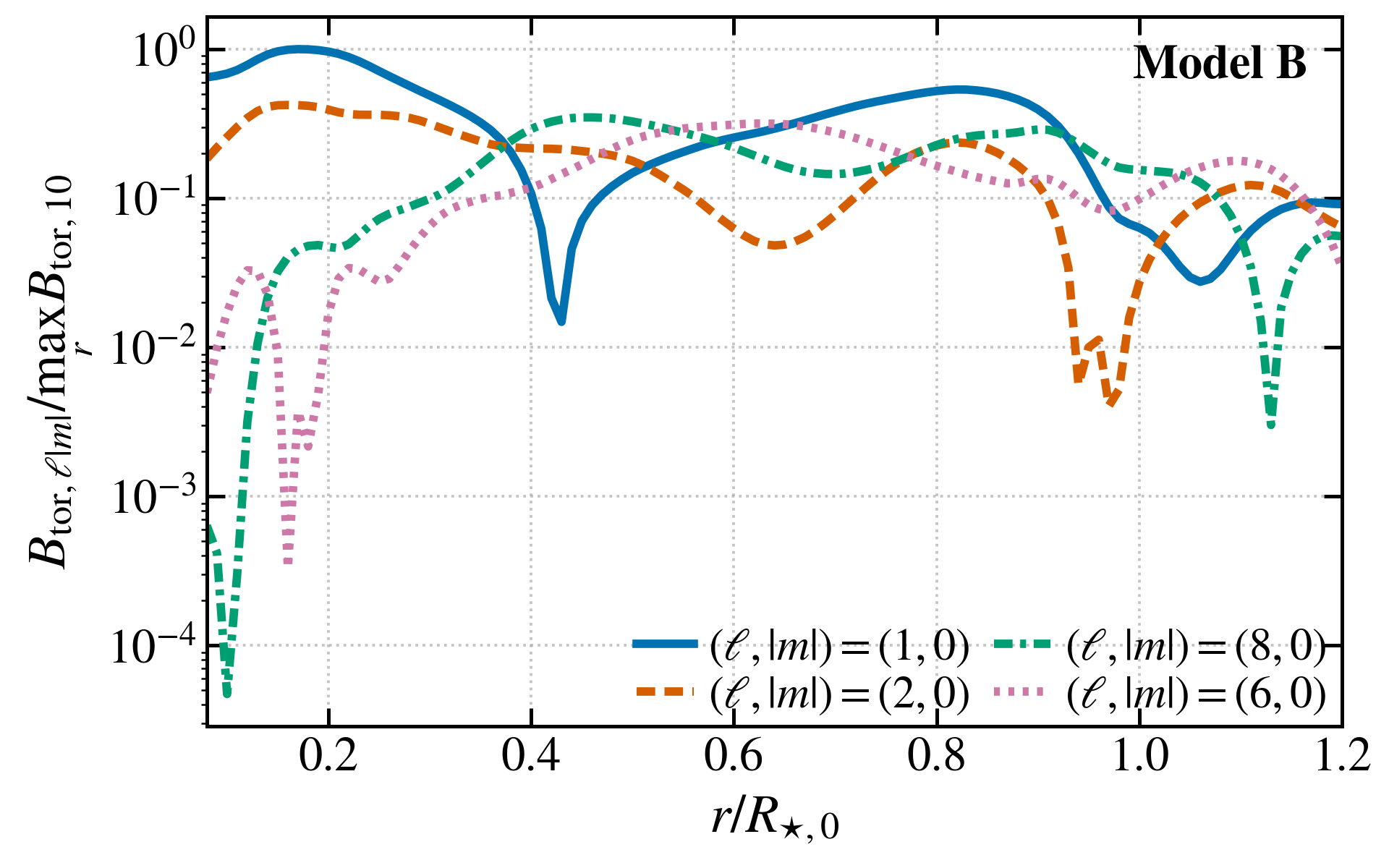}
\includegraphics[width=0.99\linewidth]{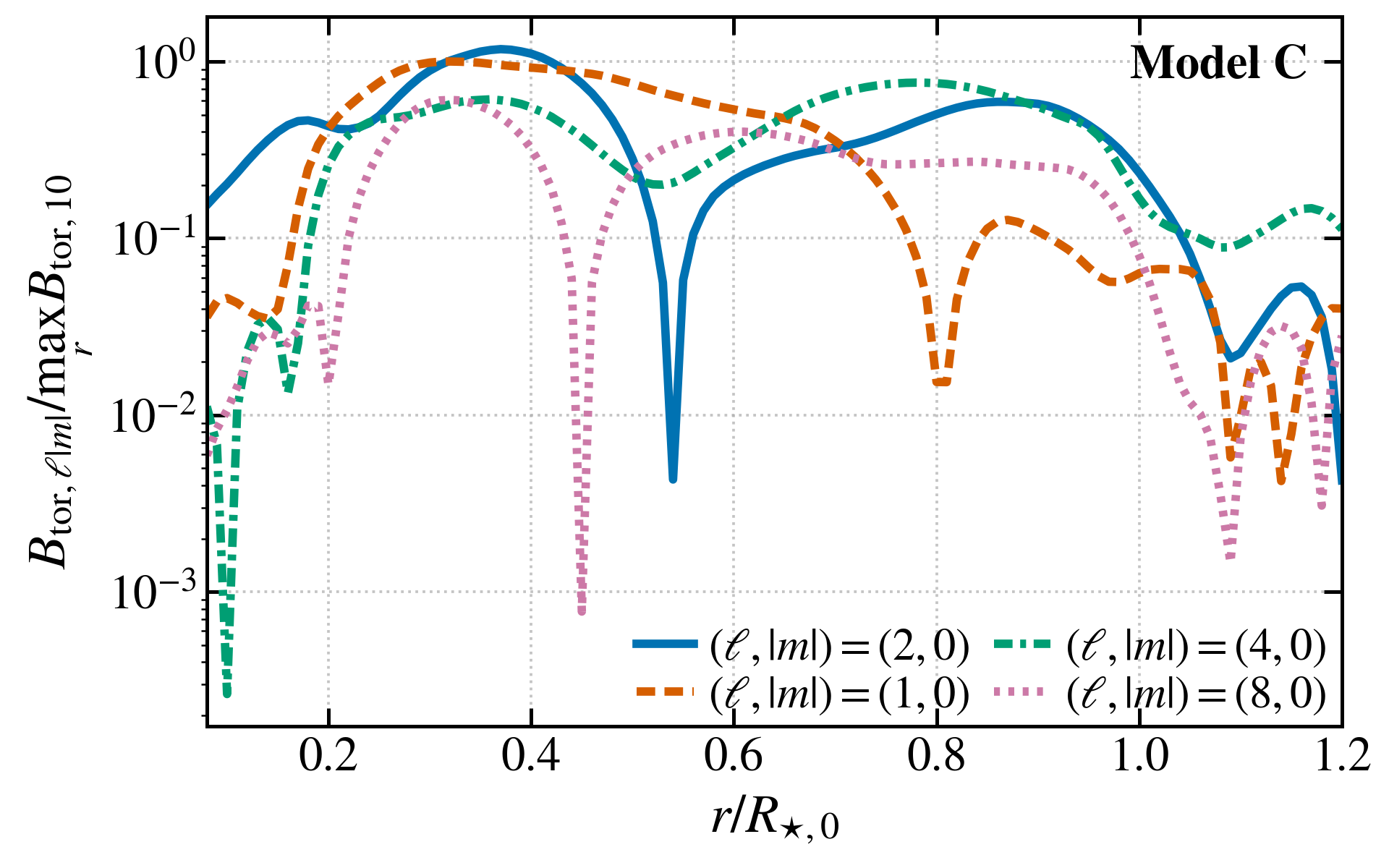}
\caption{
Dominant toroidal VSH modes in the representative late-time configurations. Model {\tt A} is evaluated at \(t\approx 24t_A\), while models {\tt B} and {\tt C} are evaluated at \(t\approx 20t_A\). Each curve shows the normalized radial amplitude \(\widehat B_{\ell m}^{({\rm tor})}\) defined in Eq.~\eqref{eq:vsh_mode_normalization}. The panels from top to bottom show models {\tt A}, {\tt B}, and {\tt C}.
}
\label{fig:vsh_toroidal_top4_ABC}
\end{figure}

The toroidal spectra in Fig.~\ref{fig:vsh_toroidal_top4_ABC} distinguish model {\tt A} particularly clearly from models {\tt B} and {\tt C}. In model {\tt A}, the four strongest toroidal contributions near \(24t_A\) are \((2,2)\), \((8,8)\), \((6,6)\), and \((1,0)\). Thus, three of the four displayed modes are nonaxisymmetric, while the leading axisymmetric \((1,0)\) mode is ranked fourth. The fifth-ranked \((9,8)\) mode has approximately \(0.89\) of the average amplitude of the fourth-ranked mode. Several additional modes with different values of \(\ell\) and \(m\) remain non-negligible. Model {\tt A} therefore possesses a broad nonaxisymmetric toroidal spectrum rather than only the four contributions displayed here.

Model {\tt B} has a predominantly axisymmetric toroidal spectrum. Its four highest-ranked contributions are \((1,0)\), \((2,0)\), \((8,0)\), and \((6,0)\). The fifth-ranked \((4,0)\) mode has approximately \(0.99\) of the average amplitude of the fourth-ranked mode, followed by the \((3,0)\) and \((5,0)\) modes. Hence, model {\tt B} contains substantial higher-order toroidal structure, but its leading contributions remain axisymmetric rather than exhibiting the broad high-\(m\) spectrum found in model {\tt A}.

Model {\tt C} is also dominated by axisymmetric toroidal modes. Its four strongest contributors are the \((2,0)\), \((1,0)\), \((4,0)\), and \((8,0)\) modes, while the fifth-ranked \((6,0)\) mode has approximately \(0.88\) of the average amplitude of the fourth-ranked contribution. At lower resolution, the first three contributions retain the ordering \((2,0)\), \((1,0)\), and \((4,0)\); their normalized average amplitudes differ from the higher-resolution values by approximately \(12.0\%\), \(13.1\%\), and \(1.04\%\), respectively. The fourth-ranked contribution changes from \((8,0)\) at higher resolution to \((6,0)\) at lower resolution. Thus, the leading toroidal hierarchy remains axisymmetric at both resolutions, but individual leading amplitudes can differ at the \(\mathcal{O}(10\%)\) level and the detailed ordering of weaker higher-order contributions is resolution sensitive.

Taken together, the poloidal and toroidal VSH results support the qualitative structures identified independently in Figs.~\ref{fig:Bfield_3d_evolved} and~\ref{fig:Bfield_2d_evolved}. The persistent \((1,0)\) poloidal mode in model {\tt A} corresponds to its large-scale axial backbone, while its broad nonaxisymmetric poloidal and toroidal spectra are consistent with its irregular equatorial structure. The predominantly axisymmetric higher-order spectra of models {\tt B} and {\tt C} support their more organized poloidal structure and closer approach to axisymmetry. The three configurations therefore retain different portions of their initial multipolar structure and do not converge toward a universal magnetic equilibrium. In particular, the twisted torus configuration does not represent a universal configuration.

The radial mode profiles characterize the global angular content of the magnetic field throughout the stellar interior and immediate near-surface region. As a complementary directional diagnostic, we evaluate the reconstructed magnetic field contributions at the pole using the center of mass defined in Eq.~\eqref{eq:vsh_coordinate_com}. We define the north pole by
\begin{equation}
\mathbf{x}_{\rm pole}
=
\mathbf{x}_{\rm c}+R_{\rm pole}\hat{\mathbf z},
\label{eq:pole_location}
\end{equation}
where \(R_{\rm pole}\) is the first outward crossing of \(\rho_b=10^{-2}\rho_{b,\max}\) along the coordinate \(+z\) direction through \(\mathbf{x}_{\rm c}\); this direction coincides with the initial rotation axis. The decomposition is performed on the coordinate sphere of radius \(R_{\rm pole}\) centered on \(\mathbf{x}_{\rm c}\).

The magnetic field amplitude measured by a normal observer at the pole is
\begin{equation}
|B_{\ell|m|}|_{\rm pole}
=
\left[
\frac{
\gamma_{ij}
\widetilde B^i_{\ell,|m|}
\widetilde B^j_{\ell,|m|}
}{\gamma}
\right]_{\rm pole}^{1/2}.
\label{eq:pole_mode_amplitude}
\end{equation}

Although the scalar harmonics \(Y_{\ell,\pm1}\) vanish on the axis, their angular derivatives entering the vector spherical harmonics have finite polar limits. Consequently, reconstructed \(|m|=1\) vector-harmonic contributions need not vanish. By contrast, contributions with \(|m|\geq2\) vanish at the exact pole. 

We consider all 2079 reconstructed real contributions with \(1\leq\ell\leq63\) and \(0\leq|m|\leq\ell\). We sum all reconstructed contributions through \(\ell_{\max}=63\) and compare the resulting normal-observer magnetic field vector directly with the independently sampled Cartesian simulation field at the pole.
The relative reconstruction errors are approximately \(\mathcal{O}(1.0)\%\), and for models {\tt A}, and {\tt B}, and \(\mathcal{O}(5.0)\%\) for {\tt C}. For each model, we use the dominant magnetic field contribution at the pole to normalize all pole contributions,
\begin{equation}
B_{\rm ref}
=
\max_{\substack{
1\leq\ell\leq63\\
0\leq|m|\leq\ell
}}
|B_{\ell|m|}|_{\rm pole},
\qquad
\mathcal{P}_{\ell|m|}
=
100
\frac{|B_{\ell|m|}|_{\rm pole}}{B_{\rm ref}}.
\label{eq:pole_mode_ratio}
\end{equation}
The percentages \(\mathcal{P}_{\ell|m|}\) are relative amplitudes. Tables~\ref{tab:pole_mode_ratios_A}--\ref{tab:pole_mode_ratios_C} list the ten largest nonzero pole contributions for each model.

The pole mode strength ordering is not required to coincide with the global radial ordering because the two diagnostics measure different quantities. The pole diagnostic and the solid-angle RMS diagnostic therefore provide complementary descriptions of the settled fields. At the coordinate north pole, the \((1,0)\) contribution is dominant in model {\tt A}  at \(t\approx 24t_A\), while the \((3,0)\) contribution remains dominant in models {\tt B} and {\tt C} near \(20t_A\). In model {\tt A}, the \((3,0)\) and \((11,0)\) contributions reach \(76.6\%\) and \(56.6\%\) of the dominant mode, respectively, while the nonaxisymmetric \((11,1)\) contribution reaches \(25.6\%\). In models {\tt B} and {\tt C}, the \((1,0)\) contribution reaches \(93.7\%\) and \(67.4\%\) of the dominant amplitude, respectively. Higher-order axisymmetric contributions also remain substantial. For example, the \((13,0)\) contribution reaches \(37.9\%\) in model {\tt B}, while the \((9,0)\) and \((13,0)\) contributions reach \(51.0\%\) and \(40.9\%\) in model {\tt C}.

\begin{table}[!t]
\centering
\caption{
Ten strongest nonzero reconstructed magnetic field amplitudes at the north pole among \(1\leq\ell\leq63\) for model {\tt A} at the closest complete snapshot to \(24t_A\). Amplitudes are measured by a normal observer; percentages are relative to the dominant \((1,0)\) contribution.
}
\label{tab:pole_mode_ratios_A}
\begingroup
\begin{ruledtabular}
\begin{tabular}{cccc}
Rank & \((\ell,|m|)\) & \(|B_{\ell|m|}|_{\rm pole}\) &
\(\mathcal{P}_{\ell|m|}\,[\%]\) \tabularnewline
\hline
1  & \((1,0)\)  & \(1.21\times10^{-3}\) & 100  \tabularnewline
2  & \((3,0)\)  & \(9.25\times10^{-4}\) & 76.6 \tabularnewline
3  & \((11,0)\) & \(6.84\times10^{-4}\) & 56.6 \tabularnewline
4  & \((4,0)\)  & \(3.67\times10^{-4}\) & 30.4 \tabularnewline
5  & \((5,0)\)  & \(3.38\times10^{-4}\) & 28.0 \tabularnewline
6  & \((11,1)\) & \(3.09\times10^{-4}\) & 25.6 \tabularnewline
7  & \((14,0)\) & \(2.42\times10^{-4}\) & 20.1 \tabularnewline
8  & \((22,0)\) & \(2.39\times10^{-4}\) & 19.8 \tabularnewline
9  & \((6,0)\)  & \(2.30\times10^{-4}\) & 19.1 \tabularnewline
10 & \((12,0)\) & \(2.30\times10^{-4}\) & 19.1 \tabularnewline
\end{tabular}
\end{ruledtabular}
\endgroup
\end{table}

\begin{table}[!t]
\centering
\caption{
Ten strongest nonzero reconstructed magnetic-field amplitudes at the north pole among \(1\leq\ell\leq63\) for model {\tt B} near \(20\,t_A\). Amplitudes are measured by a normal observer; percentages are relative to the dominant \((3,0)\) contribution.}
\label{tab:pole_mode_ratios_B}
\begingroup
\begin{ruledtabular}
\begin{tabular}{cccc}
Rank & \((\ell,|m|)\) & \(|B_{\ell|m|}|_{\rm pole}\) &
\(\mathcal{P}_{\ell|m|}\,[\%]\) \tabularnewline
\hline
1  & \((3,0)\)  & \(1.76\times10^{-4}\) & 100  \tabularnewline
2  & \((1,0)\)  & \(1.65\times10^{-4}\) & 93.7 \tabularnewline
3  & \((13,0)\) & \(6.69\times10^{-5}\) & 37.9 \tabularnewline
4  & \((4,0)\)  & \(5.36\times10^{-5}\) & 30.4 \tabularnewline
5  & \((10,0)\) & \(5.18\times10^{-5}\) & 29.4 \tabularnewline
6  & \((2,0)\)  & \(4.96\times10^{-5}\) & 28.1 \tabularnewline
7  & \((5,0)\)  & \(4.68\times10^{-5}\) & 26.6 \tabularnewline
8  & \((9,0)\)  & \(3.52\times10^{-5}\) & 19.9 \tabularnewline
9  & \((21,0)\) & \(2.26\times10^{-5}\) & 12.8 \tabularnewline
10 & \((11,0)\) & \(2.13\times10^{-5}\) & 12.1 \tabularnewline
\end{tabular}
\end{ruledtabular}
\endgroup
\end{table}

\begin{table}[!t]
\centering
\caption{
Ten strongest nonzero reconstructed magnetic-field amplitudes at the north pole among \(1\leq\ell\leq63\) for model {\tt C} near \(20\,t_A\). Amplitudes are measured by a normal observer; percentages are relative to the dominant \((3,0)\) contribution.}
\label{tab:pole_mode_ratios_C}
\begingroup
\begin{ruledtabular}
\begin{tabular}{cccc}
Rank & \((\ell,|m|)\) & \(|B_{\ell|m|}|_{\rm pole}\) &
\(\mathcal{P}_{\ell|m|}\,[\%]\) \tabularnewline
\hline
1  & \((3,0)\)  & \(3.34\times10^{-4}\) & 100  \tabularnewline
2  & \((1,0)\)  & \(2.25\times10^{-4}\) & 67.4 \tabularnewline
3  & \((9,0)\)  & \(1.70\times10^{-4}\) & 51.0 \tabularnewline
4  & \((13,0)\) & \(1.36\times10^{-4}\) & 40.9 \tabularnewline
5  & \((7,0)\)  & \(1.18\times10^{-4}\) & 35.3 \tabularnewline
6  & \((11,0)\) & \(9.27\times10^{-5}\) & 27.8 \tabularnewline
7  & \((17,0)\) & \(8.69\times10^{-5}\) & 26.1 \tabularnewline
8  & \((15,0)\) & \(7.95\times10^{-5}\) & 23.8 \tabularnewline
9  & \((5,0)\)  & \(7.35\times10^{-5}\) & 22.0 \tabularnewline
10 & \((8,0)\)  & \(5.51\times10^{-5}\) & 16.5 \tabularnewline
\end{tabular}
\end{ruledtabular}
\endgroup
\end{table}


\subsection{Pointwise toroidal-to-poloidal magnetic field structure}
\label{sec:btor_bpol_profiles}

The angle-averaged ratio \(\mathcal{R}_{\rm tp}\) defined in Eq.~\eqref{eq:vsh_total_tor_pol_ratio} is useful for characterizing the radial poloidal--toroidal balance, but it removes the angular information. To examine the anisotropic spatial structure directly, we reconstruct the poloidal and toroidal fields pointwise from Eq.~\eqref{eq:vsh_poloidal_toroidal_fields}, retaining all VSH modes through \(\ell_{\max}=63\). We note that a large value of \(B_{\rm tor}/B_{\rm pol}\) can be produced either by an enhanced toroidal field or by a local minimum of the poloidal amplitude. Although the complete-field reconstruction errors using the VSH decomposition are below \(4\%\) (see Appendix~\ref{App:VSH_Lmax_accuracy}, these errors do not provide a uniform relative-error bound on the ratio of the reconstructed poloidal and toroidal amplitudes. The ratio can be substantially more sensitive near a minimum of either component and near locations where \(B_{\rm tor}\simeq B_{\rm pol}\). Therefore, in this section we use this ratio to identify extended structures and broad regions of poloidal or toroidal dominance. 

At every Cartesian point \(\mathbf{x}\), we first calculate the centered spherical coordinates of \(\mathbf{x}-\mathbf{x}_{\rm c}(t)\), where \(\mathbf{x}_{\rm c}(t)\) is the center of mass. The radial DG representation is evaluated at the corresponding radius, the vector spherical harmonics are evaluated at the corresponding \((\theta,\phi)\), and the contributions from all \(-\ell\leq m\leq\ell\) and \(1\leq\ell\leq63\) are summed at the vector level. We thereby obtain the pointwise densitized fields \(\tilde B_{\rm pol}^i(\mathbf{x})\) and \(\tilde B_{\rm tor}^i(\mathbf{x})\).

For \(q\in\{{\rm pol},{\rm tor}\}\), the corresponding physical amplitude is
\begin{equation}
B_{(q)}(\mathbf{x})
=
\left[
\frac{
\gamma_{ij}\tilde B_{(q)}^i(\mathbf{x})\tilde B_{(q)}^j(\mathbf{x})
}{\gamma}
\right]^{1/2},
\label{eq:vsh_pointwise_amplitude}
\end{equation}
and the quantity displayed below is
\begin{equation}
\mathcal{Q}(\mathbf{x})
\equiv
\log_{10}\!\left[
\frac{B_{\rm tor}(\mathbf{x})}
{B_{\rm pol}(\mathbf{x})}
\right].
\label{eq:vsh_pointwise_ratio}
\end{equation}
Thus, \(\mathcal{Q}>0\) indicates local toroidal dominance, whereas \(\mathcal{Q}<0\) indicates local poloidal dominance.


The pointwise ratio is not displayed where the poloidal denominator is smaller than \(10^{-10}\) of the maximum reconstructed total magnetic field in the same product, or where the total magnetic field is smaller than \(10^{-12}\) of that maximum. In addition, the region \(r<0.08R_{\star,0}\) is excluded because it lies close to the coordinate singularity. This numerical exclusion is marked explicitly in the figures. The displayed color scale is balanced about \(\mathcal{Q}=0\) and clipped to \(-2\leq\mathcal{Q}\leq2\).

\begin{figure*}[!tp]
\centering
\includegraphics[width=0.98\textwidth]{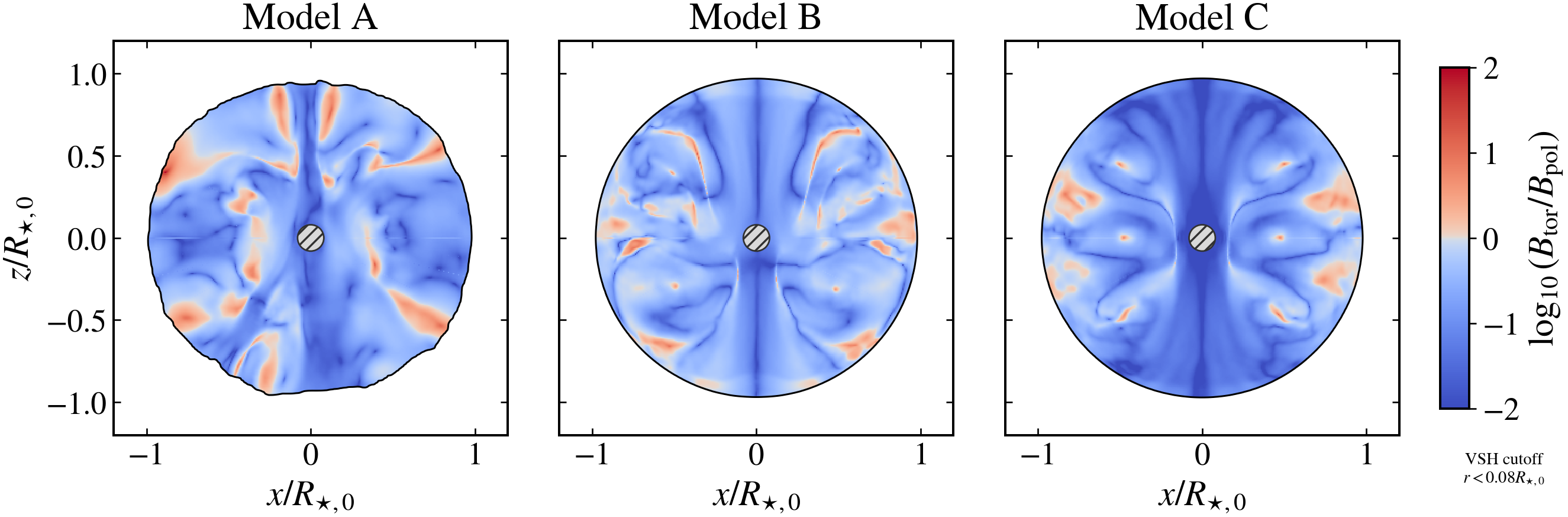}
\caption{
Pointwise toroidal-to-poloidal magnetic field amplitude ratio in the meridional plane \(y=y_c\). Model {\tt A} is shown near \(24t_A\), while models {\tt B} and {\tt C} are shown near \(20t_A\). Each pixel shows the reconstructed quantity \(\mathcal{Q}=\log_{10}(B_{\rm tor}/B_{\rm pol})\). The colors are interpreted qualitatively: red and blue identify regions of toroidal and poloidal dominance, respectively. The balanced color scale is clipped at \(\mathcal{Q}=\pm2\). The black contour marks the density-defined stellar surface, \(\rho_b=10^{-2}\rho_{b,\max}\). The hatched central circle marks the excluded region \(r<0.08R_{\star,0}\).
}
\label{fig:btor_bpol_meridional}
\end{figure*}

To visualize structures that do not lie in the meridional plane, we also construct an orthographic opacity-weighted projection of the full three-dimensional pointwise reconstruction. The star is viewed from an azimuth of \(38^\circ\) and an elevation of \(22^\circ\). At each image pixel, samples through the stellar volume are combined from front to back. Their color is determined by \(\mathcal{Q}\), while the opacity increases with both \(|\mathcal{Q}|\) and the normalized total magnetic field amplitude. This makes extended, magnetically significant regions of strong poloidal or toroidal dominance visible while retaining partial transparency through the star.

\begin{figure*}[!tp]
\centering
\includegraphics[width=0.98\textwidth]{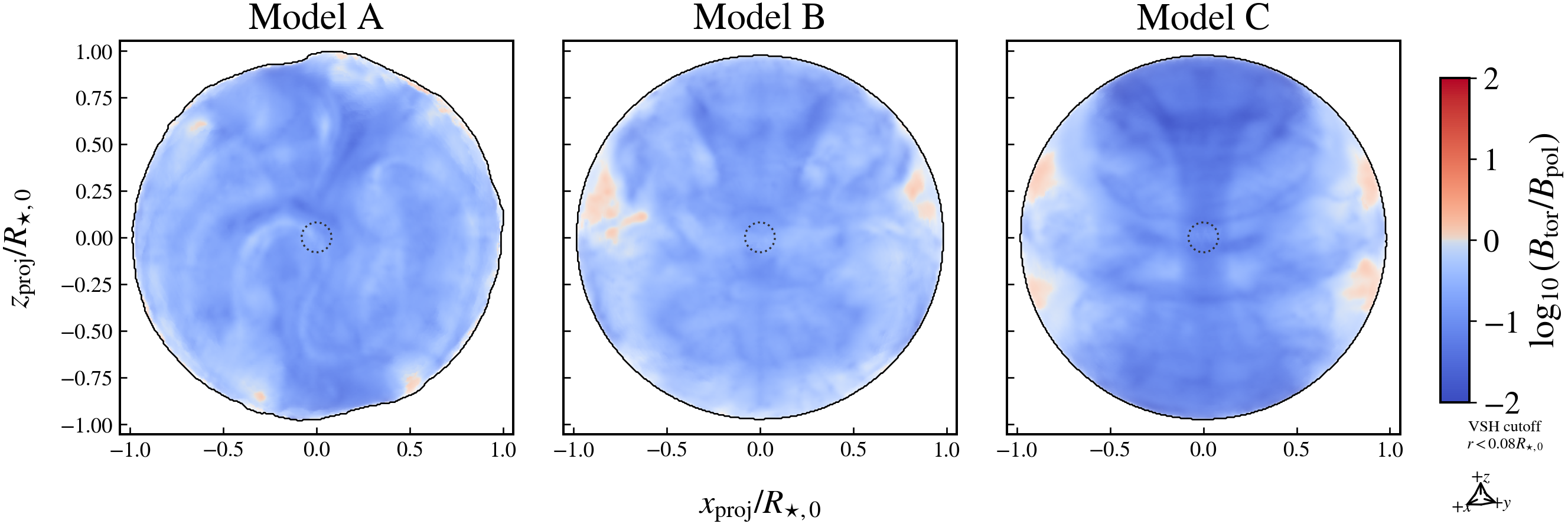}
\caption{
Orthographic opacity-weighted projection of the three-dimensional pointwise ratio \(\mathcal{Q}=\log_{10}(B_{\rm tor}/B_{\rm pol})\). Model {\tt A} is shown near \(24t_A\), while models {\tt B} and {\tt C} are shown near \(20t_A\). The viewing azimuth and elevation are \(38^\circ\) and \(22^\circ\), respectively, and the orientation triad identifies the original Cartesian directions. Red and blue denote toroidal and poloidal dominance, respectively. The opacity emphasizes regions with a pronounced ratio and appreciable total magnetic field strength. The black outline is the projected density-defined stellar surface, and the dotted central circle marks the excluded region \(r<0.08R_{\star,0}\). 
}
\label{fig:btor_bpol_3d}
\end{figure*}

Figures~\ref{fig:btor_bpol_meridional} and~\ref{fig:btor_bpol_3d} demonstrate that the poloidal--toroidal balance is strongly anisotropic. Model {\tt A} is predominantly poloidal in its central region and throughout much of its interior, consistently with its axial poloidal backbone and with the angle-averaged result \(\mathcal{R}_{\rm tp}<1\). Its localized toroidal enhancements occur mainly farther from the center. There is therefore no contradiction between the pointwise maps and the other diagnostics for model {\tt A}.

Models {\tt B} and {\tt C} also remain poloidal dominated over much of their volume, but contain coherent localized toroidal enhancements. Several of these structures, including the three localized structures identified in model {\tt C}, correspond to structures visible independently in the magnetic field line visualizations. A coherent toroidal field-line structure need not be red throughout the ratio map. A blue region means only that \(B_{\rm pol}>B_{\rm tor}\) locally, even if an organized toroidal component is present there.

\subsection{Four-current}
\label{sec:four_current_results}

In this section we examine the structure  of the four-current associated with the settled magnetic  configurations. As described in Sec.~\ref{sec:methods}, we compute the invariant
\begin{equation}
  J^2 \equiv J^\mu J_\mu ,
\end{equation}
from the numerically constructed four-current. With our sign convention, \(J^2>0\) corresponds to a spacelike four-current. This condition has a clear local meaning that the magnitude of the spatial current exceeds the charge density in an invariant sense. In flat spacetime, writing \(J^\mu=(\rho_e,\mathbf{J})\), one has
\begin{equation}
  J^\mu J_\mu = -\rho_e^2 + |\mathbf{J}|^2 ,
\end{equation}
so that \(J^\mu J_\mu>0\) implies \(|\mathbf{J}|>|\rho_e|\). Such a current cannot be carried by a single-sign charge population moving subluminally, and in a physical magnetosphere would require counter-streaming charges, pair creation, or non-ideal electric fields. However, inside the dense ideal magnetohydynamic neutron star interiors we naturally expect it to be spacelike. 

\begin{figure*}[!tp]
  \centering
  \includegraphics[width=0.9\textwidth]{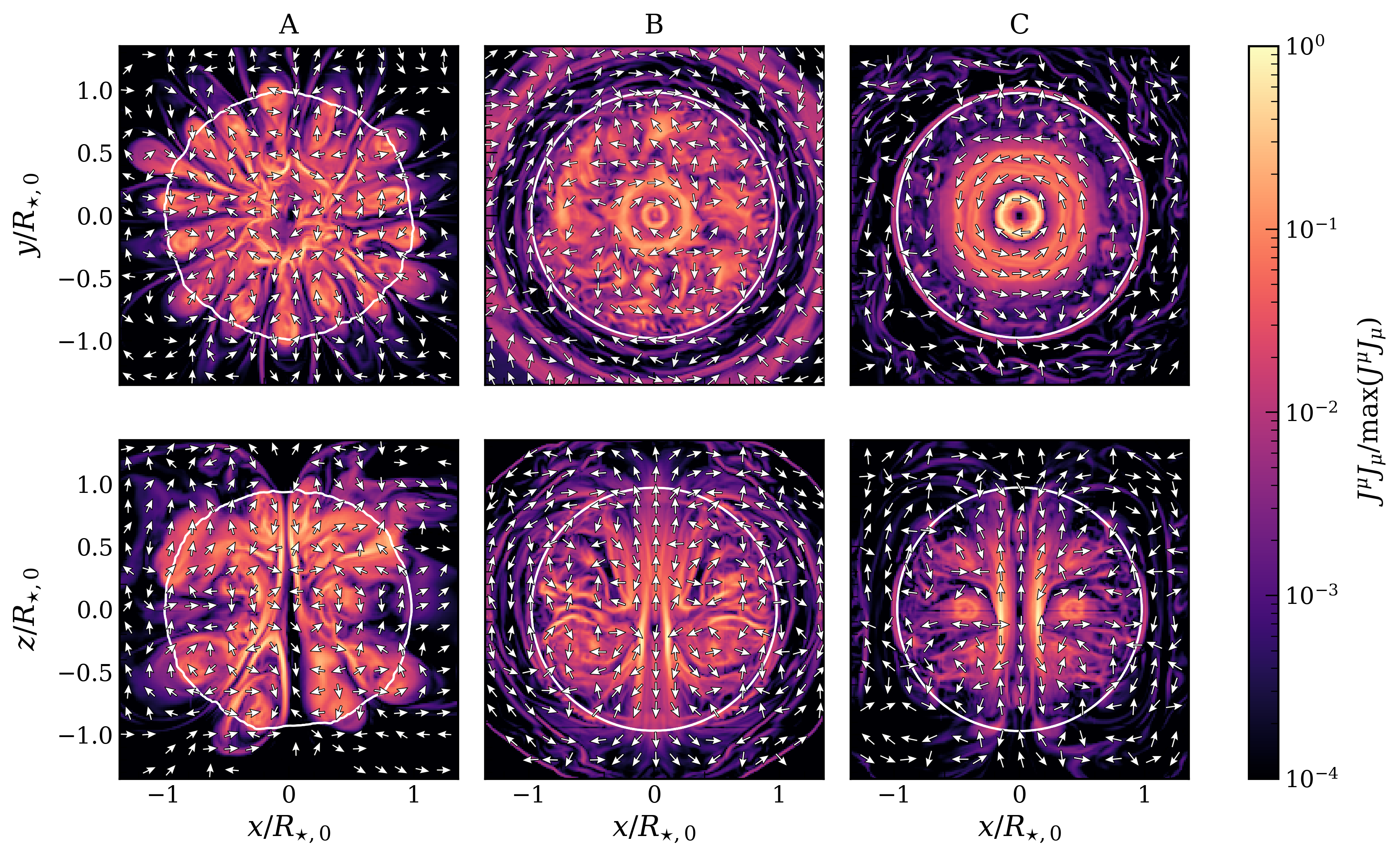}
\caption{
Normalized four-current invariant for the representative late-time configurations, shown on a logarithmic scale. Model {\tt A} is evaluated at \(t\approx 24t_A\), while models {\tt B} and {\tt C} at \(t\approx 20t_A\). Columns correspond to models {\tt A}, {\tt B}, and {\tt C}. The upper row shows the equatorial \(xy\) slices, while the lower row shows the meridional \(xz\) slices. The plotted quantity is \(J^\mu J_\mu/\max_{xy,xz}(J^\mu J_\mu)\), where the maximum positive value is computed separately for each model over its two displayed slices. White contours mark the stellar surface, \(\rho_b=10^{-2}\rho_{b,\max}\). Arrows indicate the direction of the in-plane spatial current: \((J^x,J^y)\) in the \(xy\) plane and \((J^x,J^z)\) in the \(xz\) plane.
}
  \label{fig:current_maps}
\end{figure*}

Figure~\ref{fig:current_maps} shows the four-current invariant \(J^\mu J_\mu>0\) on coordinate slices through the representative settled configurations. The arrows show the direction of the in-plane spatial current: \((J^x,J^y)\) in the \(xy\) plane and \((J^x,J^z)\) in the \(xz\) plane.
To display both the strongest concentrations and the extended weaker structures, we use a logarithmic color scale. For each model, the current invariant is normalized by the maximum positive value found over its \(xy\) and \(xz\) slices,
\begin{equation}
  \widehat{J^2}
  =
  \frac{J^\mu J_\mu}
  {\max_{xy,xz}\left(J^\mu J_\mu\right)} .
\end{equation}
This choice allows the spatial morphology and dynamic range of the current structure to be compared clearly within and among the models, but the colors should not be used to compare their absolute current magnitudes.

The current morphology is strongly model dependent providing an additional picture that there is no unique dynamically stable configuration for the magnetic fields of neutron stars. Model {\tt A} has the most visibly non-axisymmetric equatorial structure. Its \(xy\) slice contains multiple filamentary, petal-like current concentrations extending through the stellar interior and into the near-surface region. In the \(xz\) slice, the strongest currents are concentrated in two broad off-axis lobes separated by a comparatively weak central vertical region. 

Model {\tt B} exhibits a more layered structure. Its equatorial slice contains a structured central region surrounded by several approximately concentric current shells. The meridional slice is more symmetric, with a prominent current concentration near the central vertical direction and additional shell-like structures extending toward and beyond the stellar surface.

Model {\tt C} is the most nearly axisymmetric in the equatorial plane. Its \(xy\) slice is dominated by concentric and loop-like current structures centered on the rotation axis. The \(xz\) slice contains a strong central vertical current region together with approximately symmetric off-axis structures in the stellar interior.

Overall, the logarithmic current maps reinforce the same model-to-model diversity identified independently by the magnetic-field diagnostics. Model {\tt A} shows the strongest nonaxisymmetric current morphology, model {\tt B} exhibits a more layered and organized structure, model {\tt C} remains the most nearly axisymmetric. 

\section{Discussion and Conclusions}
\label{sec:discussion}

We have performed magnetohydrodynamic simulations in full general relativity of slowly rotating neutron stars endowed with self-consistent mixed poloidal and toroidal magnetic fields. The initial configurations differ in their magnetization and magnetic field geometry, allowing us to test whether dynamically settled magnetic equilibria converge toward a universal configuration.

Our diagnostics, including a vector spherical harmonic decomposition, magnetic field line visualizations, and four-current distributions, consistently show that the late-time configurations retain at least partial memory of their initial magnetic fields -- the leading large-scale poloidal contribution remains recognizable in each model, but the detailed higher-order hierarchy can be substantially rearranged. This is especially clear in model {\tt A}: its poloidal spectrum near \(24t_A\) remains dominated by the initial \((1,0)\) contribution, while its next strongest contributions are nonaxisymmetric. The late-time toroidal magnetic field component is significantly enhanced compared with the initial data in all configurations and reaches an average amplitude of approximately \(20\)--\(40\%\) of the poloidal component through much of the stellar interior.

While our simulations show that a unique dynamically stable magnetic field geometry does not emerge, the settled configurations exhibit a limited degree of universality in the angle-averaged radial profile of the magnetic toroidal and poloidal components. In all three models, \(B_{\rm tor}^{\rm rms}/B_{\rm pol}^{\rm rms}\) exhibits a similar radial profile with values $20\%-40\%$ in the bulk of the star. However, the precise amplitudes of these features and the underlying multipolar and three-dimensional magnetic field structures remain model dependent. As a result, our simulations support the existence of a diverse family of stable mixed poloidal--toroidal and multipolar magnetic fields that share certain radial structural features.

A systematic resolution study is necessary before more precise quantitative statements can be made about the individual multipolar amplitudes. This is most easily seen by our limited resolution study of model {\tt C}; while the qualitative aspects of our result (in particular the axisymmetric magnetic field modes) are robust against resolution, we can see quantitative differences in other modes. Therefore, the highly multipolar structure found in model {\tt A} can be affected by the grid and be resolution dependent. However, we note that such multipolar structures have been seen in analogue Newtonian calculations of magnetized stars before with different codes and grid setups~\cite{Sur:2020hwn}. Therefore, we expect that our qualitative conclusions are robust even though the precise multipolar hierarchy may not be fully converged. A systematic resolution study goes beyond the scope of the current paper, but will be the topic of future work.

The stars considered here are initially modeled as simple zero-temperature polytropes and span a limited range of compactness. Additional effects may arise when realistic equations of state and different stellar compactness, rotation rates, finite temperature or superconductivity (the Meissner effect) are considered. However, these additional variations are more likely to broaden the family of possible stable magnetic configurations than to produce a unique equilibrium, and strengthen our main conclusions. Regarding the Meissner effect, its effectiveness depends on the  strength of the magnetic field, and
recent work has shown that the magnetic flux expulsion from the bulk of the neutron star takes a very long time: $10^6-10^7$ yr~\cite{Ho:2017bia}. Also, according to the framework of~\cite{Lander:2024kye} such expulsion can never be fully realized, and it is only one of four possible outcomes of the effect.

Finally, the exterior solutions in our simulations are not force-free. We therefore cannot use the present magnetic fields and four-current distributions to make quantitative statements about thermal hot spots. Force-free simulations of the corresponding pulsar magnetospheres are necessary for such predictions, and we plan to perform these calculations in forthcoming work.

\begin{acknowledgments}
This work was supported in part by National Science Foundation (NSF) Grants PHY-2308242 and OAC-2310548 to the University of Illinois at Urbana-Champaign, by NASA grants 80NSSC24K0771 80NSSC26K0343, and NSF grant PHY-2145421 to the University of Arizona, and by JSPS Grant-in-Aid for Scientific Research(C) 22K03636, 18K03624, 25K07274, 18K03606, 
24K07053, 21K03556, 20H04728 to the University of the Ryukyus, Okinawa 903-0213, Japan. A.T. acknowledges support from the National Center for Supercomputing Applications (NCSA) at the University of Illinois at Urbana-Champaign through the NCSA Fellows program. 
The work was also supported by allocation PHY190020 from the Advanced Cyberinfrastructure Coordination Ecosystem: Services \& Support (ACCESS) program, which is supported by U.S. National Science Foundation grants 2138259, 2138286, 2138307, 2137603, and 2138296.
This work was  performed in part at the Aspen Center for Physics, which is supported by National Science Foundation grant PHY-2210452.

\end{acknowledgments}

\appendix

\section{An ideal-GRMHD analytic test for the four-current}\label{App:4curr}

We validate the four-current diagnostic with an analytic solution on the spacetime of a constant-density, spherically symmetric star.

Using spherical coordinates $(t,r,\theta,\phi)$, the metric is given by
\begin{equation}
ds^2
=
-\alpha(r)^2 dt^2
+
f(r)^2 dr^2
+
r^2 d\theta^2
+
r^2\sin^2\theta\, d\phi^2 .
\end{equation}

For \(0\leq r\leq R\), the interior Schwarzschild metric functions are
\begin{equation}
f(r)^2
=
\left(
1-\frac{2Mr^2}{R^3}
\right)^{-1},
\qquad
0\le r\le R,
\end{equation}
and
\begin{equation}
\alpha(r)
=
\frac12
\left[
3\sqrt{1-\frac{2M}{R}}
-
\sqrt{1-\frac{2Mr^2}{R^3}}
\right].
\end{equation}

The exterior Schwarzschild metric functions , for $r\ge R$, are
\begin{equation}
\alpha(r)^2 = 1-\frac{2M}{r},
\qquad
f(r)^2 = \left(1-\frac{2M}{r}\right)^{-1}.
\end{equation}

A regular stellar solution requires that the compactness $C=M/R$ satisfy
\begin{equation}
 C<\frac49  .
\end{equation}

A vector potential that gives rise to the analytic ideal GRMHD current is the following

\begin{equation}
A_t
=
-\alpha(r)\Omega(t)A_\phi ,
\end{equation}
where $\Omega(t)$ gives rise to the (time-dependent) angular velocity of the plasma.

\begin{equation}
A_r
=
-\alpha'(r)I(t)A_\phi ,
\end{equation}
where $\dot I(t)=\Omega(t)$, 
\begin{equation}
A_\theta=0, 
\end{equation}
and
\begin{equation}
A_\phi
=
\frac12 B_c r^2\sin^2\theta ,
\end{equation}

For this potential the non-zero (independent) components of the Faraday tensor are
\begin{align}
\label{eq:Fmunucomps}
F_{tr}
& =
\alpha\Omega B_c r\sin^2\theta ,\nonumber
\\
F_{t\theta}
& =
\alpha\Omega B_c r^2\sin\theta\cos\theta ,\nonumber
\\
F_{r\theta}
&=
\alpha' I B_c r^2\sin\theta\cos\theta ,
\\
F_{r\phi}
& =
B_c r\sin^2\theta ,\nonumber
\\
F_{\theta\phi}
& =
B_c r^2\sin\theta\cos\theta .\nonumber
\end{align}

We choose a purely azimuthal velocity
\begin{equation}\label{eq:umu}
u^\mu
=
W\left(
\frac1\alpha,0,0,\Omega(t)
\right),
\end{equation}
with Lorentz factor (measured by a normal observer) given by
\begin{equation}
W
=
\frac{1}
{\sqrt{1-\Omega(t)^2 r^2\sin^2\theta}}.
\end{equation}

The solution is subluminal if
\begin{equation}
\Omega(t)^2r^2\sin^2\theta <1 .
\end{equation}
Since $\sin\theta < 1$, this condition implies that the size of the computational domain cannot exceed
\begin{equation}
r < \frac{1}{\Omega(t)}. 
\end{equation}

Because \(F=\dd A\), the homogeneous Maxwell equation \(\dd F=0\) is satisfied identically. Equations~\eqref{eq:Fmunucomps} and~\eqref{eq:umu} also imply the following:
\begin{itemize}
    \item Ideal GRMHD condition:
\begin{equation}
F_{\mu\nu}u^\nu=0 .
\end{equation}

\item The degeneracy condition:
\begin{equation}
^*F^{\mu\nu}F_{\mu\nu}=0 .
\end{equation}

\item If the plasma velocity is subluminal, it can be shown that the solution is magnetically dominated, i.e., 
\begin{equation}
F^{\mu\nu}F_{\mu\nu}>0.
\end{equation}
\end{itemize}

The current is computed from the inhomogeneous Maxwell equation (in Heaviside-Lorentz units)
\begin{equation}
J^\mu = \nabla_\nu F^{\mu\nu},
\end{equation}
and its components are given by
\begin{align}
J^t
=
\frac{B_c\Omega}
{\alpha f^3}
\left[
rf'\sin^2\theta
-
3f\sin^2\theta
-
f^3
\left(
2\cos^2\theta-\sin^2\theta
\right)
\right].
\end{align}

\begin{equation}
J^r
=
\frac{B_c}
{\alpha f^2}
\left[
r\dot{\Omega}\sin^2\theta
+
\alpha\alpha'I
\left(
2\cos^2\theta-\sin^2\theta
\right)
\right].
\end{equation}

\begin{equation}
J^\theta
=
\frac{B_c\sin\theta\cos\theta}
{\alpha f r^2}
\left[
fr^2\dot{\Omega}
-
\frac{d}{dr}
\left(
\frac{\alpha\alpha'I r^2}{f}
\right)
\right].
\end{equation}

\begin{equation}
J^\phi
=
\frac{B_c}
{\alpha f r^2}
\left[
\alpha f
-
\frac{d}{dr}
\left(
\frac{\alpha r}{f}
\right)
\right].
\end{equation}

The solution for the current is strictly speaking valid only up to the stellar surface. This is because $f(r)$ has a discontinuous derivative across the stellar surface due to the energy density discontinuity at the surface for this solution.

In our code we evolve Cartesian components of all vectorial and tensorial quantities, so we convert the above formulae to Cartesian coordinates using the standard coordinate transformation. 

To test our implementation of the computation of the current based on the electric and magnetic fields measured by a normal observer, we make the following simple choice 
\begin{equation}
\Omega(t)=\Omega_0\sin(\omega t).
\end{equation}
with
\begin{equation}
I(t)=-\frac{\Omega_0}{\omega}\cos(\omega t).
\end{equation}
While this choice is unphysical, because it makes the star change its angular momentum as a function of time, it is important to have time-dependence in $\Omega$, so the ``displacement current'' is non-zero, in other words for time derivatives of the E-field (or $F^{\mu\nu}$) to be non-vanishing. This analytic test is designed to simply test the implementation of computation of the 4-current in our code.

\begin{figure*}[!tp]
  \centering
  \includegraphics[width=\columnwidth]{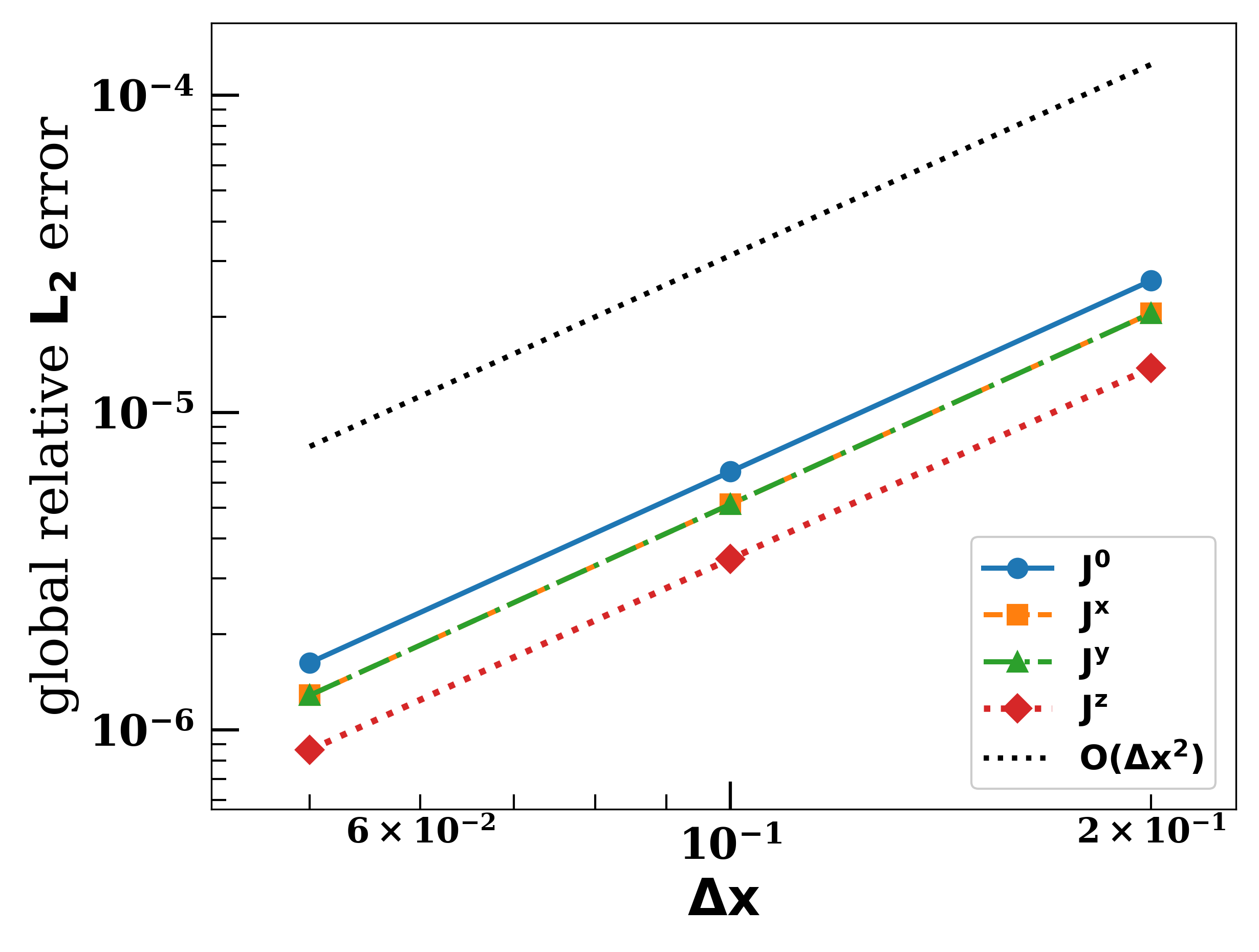}
  \includegraphics[width=\columnwidth]{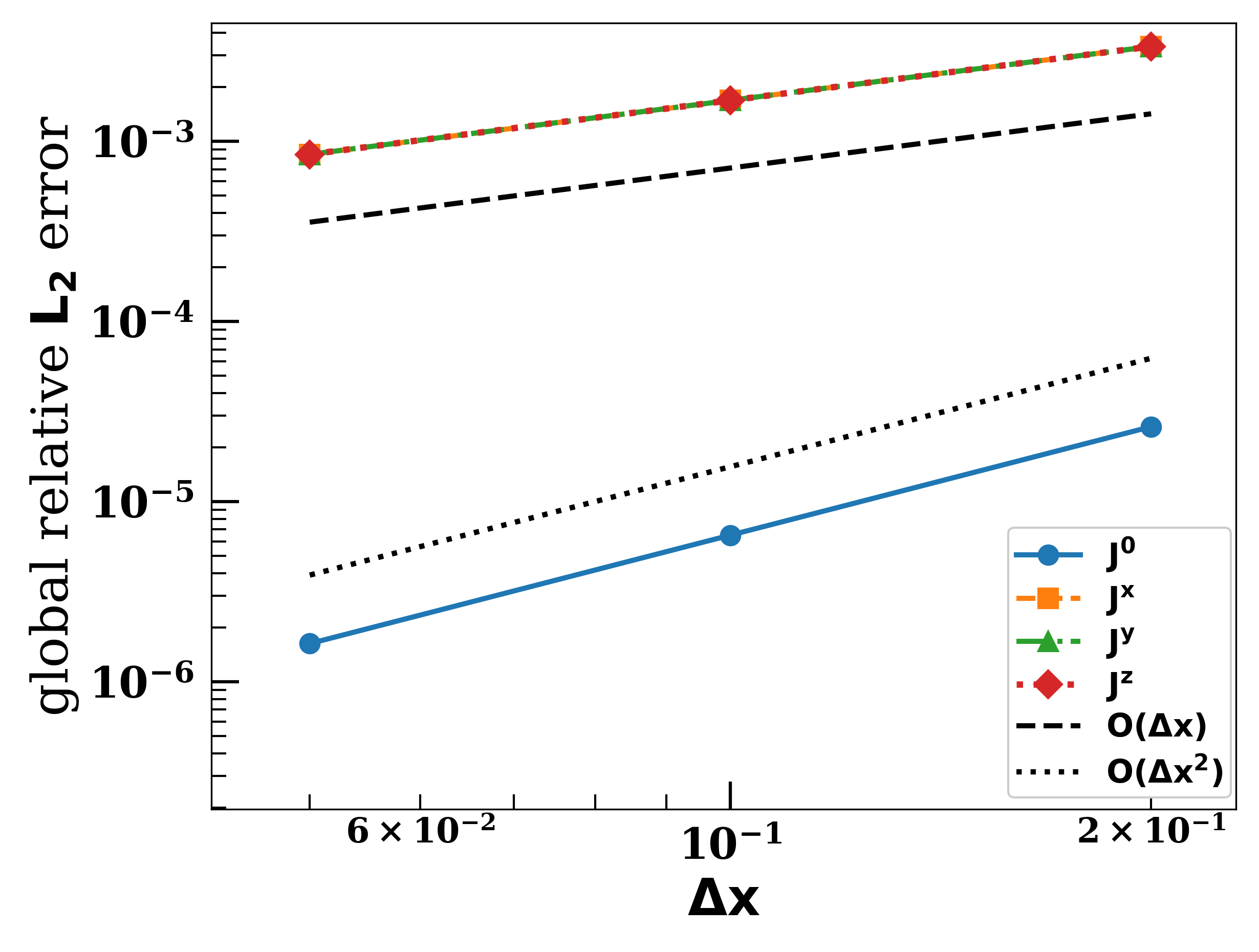}
    \caption{
    Convergence of the four-current components for the analytic-current test with \(M/R=0.1\) and \(\Omega_0R=0.1\). The left and right panels correspond to \(\omega/\Omega_0=0.1\) and \(1000\), respectively. The plotted error is the global relative \(L_2\) norm, \(\|J^\mu_{\rm num}-J^\mu_{\rm exact}\|_2/\|J^\mu_{\rm exact}\|_2\), computed over \(0.01R<r<0.9R\). For slow time variation, the \(J^i\) error is dominated by the centered spatial differences and converges at second order. For rapid variation, the backward time difference dominates \(J^i\), producing first-order convergence. The charge component \(J^t\) remains second-order because it contains only spatial derivatives.    }\label{fig:current_convergence_smallw_currents}
\end{figure*}

We set \(M/R=0.1\), \(\Omega_0R=0.1\), and consider \(\omega/\Omega_0=0.1\) and \(1000\). In the slowly varying case, second-order spatial differencing should dominate the error. In the rapidly varying case, the first-order backward time difference should dominate the spatial current components. The outer boundary is at \(12.5M\), and error norms are evaluated over \(0.01R<r<0.9R\) to exclude both the coordinate origin and the density discontinuity at the surface. We construct
\begin{equation}
B^\mu=\frac12\epsilon^{\mu\nu\rho\sigma}n_\nu F_{\sigma\rho}
\end{equation}
from Eq.~\eqref{eq:Fmunucomps}, use the four-velocity in Eq.~\eqref{eq:umu}, and compare the numerical current with the analytic expressions above. The results are shown in Fig.~\ref{fig:current_convergence_smallw_currents}.

\section{Vector spherical harmonic decomposition}
\label{App:VSH}

Given the spherical unit vectors \(\hat{\mathbf r}\), \(\hat{\boldsymbol\theta}\), and \(\hat{\boldsymbol\phi}\), the angular dependence of a vector field \(\mathbf F(r,\theta,\phi)\) can be expanded in vector spherical harmonics defined from \(Y_{\ell m}\) as~\cite{Barrera_1985,Carrascal_1991}.
\begin{align}
\mathbf{Y}_{\ell m}(\theta,\phi) &= Y_{\ell m}(\theta,\phi)\, \hat{\mathbf{r}}, \\
\mathbf{\Psi}_{\ell m}(\theta,\phi) &= \nabla_{\Omega} Y_{\ell m}(\theta,\phi), \\
\mathbf{\Phi}_{\ell m}(\theta,\phi) &= \hat{\mathbf{r}} \times \nabla_{\Omega} Y_{\ell m}(\theta,\phi).
\end{align}
Here $\nabla_{\Omega}$ is the tangential (angular) gradient on the unit sphere
\begin{align}
\nabla_{\Omega} Y_{\ell m} = \hat{\boldsymbol{\theta}}\,\frac{\partial Y_{\ell m}}{\partial \theta} + \hat{\boldsymbol{\phi}}\, \frac{1}{\sin\theta} \frac{\partial Y_{\ell m}}{\partial \phi}.
\end{align}

The toroidal harmonics are given by
\begin{align}
\mathbf{\Phi}_{\ell m} = \hat{\mathbf{r}} \times \nabla_{\Omega} Y_{\ell m} = 
-\hat{\boldsymbol\theta}\,\frac{1}{\sin\theta}\frac{\partial Y_{\ell m}}{\partial\phi}+\hat{\boldsymbol\phi}\,\frac{\partial Y_{\ell m}}{\partial\theta}.
\end{align}

Given these definitions, 
a general vector field can be written as\footnote{Note that our definition of  $\mathbf{\Psi}_{\ell m}(\theta,\phi)$ is the same as in~\cite{Barrera_1985} even though we don't multiply the angular gradient by $r$. The difference is that we use the angular gradient on the unit sphere.}

\begin{widetext}
\begin{align}
\mathbf{F}(r,\theta,\phi) =
\sum_{\ell=0}^{\infty} \sum_{m=-\ell}^{\ell} 
\Big[
 a_{\ell m}(r)\, \mathbf{Y}_{\ell m}
+ b_{\ell m}(r)\, \mathbf{\Psi}_{\ell m}
+ c_{\ell m}(r)\, \mathbf{\Phi}_{\ell m}
\Big].
\end{align}
\end{widetext}

The coefficients of the expansion can be derived using the vector spherical harmonics orthogonality relations~\cite{Barrera_1985}.
First one splits the field into radial and tangential parts:
\begin{align}
F_r &= \mathbf{F} \cdot \hat{\mathbf{r}}, \\
\mathbf{F}_\perp &= \mathbf{F} - F_r\, \hat{\mathbf{r}}.
\end{align}
Then the coefficients are given as follows:

\begin{align}
a_{\ell m}(r) = 
\int_{S^2} Y_{\ell m}^*(\hat{\mathbf{r}})
\big( \mathbf{F} \cdot \hat{\mathbf{r}} \big)
\, d\Omega.
\end{align}

\begin{align}
b_{\ell m}(r) = 
\frac{1}{\ell(\ell+1)} 
\int_{S^2} 
(\nabla_{\Omega} Y_{\ell m}^*) \cdot \mathbf{F}_\perp \, d\Omega.
\end{align}

\begin{align}
c_{\ell m}(r) = 
\frac{1}{\ell(\ell+1)} 
\int_{S^2} 
\big( \hat{\mathbf{r}} \times \nabla_{\Omega} Y_{\ell m}^* \big)
 \cdot \mathbf{F}_\perp
\, d\Omega.
\end{align}

These three coefficient sets completely characterize the vector field on the sphere at radius \(r\). For the simulation data, the spheres are centered on the snapshot-dependent \(\mathbf{x}_{\rm c}(t)\) defined in Eq.~\eqref{eq:vsh_centered_coordinates}; the analytic tests use the natural origin of the prescribed fields. We decompose the spatial vector potential and the densitized magnetic field measured by a normal observer, \(\tilde B^i=\sqrt{\gamma}B^i\), in vector spherical harmonics at multiple radii and then reconstruct the radial dependence using a discontinuous Galerkin expansion based on Legendre polynomials.

\begin{figure}
  \centering
  \includegraphics[width=0.48\textwidth]{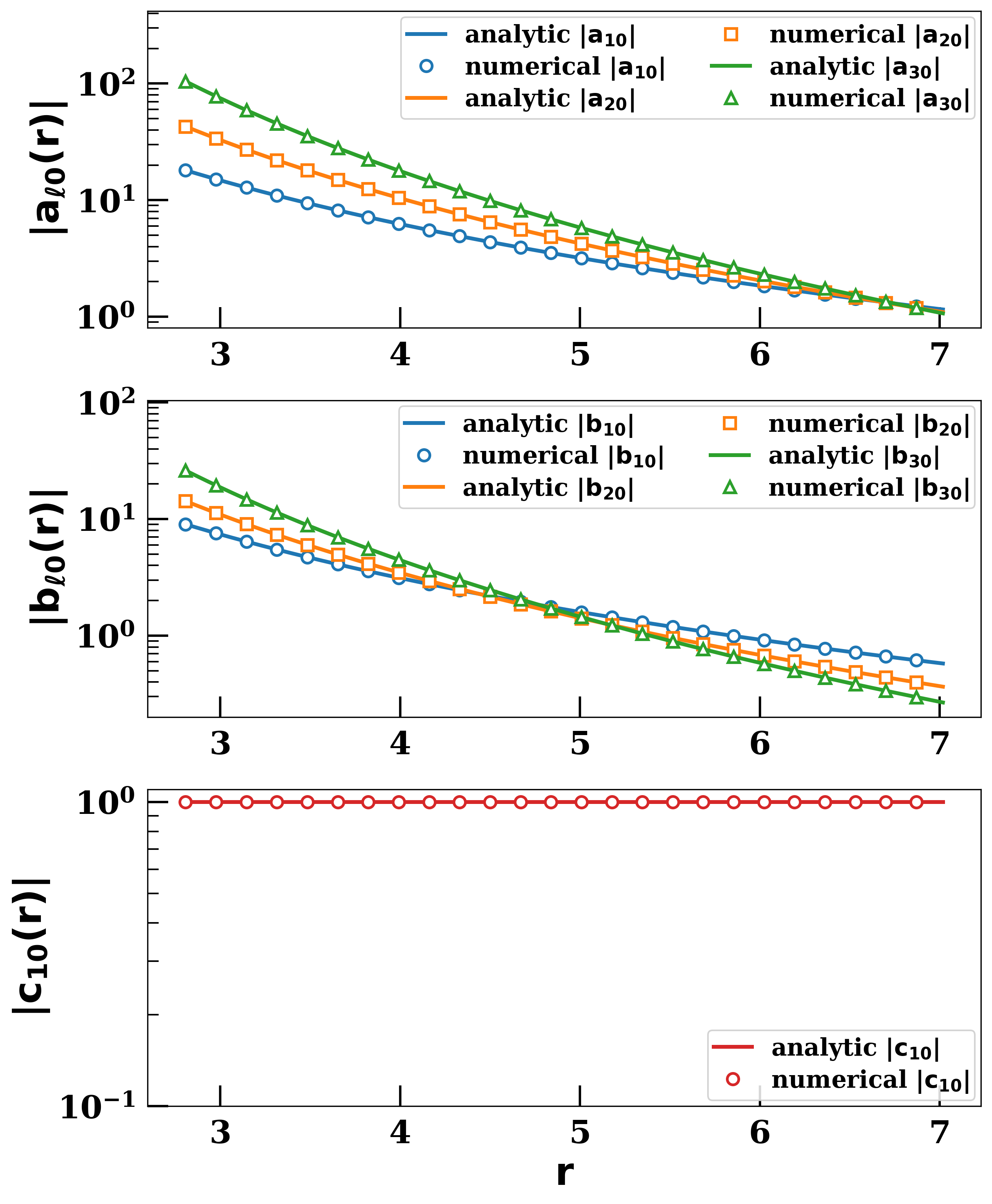}
  \caption{
  Validation of the vector spherical harmonic decomposition for a superposed analytic field containing $\ell=1,2,3$ axisymmetric poloidal multipoles and a purely toroidal $\ell=1$, $m=0$ component. We show the magnitudes of the
  recovered coefficients as functions of radius. The solid curves are the analytic coefficients, while the open markers are the numerical coefficients obtained by applying our decomposition and radial DG reconstruction pipeline
  to the sampled Cartesian field.
  }
  \label{fig:vsh_superposition_coeffs}
\end{figure}

\begin{figure}
  \centering
  \includegraphics[width=0.48\textwidth]{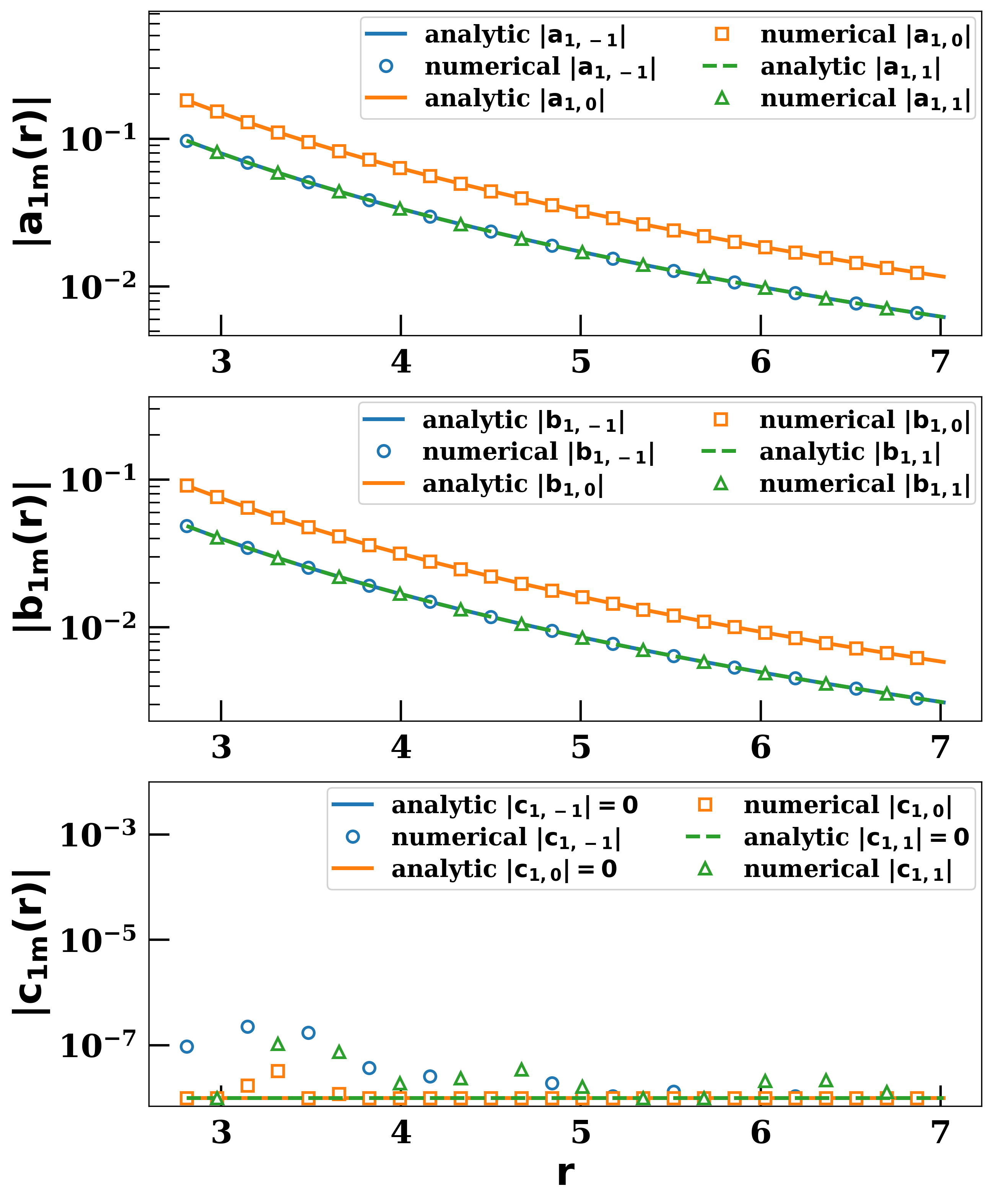}
  \caption{
  Validation of the vector spherical harmonic decomposition for a rotated analytic dipole. A rotation of a dipole leaves the field in the $\ell=1$ sector, but excites the $m=-1,0,1$ components. We show the magnitudes of the recovered $a_{1m}$, $b_{1m}$, and $c_{1m}$ coefficients as functions of radius. The solid curves are the analytic coefficients, while the open markers are the numerical coefficients obtained by applying our decomposition and
  radial DG reconstruction pipeline to the sampled Cartesian field.The $c_{1m}$ coefficients are consistent with zero, confirming that the rotated dipole is recovered as a purely poloidal field.
  }
  \label{fig:vsh_rotated_dipole_coeffs}
\end{figure}

\begin{figure*}[!tp]
\centering
\includegraphics[width=0.49\textwidth]{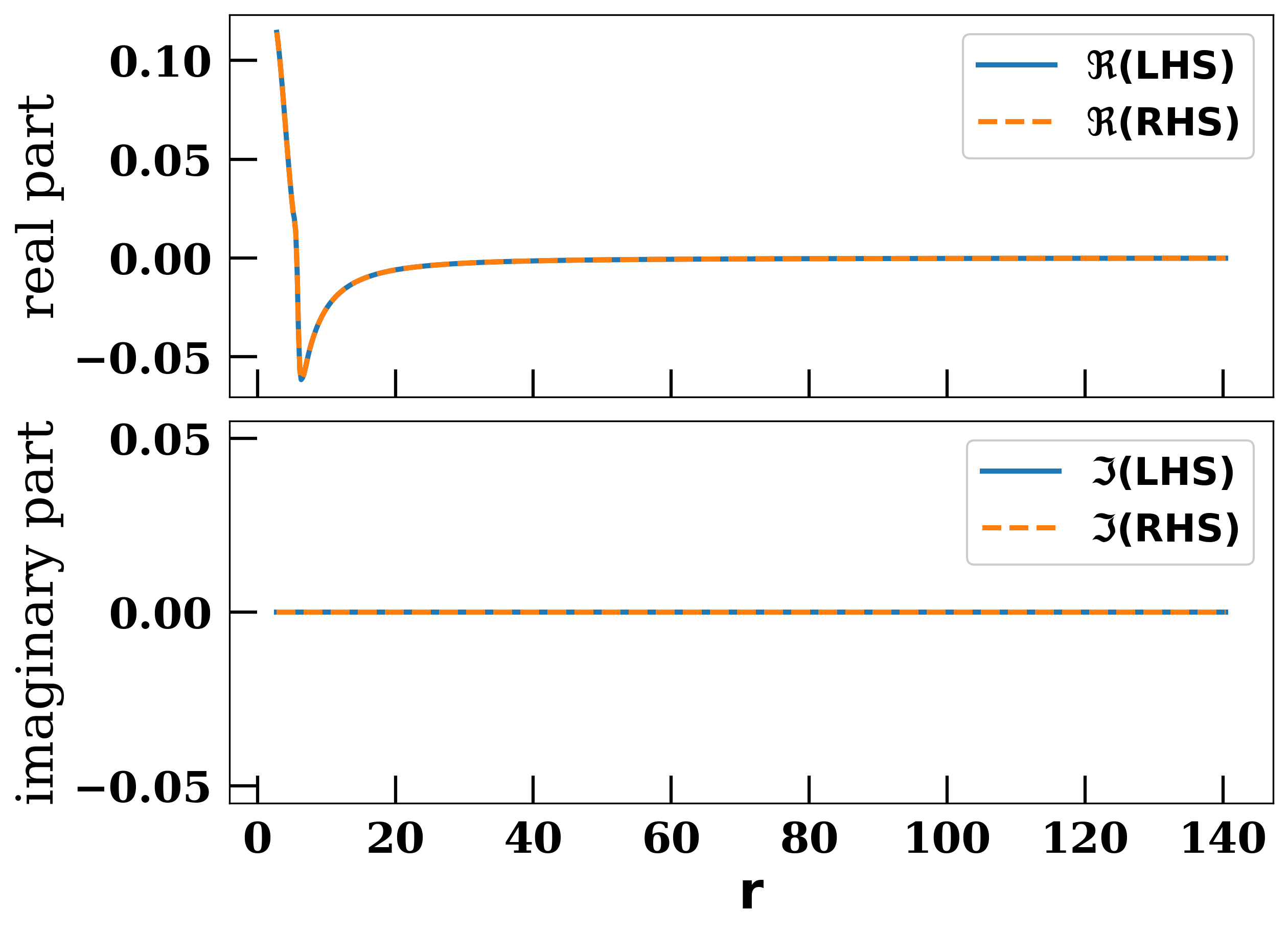}
\includegraphics[width=0.49\textwidth]{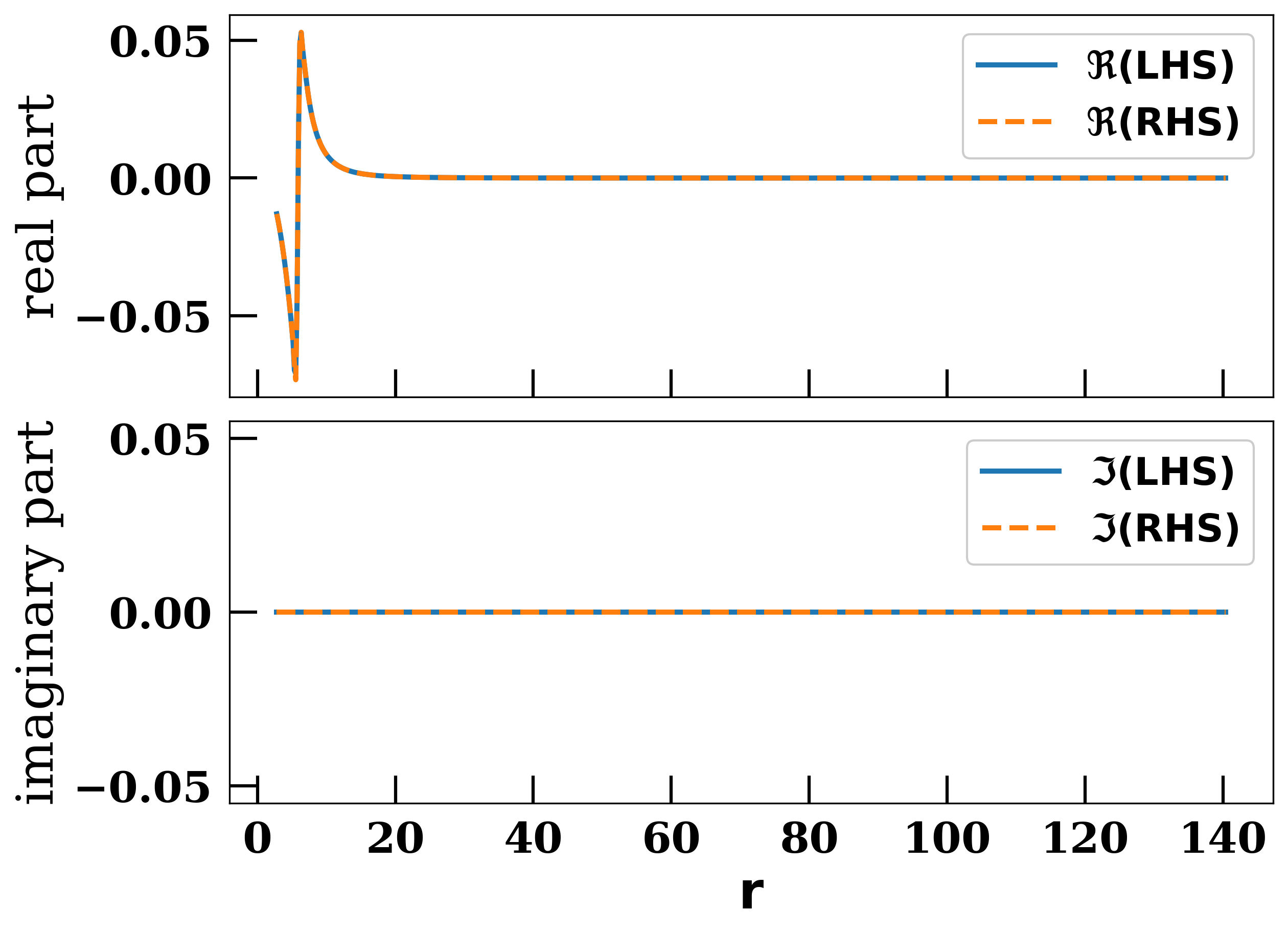}
\caption{Divergence-free consistency check for the densitized magnetic field $\tilde{B}^i=\sqrt{\gamma}B^i$ in the dominant axisymmetric VSH modes. The left panel shows the $(\ell,m)=(1,0)$ mode and the right panel shows the $(\ell,m)=(3,0)$ mode. For each mode, we compare the real and imaginary parts of the two sides of Eq.~\eqref{eq:vsh_divfree_main}. Their close agreement demonstrates that the dominant densitized magnetic field modes satisfy the VSH divergence-free relation.}
\label{fig:divfree_dominant_modes}
\end{figure*}

As an additional consistency check, we verify that the densitized magnetic field $\tilde{B}^i=\sqrt{\gamma}B^i$ remains divergence-free in the vector spherical harmonic representation. For a divergence-free vector field, the VSH coefficients must satisfy
\begin{equation}
\frac{d}{dr}\left(r^2a_{\ell m}(r)\right)=\ell(\ell+1)r\,b_{\ell m}(r),
\label{eq:vsh_divfree_main}
\end{equation}
which can also be used to determine $b_{\ell m}$ from $a_{\ell m}$. We evaluate the two sides of this relation using the radial discontinuous Galerkin representation of the coefficients. Since the decomposition is dominated by the axisymmetric $(\ell,m)=(1,0)$ and $(3,0)$ modes, we show the check for these two modes in Fig.~\ref{fig:divfree_dominant_modes}. The close agreement between the left- and right-hand sides confirms that the dominant components of $\tilde{B}^i$ satisfy the divergence-free condition.

\subsection{Accuracy of the VSH reconstruction}
\label{App:VSH_Lmax_accuracy}

We test the final reconstruction directly against the original Cartesian magnetic field. At one fixed noncollocation radius inside each of the 16 radial DG elements, corresponding to the reference-element coordinate \(\xi=0.37\), we interpolate the original simulation magnetic field and spatial metric onto the same \(N_\theta\times N_\phi=64\times128\) angular grid used for the VSH decomposition. If \(r_e^-\) and \(r_e^+\) are the inner and outer radii of element \(e\), then
\begin{equation}
r(\xi)
=
\frac{1-\xi}{2}r_e^-
+
\frac{1+\xi}{2}r_e^+,
\qquad
-1\leq\xi\leq1.
\end{equation}
The value \(\xi=0.37\) is chosen randomly and lies away from the DG construction nodes and element boundaries, so the comparison tests interpolation within every element. All sampling spheres are centered on the center of mass. We denote the directly sampled densitized field by \(\tilde{\mathbf B}_{\rm raw}\). At the same radius, we first project this field directly through \(\ell_{\max}=63\), giving \(\tilde{\mathbf B}_{63}\), and independently evaluate the order-seven radial-DG representation of the same VSH coefficients, giving \(\tilde{\mathbf B}_{\rm DG}\). This separates angular truncation from radial interpolation.

For a densitized vector field, we define the metric-weighted solid-angle root-mean-square norm
\begin{equation}
\left\lVert\tilde{\mathbf F}\right\rVert_\gamma^2
\equiv
\frac{1}{4\pi}
\int_{S^2}
\frac{\gamma_{ij}\tilde F^i\tilde F^j}{\gamma}
\,d\Omega .
\label{eq:vsh_field_error_norm}
\end{equation}

We define the angular-truncation and radial-DG interpolation errors by
\begin{align}
\epsilon_{\rm ang}(r)
&\equiv
\frac{
\left\lVert
\tilde{\mathbf B}_{63}
-
\tilde{\mathbf B}_{\rm raw}
\right\rVert_\gamma
}{
\left\lVert
\tilde{\mathbf B}_{\rm raw}
\right\rVert_\gamma
},
\\
\epsilon_{\rm DG}(r)
&\equiv
\frac{
\left\lVert
\tilde{\mathbf B}_{\rm DG}
-
\tilde{\mathbf B}_{63}
\right\rVert_\gamma
}{
\left\lVert
\tilde{\mathbf B}_{63}
\right\rVert_\gamma
}.
\label{eq:vsh_separated_errors}
\end{align}

The total relative magnetic field reconstruction error is
\begin{equation}
\epsilon_B(r)
\equiv
\frac{
\left\lVert
\tilde{\mathbf B}_{\rm DG}
-
\tilde{\mathbf B}_{\rm raw}
\right\rVert_\gamma
}{
\left\lVert
\tilde{\mathbf B}_{\rm raw}
\right\rVert_\gamma
}.
\label{eq:vsh_field_error}
\end{equation}
These tests compare all three components using the physical spatial metric.

\begin{figure}[!tp]
\centering
\includegraphics[width=\linewidth]{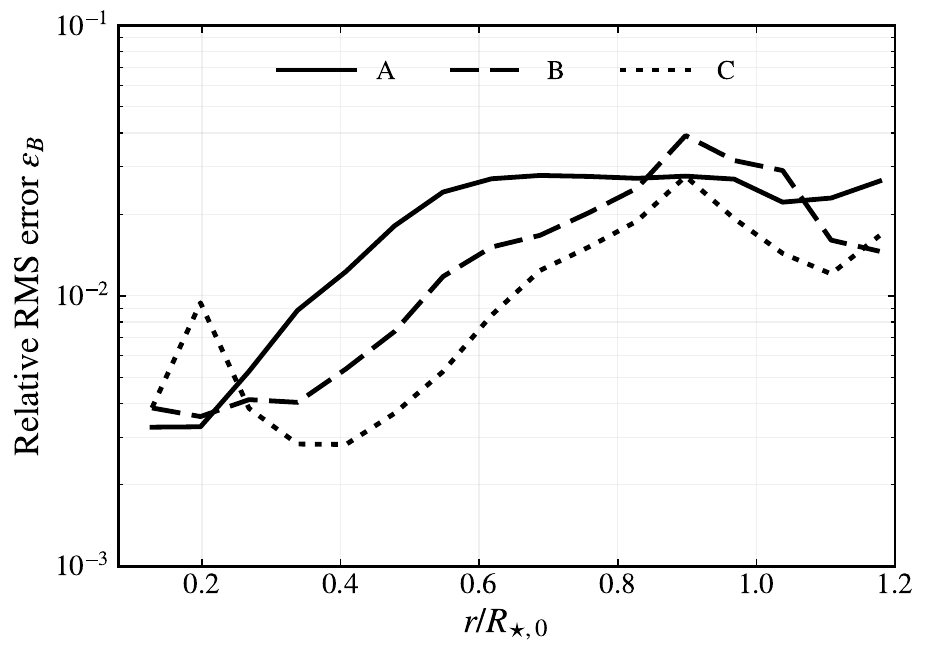}
\caption{
Total relative error of the \(\ell_{\max}=63\), order-seven radial-DG magnetic field reconstruction for models {\tt A}, {\tt B}, and {\tt C}. At one fixed radius inside each radial DG element, the complete reconstructed magnetic field is compared directly with the original Cartesian simulation field using the metric-weighted solid-angle RMS norm in Eq.~\eqref{eq:vsh_field_error}.
}
\label{fig:vsh_Lmax63_field_error}
\end{figure}

Figure~\ref{fig:vsh_Lmax63_field_error} shows \(\epsilon_B\) at the 16 representative radii samples. The maximum total reconstruction errors are \(2.8\%\), \(3.9\%\), and \(2.8\%\) for models {\tt A}, {\tt B}, and {\tt C}, respectively. The corresponding maximum angular-truncation errors are \(2.8\%\), \(3.9\%\), and \(2.7\%\), whereas the maximum radial-DG interpolation errors are \(0.7\%\), \(1.2\%\), and \(1.0\%\). Thus, the total reconstruction error is dominated primarily by the finite angular truncation at \(\ell_{\max}=63\).

This is a direct field-level reconstruction test, but it is not a direct error estimate for the derived pointwise ratio \(B_{\rm tor}/B_{\rm pol}\).

\subsection{Examples: magnetic multipoles}

We tested our implementation using vacuum magnetic fields with $\ell=1$ (dipole), $\ell=2$ (quadrupole), and $\ell=3$ ( octupole) structure. We consider axisymmetric ($m=0$) fields aligned with the $z$--axis, but also rotate the solutions off of the $z$--axis to have non-zero $m$--modes contribute.

\subsubsection{General axisymmetric poloidal multipole}

In vacuum flat spacetime, an axisymmetric poloidal magnetic field can be written in terms of a scalar potential
\begin{align}
\Phi_{\ell}(r,\theta) = \frac{M_{\ell}}{r^{\ell+1}} P_{\ell}(\cos\theta),
\end{align}
where $P_{\ell}$ is the Legendre polynomial and $M_{\ell}$ is the multipole strength. The magnetic field is
\begin{align}
\mathbf{B}_{\ell} &= -\nabla \Phi_{\ell} \\
&= B_{r}^{(\ell)}\, \hat{\mathbf{r}} + B_{\theta}^{(\ell)}\, \hat{\boldsymbol{\theta}},
\end{align}
with
\begin{align}
B_{r}^{(\ell)}(r,\theta) &= (\ell+1)\, \frac{M_{\ell}}{r^{\ell+2}} P_{\ell}(\cos\theta), \\
B_{\theta}^{(\ell)}(r,\theta) &= -\frac{M_{\ell}}{r^{\ell+2}} \frac{\partial P_{\ell}(\cos\theta)}{\partial \theta}.
\end{align}
This field is purely poloidal: there is no $\hat{\boldsymbol{\phi}}$ component, so all toroidal coefficients $c_{\ell m}(r)$ vanish.

Using $Y_{\ell 0}(\theta,\phi) = \sqrt{\frac{2\ell+1}{4\pi}}\,P_{\ell}(\cos\theta)$, we can write 
\begin{align}
B_{r}^{(\ell)}(r,\theta,\phi) &= A_{\ell}(r)\,Y_{\ell 0}(\theta,\phi), \\
\mathbf{B}_{\perp}^{(\ell)}(r,\theta,\phi) &= B_{\ell}(r)\, \nabla_{\Omega} Y_{\ell 0}(\theta,\phi),
\end{align}
for suitable radial functions $A_{\ell}(r), B_{\ell}(r)$. Comparing with the general expansion,
\begin{align}
\mathbf{B}_{\ell}(r,\theta,\phi) = a_{\ell 0}(r)\, \mathbf{Y}_{\ell 0}(\theta,\phi) + b_{\ell 0}(r)\, \mathbf{\Psi}_{\ell 0}(\theta,\phi),
\end{align}
we see that for each $\ell$,
\begin{align}
a_{\ell 0}(r) &\propto \frac{M_{\ell}}{r^{\ell+2}}, &
b_{\ell 0}(r) &\propto \frac{M_{\ell}}{r^{\ell+2}}, &
c_{\ell 0}(r) &= 0.
\end{align}

\subsubsection{Purely Toroidal Magnetic Field}

In vacuum flat spacetime a purely toroidal magnetic field is of the form 
\begin{align}
\mathbf{B}(r,\theta,\phi) = c_{\ell m}(r)\, \mathbf{\Phi}_{\ell m}(\theta,\phi)
\end{align}
produces a divergence-free tangential field on the sphere.

To test our implementation we specifically take a purely toroidal dipole ($l=1$, $m=0$) field 
\begin{align}
\mathbf{B}(r,\theta,\phi) = T(r)\, \mathbf{\Phi}_{10}(\theta,\phi),
\end{align}
where $T(r)$ is an arbitrary radial profile, and 
\begin{align}
\mathbf{\Phi}_{10}(\theta,\phi) = -\sqrt{\frac{3}{4\pi}}\sin\theta\, \hat{\boldsymbol{\phi}}.
\end{align}

This field trivially satisfies 
$a_{10}(r) = 0,  b_{10}(r) = 0$, and 
$c_{10}(r) = T(r)$. All other $(\ell,m)$ coefficients vanish.

In our testsuite we set the magnetic field in the code to be a linear
combination of $\ell\leq 3$ poloidal modes and the $\ell=1$ toroidal mode,
and verify that the vector spherical harmonic decomposition recovers the
expected expansion coefficients. We also test a rotated dipole, which remains a
pure $\ell=1$ poloidal field but excites the $m=-1,0,1$ components. In both
tests the analytic Cartesian field is first sampled on a three-dimensional grid,
then passed through the same vector-spherical-harmonic decomposition and radial
DG reconstruction pipeline used for the simulation data. Thus the comparison
tests both the angular decomposition and the radial interpolation of the
coefficients.

\section{{Energy diagnostics of the initial equilibria}}
\label{App:initial_energies}
The energy ratios in Table~\ref{tab:initial_data} use the quantities entering the COCAL virial relation. Let $\alpha$ be the lapse, $\gamma_{ij}$ the spatial metric, and $dV=\sqrt{\gamma}\,d^3x$ the proper volume element on a spatial slice $\Sigma$. We write $\gamma_{ij}=\psi^4\tilde{\gamma}_{ij}$, where $\psi$ is the conformal factor and $\tilde{\gamma}_{ij}$ is the conformal metric. The derivative $\tilde D_i$ and scalar curvature $\tilde R$ are associated with $\tilde{\gamma}_{ij}$.
The kinetic energy and pressure integral are

\begin{align}
T&=\frac12\int_{\rm star}\rho_b h
\left[(\alpha u^t)^2-1\right]\,dV,
\label{eq:initial_virial_T}\\
\Pi&=\int_{\rm star}P\,dV,
\label{eq:initial_virial_Pi}
\end{align}
where $h=1+\epsilon+P/\rho_b$ is the specific enthalpy and $\epsilon$ is the specific internal energy. For the maximal slicing used in the initial data, $\gamma^{ij}K_{ij}=0$, the gravitational contribution is

\begin{equation}
\begin{aligned}
W=\frac{1}{4\pi}\int_{\Sigma}\Bigl\{
&\psi^{-4}\Bigl[
2\tilde D_i\ln\psi\,\tilde D^i\ln\psi
\\
&-\tilde D_i\ln\alpha\,\tilde D^i\ln\alpha
+\frac14\tilde R
\Bigr]
+\frac34 K_{ij}K^{ij}
\Bigr\}\,dV.
\end{aligned}
\label{eq:initial_virial_W}
\end{equation}

The quantity $W$ is negative for the models considered here. For the electromagnetic energies, we use the Gaussian electromagnetic convention employed by COCAL. Let $E_i$ be the electric field measured by the observer normal to the slice, and let $F_{ij}$ denote the spatial projection of the Faraday tensor. In the following expressions, $F^{ij}=\gamma^{ik}\gamma^{jl}F_{kl}$ and $E^i=\gamma^{ij}E_j$. The total electromagnetic energy is

\begin{equation}
\mathcal M=\frac{1}{8\pi}\int_{\Sigma}
\left(E_iE^i+\frac12F_{ij}F^{ij}\right)dV.
\label{eq:initial_em_energy}
\end{equation}

For the axisymmetric initial equilibria, we use spherical coordinates \((r,\theta,\phi)\), with the polar axis aligned with the rotation axis and meridional indices \(A,B\in\{r,\theta\}\). COCAL evaluates the toroidal and poloidal magnetic energies as

\begin{align}
\mathcal M_{\rm tor}
&=\frac{1}{16\pi}\int_{\Sigma}F_{AB}F^{AB}\,dV,
\label{eq:initial_toroidal_energy}\\
\mathcal M_{\rm pol}
&=\frac{1}{8\pi}\int_{\Sigma}F_{A\phi}F^{A\phi}\,dV,
\label{eq:initial_poloidal_energy}
\end{align}

where repeated meridional indices are summed. The electric contribution and the total energy satisfy

\begin{align}
\mathcal M_{\rm ele}
&=\frac{1}{8\pi}\int_{\Sigma}E_iE^i\,dV,
\label{eq:initial_electric_energy}\\
\mathcal M_{\phantom{\rm ele}}
&=\mathcal M_{\rm tor}+\mathcal M_{\rm pol}
+\mathcal M_{\rm ele}.
\label{eq:initial_energy_sum}
\end{align}

The gravitational and electromagnetic integrals extend over the COCAL gravitational computational domain, including the stellar exterior, while the matter integrals extend over the star. The equilibrium virial relation is

\begin{equation}
2T+3\Pi+\mathcal M+W=0,
\label{eq:initial_virial_relation}
\end{equation}
up to numerical residuals. All four energy ratios in Table~\ref{tab:initial_data} use the same $|W|$ defined in Eq.~\eqref{eq:initial_virial_W}.
The poloidal and toroidal energies above use component contractions of the spatial Faraday tensor with the full spatial metric. They differ from the spherical-harmonic diagnostics used for the evolved magnetic field, which reconstruct poloidal and toroidal fields on spherical shells and measure their angular RMS amplitudes. Those shell amplitudes are not the volume-integrated energies defined here.

\bibliography{MagnetizedNS}

@article{Gourgouliatos:2016fnl,
    author = "Gourgouliatos, Konstantinos N. and Wood, Toby and Hollerbach, Rainer",
    title = "{Magnetic field evolution in magnetar crusts through three dimensional simulations}",
    eprint = "1604.01399",
    archivePrefix = "arXiv",
    primaryClass = "astro-ph.SR",
    doi = "10.1073/pnas.1522363113",
    journal = "Proc. Nat. Acad. Sci.",
    volume = "113",
    pages = "3944",
    year = "2016"
}

@article{Ciolfi:2009bv,
    author = "Ciolfi, R. and Ferrari, V. and Gualtieri, L. and Pons, J. A.",
    title = "{Relativistic models of magnetars: the twisted-torus magnetic field configuration}",
    eprint = "0903.0556",
    archivePrefix = "arXiv",
    primaryClass = "astro-ph.SR",
    doi = "10.1111/j.1365-2966.2009.14990.x",
    journal = "Mon. Not. Roy. Astron. Soc.",
    volume = "397",
    pages = "913",
    year = "2009"
}

@article{Kundu:2026guq,
    author = "Kundu, Anu and Kalapotharakos, Constantinos and Wadiasingh, Zorawar and Olmschenk, Greg and Wallace, Wendy F. and Harding, Alice K. and Venter, Christo and Kazanas, Demosthenes",
    title = "{The swept-back multipolar magnetic field of neutron stars: Application to NICER MSP J0030+0451}",
    eprint = "2604.19534",
    archivePrefix = "arXiv",
    primaryClass = "astro-ph.HE",
    doi = "10.3847/1538-4357/ae8e6f",
    journal = "Astrophys. J.",
    volume = "1008",
    number = "2",
    pages = "215",
    year = "2026"
}

@article{Lander:2024kye,
    author = "Lander, S. K.",
    title = "{The Meissner effect in neutron stars}",
    eprint = "2411.08021",
    archivePrefix = "arXiv",
    primaryClass = "astro-ph.HE",
    doi = "10.1093/mnras/stae2453",
    journal = "Mon. Not. Roy. Astron. Soc.",
    volume = "535",
    number = "3",
    pages = "2449--2468",
    year = "2024"
}

@article{Capobianco:2026ots,
    author = "Capobianco, Aurora and Cook, William and Bernuzzi, Sebastiano and Haskell, Brynmor and Fields, Jacob",
    title = "{Magnetic field dynamics in isolated neutron stars with an external dipole field}",
    eprint = "2605.22921",
    archivePrefix = "arXiv",
    primaryClass = "astro-ph.HE",
    journal = "arXiv e-prints",
    month = "5",
    year = "2026"
}

@article{Duez:2005sf,
    author = "Duez, Matthew D. and Liu, Yuk Tung and Shapiro, Stuart L. and Stephens, Branson C.",
    title = "{Relativistic magnetohydrodynamics in dynamical spacetimes: Numerical methods and tests}",
    eprint = "astro-ph/0503420",
    archivePrefix = "arXiv",
    doi = "10.1103/PhysRevD.72.024028",
    journal = "Phys. Rev. D",
    volume = "72",
    pages = "024028",
    year = "2005"
}

@article{Sur:2020hwn,
    author = "Sur, Ankan and Haskell, Brynmor and Kuhn, Emily",
    title = "{Magnetic field configurations in neutron stars from MHD simulations}",
    eprint = "2002.10357",
    archivePrefix = "arXiv",
    primaryClass = "astro-ph.HE",
    doi = "10.1093/mnras/staa1212",
    journal = "Mon. Not. Roy. Astron. Soc.",
    volume = "495",
    number = "1",
    pages = "1360--1371",
    year = "2020"
}

@ARTICLE{2015MNRAS.447.1213M,
       author = {{Mitchell}, J.~P. and {Braithwaite}, J. and {Reisenegger}, A. and {Spruit}, H. and {Valdivia}, J.~A. and {Langer}, N.},
        title = "{Instability of magnetic equilibria in barotropic stars}",
      journal = {Monthly Notices of the Royal Astronomical Society},
         year = 2015,
        month = feb,
       volume = {447},
       number = {2},
        pages = {1213-1223},
          doi = {10.1093/mnras/stu2514},
archivePrefix = {arXiv},
       eprint = {1411.7252},
 primaryClass = {astro-ph.SR},
       adsurl = {https://ui.adsabs.harvard.edu/abs/2015MNRAS.447.1213M}
}

@ARTICLE{2022MNRAS.517..560B,
       author = {{Becerra}, Laura and {Reisenegger}, Andreas and {Valdivia}, Juan Alejandro and {Gusakov}, Mikhail},
        title = "{Stability of axially symmetric magnetic fields in stars}",
      journal = {Monthly Notices of the Royal Astronomical Society},
         year = 2022,
        month = nov,
       volume = {517},
       number = {1},
        pages = {560-568},
          doi = {10.1093/mnras/stac2704},
archivePrefix = {arXiv},
       eprint = {2209.01042},
 primaryClass = {astro-ph.SR},
       adsurl = {https://ui.adsabs.harvard.edu/abs/2022MNRAS.517..560B}
}

@article{Ho:2017bia,
    author = "Ho, Wynn C. G. and Andersson, Nils and Graber, Vanessa",
    title = "{Dynamical onset of superconductivity and retention of magnetic fields in cooling neutron stars}",
    eprint = "1711.08480",
    archivePrefix = "arXiv",
    primaryClass = "astro-ph.HE",
    doi = "10.1103/PhysRevC.96.065801",
    journal = "Phys. Rev. C",
    volume = "96",
    number = "6",
    pages = "065801",
    year = "2017"
}

@article{Chen:2020rud,
    author = "Chen, Alexander Y. and Yuan, Yajie and Vasilopoulos, Georgios",
    title = "{A Numerical Model for the Multiwavelength Lightcurves of PSR J0030+0451}",
    eprint = "2002.06104",
    archivePrefix = "arXiv",
    primaryClass = "astro-ph.HE",
    doi = "10.3847/2041-8213/ab85c5",
    journal = "Astrophys. J. Lett.",
    volume = "893",
    number = "2",
    pages = "L38",
    year = "2020"
}

@article{Cao:2026xnp,
    author = "Cao, Gang and Yang, Xiongbang",
    title = "{The Combined X-Ray and {\ensuremath{\gamma}}-Ray Modeling of Millisecond Pulsar PSR J0030+0451 in the Dissipative Magnetospheres}",
    eprint = "2601.07491",
    archivePrefix = "arXiv",
    primaryClass = "astro-ph.HE",
    doi = "10.3847/1538-4357/ae3467",
    journal = "Astrophys. J.",
    volume = "997",
    number = "2",
    pages = "306",
    year = "2026"
}

@article{Kaspi:2017fwg,
    author = "Kaspi, Victoria M. and Beloborodov, Andrei",
    title = "{Magnetars}",
    eprint = "1703.00068",
    archivePrefix = "arXiv",
    primaryClass = "astro-ph.HE",
    doi = "10.1146/annurev-astro-081915-023329",
    journal = "Ann. Rev. Astron. Astrophys.",
    volume = "55",
    pages = "261--301",
    year = "2017"
}

@ARTICLE{PhilippovAAReview,
       author = {{Philippov}, A. and {Kramer}, M.},
        title = "{Pulsar Magnetospheres and Their Radiation}",
      journal = {Ann. Rev. of Astron. \& Astroph.},
         year = 2022,
        month = aug,
       volume = {60},
        pages = {495-558},
          doi = {10.1146/annurev-astro-052920-112338},
       adsurl = {https://ui.adsabs.harvard.edu/abs/2022ARA&A..60..495P}
}

@ARTICLE{BeskinPulsarReview2018,
       author = {{Beskin}, V.~S.},
        title = "{Radio pulsars: already fifty years!}",
      journal = {Physics Uspekhi},
         year = 2018,
        month = apr,
       volume = {61},
       number = {4},
        pages = {353-380},
          doi = {10.3367/UFNe.2017.10.038216},
archivePrefix = {arXiv},
       eprint = {1807.08528},
 primaryClass = {astro-ph.HE},
       adsurl = {https://ui.adsabs.harvard.edu/abs/2018PhyU...61..353B}
}

@article{Zhang:2022uzl,
    author = "Zhang, Bing",
    title = "{The physics of fast radio bursts}",
    eprint = "2212.03972",
    archivePrefix = "arXiv",
    primaryClass = "astro-ph.HE",
    doi = "10.1103/RevModPhys.95.035005",
    journal = "Rev. Mod. Phys.",
    volume = "95",
    number = "3",
    pages = "035005",
    year = "2023"
}

@article{Riley:2019yda,
    author = "Riley, Thomas E. and others",
    title = "{A $NICER$ View of PSR J0030+0451: Millisecond Pulsar Parameter Estimation}",
    eprint = "1912.05702",
    archivePrefix = "arXiv",
    primaryClass = "astro-ph.HE",
    doi = "10.3847/2041-8213/ab481c",
    journal = "Astrophys. J. Lett.",
    volume = "887",
    number = "1",
    pages = "L21",
    year = "2019"
}

@article{Miller:2021qha,
    author = "Miller, M. C. and others",
    title = "{The Radius of PSR J0740+6620 from NICER and XMM-Newton Data}",
    eprint = "2105.06979",
    archivePrefix = "arXiv",
    primaryClass = "astro-ph.HE",
    doi = "10.3847/2041-8213/ac089b",
    journal = "Astrophys. J. Lett.",
    volume = "918",
    number = "2",
    pages = "L28",
    year = "2021"
}

@article{Salmi:2024bss,
    author = "Salmi, Tuomo and others",
    title = "{A NICER View of PSR J1231{\ensuremath{-}}1411: A Complex Case}",
    eprint = "2409.14923",
    archivePrefix = "arXiv",
    primaryClass = "astro-ph.HE",
    doi = "10.3847/1538-4357/ad81d2",
    journal = "Astrophys. J.",
    volume = "976",
    number = "1",
    pages = "58",
    year = "2024"
}

@article{Miller:2025qfq,
    author = "Miller, M. C. and others",
    title = "{The Radius of PSR J0437{\textendash}4715 from NICER Data}",
    eprint = "2512.08790",
    archivePrefix = "arXiv",
    primaryClass = "astro-ph.HE",
    doi = "10.3847/2041-8213/ae5057",
    journal = "Astrophys. J. Lett.",
    volume = "1000",
    number = "2",
    pages = "L48",
    year = "2026"
}

@article{Mauviard:2025dmd,
    author = "Mauviard, Lucien and others",
    title = "{A NICER View of the 1.4 M$_{⊙}$ Edge-on Pulsar PSR J0614-3329}",
    eprint = "2506.14883",
    archivePrefix = "arXiv",
    primaryClass = "astro-ph.HE",
    doi = "10.3847/1538-4357/ae145d",
    journal = "Astrophys. J.",
    volume = "995",
    number = "1",
    pages = "60",
    year = "2025"
}

@article{Kini:2026rjx,
    author = "Kini, Yves and others",
    title = "{A NICER View of PSR J0030+0451: Updated Constraints from Six Years of NICER Observations}",
    eprint = "2602.23743",
    archivePrefix = "arXiv",
    primaryClass = "astro-ph.HE",
    doi = "10.3847/1538-4357/ae733e",
    journal = "Astrophys. J.",
    volume = "1005",
    number = "2",
    pages = "201",
    year = "2026"
}

@article{Bilous:2019knh,
    author = "Bilous, Anna V. and others",
    title = "{A $NICER$ View of PSR J0030+0451: Evidence for a Global-scale Multipolar Magnetic Field}",
    eprint = "1912.05704",
    archivePrefix = "arXiv",
    primaryClass = "astro-ph.HE",
    doi = "10.3847/2041-8213/ab53e7",
    journal = "Astrophys. J. Lett.",
    volume = "887",
    number = "1",
    pages = "L23",
    year = "2019"
}

@article{Barrera_1985,
doi = {10.1088/0143-0807/6/4/014},
url = {https://doi.org/10.1088/0143-0807/6/4/014},
year = {1985},
month = {oct},
publisher = {},
volume = {6},
number = {4},
pages = {287},
author = {R G Barrera and G A Estevez and J Giraldo},
title = {Vector spherical harmonics and their application to magnetostatics},
journal = {European Journal of Physics}
}

@article{Carrascal_1991,
doi = {10.1088/0143-0807/12/4/007},
url = {https://doi.org/10.1088/0143-0807/12/4/007},
year = {1991},
month = {jul},
publisher = {},
volume = {12},
number = {4},
pages = {184},
author = {B Carrascal and G A Estevez and Peilian Lee and V Lorenzo},
title = {Vector spherical harmonics and their application to classical electrodynamics},
journal = {European Journal of Physics}
}

@article{Baubock:2013gna,
    author = {Baub{\"o}ck, Michi and Berti, Emanuele and Psaltis, Dimitrios and {\"O}zel, Feryal},
    title = "{Relations Between Neutron-Star Parameters in the Hartle-Thorne Approximation}",
    eprint = "1306.0569",
    archivePrefix = "arXiv",
    primaryClass = "astro-ph.HE",
    doi = "10.1088/0004-637X/777/1/68",
    journal = "Astrophys. J.",
    volume = "777",
    pages = "68",
    year = "2013"
}

@article{Gralla:2016fix,
    author = "Gralla, Samuel E. and Lupsasca, Alexandru and Philippov, Alexander",
    title = "{Pulsar Magnetospheres: Beyond the Flat Spacetime Dipole}",
    eprint = "1604.04625",
    archivePrefix = "arXiv",
    primaryClass = "astro-ph.HE",
    doi = "10.3847/1538-4357/833/2/258",
    journal = "Astrophys. J.",
    volume = "833",
    number = "2",
    pages = "258",
    year = "2016"
}

@article{Gralla:2017nbw,
    author = "Gralla, Samuel E. and Lupsasca, Alexandru and Philippov, Alexander",
    title = "{Inclined Pulsar Magnetospheres in General Relativity: Polar Caps for the Dipole, Quadrudipole and Beyond}",
    eprint = "1704.05062",
    archivePrefix = "arXiv",
    primaryClass = "astro-ph.HE",
    doi = "10.3847/1538-4357/aa978d",
    journal = "Astrophys. J.",
    volume = "851",
    number = "2",
    pages = "137",
    year = "2017"
}

@article{Lockhart:2019nch,
    author = {Lockhart, Will and Gralla, Samuel E. and {\"O}zel, Feryal and Psaltis, Dimitrios},
    title = "{X-ray light curves from realistic polar cap models: inclined pulsar magnetospheres and multipole fields}",
    eprint = "1904.11534",
    archivePrefix = "arXiv",
    primaryClass = "astro-ph.HE",
    doi = "10.1093/mnras/stz2524",
    journal = "Mon. Not. Roy. Astron. Soc.",
    volume = "490",
    number = "2",
    pages = "1774--1783",
    year = "2019"
}

@article{Thompson:2016dkd,
    author = "Thompson, Christopher and Yang, Huan and Ortiz, N{\'e}stor",
    title = "{Global Crustal Dynamics of Magnetars in Relation to their Bright X-ray Outbursts}",
    eprint = "1608.02633",
    archivePrefix = "arXiv",
    primaryClass = "astro-ph.HE",
    doi = "10.3847/1538-4357/aa6c30",
    journal = "Astrophys. J.",
    volume = "841",
    number = "1",
    pages = "54",
    year = "2017"
}

@article{Chamel:2012uz,
    author = "Chamel, N. and Pavlov, R. L. and Mihailov, L. M. and Velchev, Ch. J. and Stoyanov, Zh. K. and Mutafchieva, Y. D. and Ivanovich, M. D. and Pearson, J. M. and Goriely, S.",
    title = "{Outer crust of strongly magnetized neutron stars for Hartree-Fock-Bogoliubov atomic mass models}",
    eprint = "1210.5874",
    archivePrefix = "arXiv",
    primaryClass = "astro-ph.HE",
    doi = "10.1103/PhysRevC.86.055804",
    journal = "Phys. Rev. C",
    volume = "86",
    pages = "055804",
    year = "2012"
}

@article{Pons:2019zyc,
    author = "Pons, Jos{\'e} A. and Vigan{\`o}, Daniele",
    title = "{Magnetic, thermal and rotational evolution of isolated neutron stars}",
    eprint = "1911.03095",
    archivePrefix = "arXiv",
    primaryClass = "astro-ph.HE",
    doi = "10.1007/s41115-019-0006-7",
    journal = "Liv. Rev. Comput. Astrophys.",
    volume = "5",
    number = "1",
    pages = "3",
    year = "2019"
}

@article{Tayler_1957,
doi = {10.1088/0370-1301/70/11/305},
url = {https://doi.org/10.1088/0370-1301/70/11/305},
year = {1957},
month = {nov},
publisher = {},
volume = {70},
number = {11},
pages = {1049},
author = {R J Tayler},
title = {The Influence of an Axial Magnetic Field on the Stability of a Constricted Gas Discharge},
journal = {Proceedings of the Physical Society. Section B}
}

@article{10.1093/mnras/161.4.365,
    author = {Tayler, R. J.},
    title = { The Adiabatic Stability of Stars Containing Magnetic Fields–I: T OROIDAL F IELDS},
    journal = {Monthly Notices of the Royal Astronomical Society},
    volume = {161},
    number = {4},
    pages = {365-380},
    year = {1973},
    month = {04},
    issn = {0035-8711},
    doi = {10.1093/mnras/161.4.365},
    url = {https://doi.org/10.1093/mnras/161.4.365},
    eprint = {https://academic.oup.com/mnras/article-pdf/161/4/365/8076545/mnras161-0365.pdf},
}

@ARTICLE{1973MNRAS.163...77M,
       author = {{Markey}, P. and {Tayler}, R.~J.},
        title = "{The adiabatic stability of stars containing magnetic fields. II. Poloidal fields}",
      journal = {Monthly Notices of the Royal Astronomical Society},
         year = 1973,
        month = mar,
       volume = {163},
        pages = {77-91},
          doi = {10.1093/mnras/163.1.77},
       adsurl = {https://ui.adsabs.harvard.edu/abs/1973MNRAS.163...77M}
}

@article{10.1093/mnras/162.4.339,
    author = {Wright, G. A. E.},
    title = {Pinch Instabilities in Magnetic Stars},
    journal = {Monthly Notices of the Royal Astronomical Society},
    volume = {162},
    number = {4},
    pages = {339-358},
    year = {1973},
    month = {06},
    issn = {0035-8711},
    doi = {10.1093/mnras/162.4.339},
    url = {https://doi.org/10.1093/mnras/162.4.339},
    eprint = {https://academic.oup.com/mnras/article-pdf/162/4/339/8073447/mnras162-0339.pdf},
}

@ARTICLE{1977ApJ...215..302F,
       author = {{Flowers}, E. and {Ruderman}, M.~A.},
        title = "{Evolution of pulsar magnetic fields.}",
      journal = {The Astrophysical Journal},
         year = 1977,
        month = jul,
       volume = {215},
        pages = {302-310},
          doi = {10.1086/155359},
       adsurl = {https://ui.adsabs.harvard.edu/abs/1977ApJ...215..302F}
}

@ARTICLE{2007A&A...469..275B,
       author = {{Braithwaite}, J.},
        title = "{The stability of poloidal magnetic fields in rotating stars}",
      journal = {Astronomy \& Astrophysics},
         year = 2007,
        month = jul,
       volume = {469},
       number = {1},
        pages = {275-284},
          doi = {10.1051/0004-6361:20065903},
archivePrefix = {arXiv},
       eprint = {0705.0185},
 primaryClass = {astro-ph},
       adsurl = {https://ui.adsabs.harvard.edu/abs/2007A&A...469..275B}
}

@article{Lasky:2012ju,
    author = "Lasky, Paul D. and Zink, Burkhard and Kokkotas, Kostas D.",
    title = "{Gravitational Waves and Hydromagnetic Instabilities in Rotating Magnetized Neutron Stars}",
    eprint = "1203.3590",
    archivePrefix = "arXiv",
    primaryClass = "astro-ph.SR",
    journal = "arXiv e-prints",
    month = "3",
    year = "2012"
}

@ARTICLE{2011MNRAS.412.1730L,
       author = {{Lander}, S.~K. and {Jones}, D.~I.},
        title = "{Oscillations and instabilities in neutron stars with poloidal magnetic fields}",
      journal = {Monthly Notices of the Royal Astronomical Society},
         year = 2011,
        month = apr,
       volume = {412},
       number = {3},
        pages = {1730-1740},
          doi = {10.1111/j.1365-2966.2010.18009.x},
archivePrefix = {arXiv},
       eprint = {1010.0614},
 primaryClass = {astro-ph.SR},
       adsurl = {https://ui.adsabs.harvard.edu/abs/2011MNRAS.412.1730L}
}

@article{Lasky:2011un,
    author = "Lasky, Paul D. and Zink, Burkhard and Kokkotas, Kostas D. and Glampedakis, Kostas",
    title = "{Hydromagnetic Instabilities in Neutron Stars}",
    eprint = "1105.1895",
    archivePrefix = "arXiv",
    primaryClass = "astro-ph.SR",
    doi = "10.1088/2041-8205/735/1/L20",
    journal = "Astrophys. J. Lett.",
    volume = "735",
    pages = "L20",
    year = "2011"
}

@article{Ciolfi:2011xa,
    author = "Ciolfi, Riccardo and Lander, Samuel K. and Manca, Gian Mario and Rezzolla, Luciano",
    title = "{Instability-driven evolution of poloidal magnetic fields in relativistic stars}",
    eprint = "1105.3971",
    archivePrefix = "arXiv",
    primaryClass = "gr-qc",
    doi = "10.1088/2041-8205/736/1/L6",
    journal = "Astrophys. J. Lett.",
    volume = "736",
    pages = "L6",
    year = "2011"
}

@article{Ciolfi:2012en,
    author = "Ciolfi, Riccardo and Rezzolla, Luciano",
    title = "{Poloidal-Field Instability in Magnetized Relativistic Stars}",
    eprint = "1206.6604",
    archivePrefix = "arXiv",
    primaryClass = "astro-ph.SR",
    doi = "10.1088/0004-637X/760/1/1",
    journal = "Astrophys. J.",
    volume = "760",
    pages = "1",
    year = "2012"
}

@article{Ciolfi:2013dta,
    author = "Ciolfi, Riccardo and Rezzolla, Luciano",
    title = "{Twisted-torus configurations with large toroidal magnetic fields in relativistic stars}",
    eprint = "1306.2803",
    archivePrefix = "arXiv",
    primaryClass = "astro-ph.SR",
    doi = "10.1093/mnrasl/slt092",
    journal = "Mon. Not. Roy. Astron. Soc.",
    volume = "435",
    pages = "L43--L47",
    year = "2013"
}

@article{Tsokaros:2021pkh,
    author = "Tsokaros, Antonios and Ruiz, Milton and Shapiro, Stuart L. and Ury{\={u}}, K{\={o}}ji",
    title = "{Magnetohydrodynamic Simulations of Self-Consistent Rotating Neutron Stars with Mixed Poloidal and Toroidal Magnetic Fields}",
    eprint = "2111.00013",
    archivePrefix = "arXiv",
    primaryClass = "gr-qc",
    doi = "10.1103/PhysRevLett.128.061101",
    journal = "Phys. Rev. Lett.",
    volume = "128",
    number = "6",
    pages = "061101",
    year = "2022"
}

@article{Sur:2021awe,
    author = "Sur, Ankan and Cook, William and Radice, David and Haskell, Brynmor and Bernuzzi, Sebastiano",
    title = "{Long-term general relativistic magnetohydrodynamics simulations of magnetic field in isolated neutron stars}",
    eprint = "2108.11858",
    archivePrefix = "arXiv",
    primaryClass = "astro-ph.HE",
    doi = "10.1093/mnras/stac353",
    journal = "Mon. Not. Roy. Astron. Soc.",
    volume = "511",
    number = "3",
    pages = "3983--3993",
    year = "2022"
}

@article{Pinas:2025bpq,
    author = "Pi{\~n}as, Fabrizio Venturi and Yip, Anson Ka Long and Cheong, Patrick Chi-Kit and Ruiz, Milton",
    title = "{Impact of Rotation on Magnetic Field Stability and Orientation in Isolated Neutron Stars}",
    eprint = "2508.20220",
    archivePrefix = "arXiv",
    primaryClass = "astro-ph.HE",
    doi = "10.3847/1538-4357/ae4c3f",
    journal = "Astrophys. J.",
    volume = "1001",
    number = "1",
    pages = "49",
    year = "2026"
}

@article{Loffler:2011ay,
  author = {L{\"o}ffler, Frank and Faber, Joshua and Bentivegna, Eloisa and Bode, Tanja and Diener, Peter and Haas, Roland and Hinder, Ian and Mundim, Bruno C. and Ott, Christian D. and Schnetter, Erik and Allen, Gabrielle and Campanelli, Manuela and Laguna, Pablo},
  title = {The Einstein Toolkit: A Community Computational Infrastructure for Relativistic Astrophysics},
  journal = {Classical and Quantum Gravity},
  volume = {29},
  pages = {115001},
  year = {2012},
  eprint = {1111.3344},
  archivePrefix = {arXiv},
  primaryClass = {gr-qc}
}

@article{Zilhao:2013hia,
  author = {Zilh{\~a}o, Miguel and L{\"o}ffler, Frank},
  title = {An Introduction to the Einstein Toolkit},
  journal = {International Journal of Modern Physics A},
  volume = {28},
  pages = {1340014},
  year = {2013},
  eprint = {1305.5299},
  archivePrefix = {arXiv},
  primaryClass = {gr-qc}
}

@article{Shibata:1995we,
  author = {Shibata, Masaru and Nakamura, Takashi},
  title = {Evolution of three-dimensional gravitational waves: Harmonic slicing case},
  journal = {Physical Review D},
  volume = {52},
  pages = {5428--5444},
  year = {1995}
}

@article{Baumgarte:1998te,
  author = {Baumgarte, Thomas W. and Shapiro, Stuart L.},
  title = {On the numerical integration of Einstein's field equations},
  journal = {Physical Review D},
  volume = {59},
  pages = {024007},
  year = {1998},
  eprint = {gr-qc/9810065},
  archivePrefix = {arXiv}
}

@article{Brown:2008sb,
  author = {Brown, J. David},
  title = {Covariant formulations of BSSN and the standard gauge},
  journal = {Physical Review D},
  volume = {79},
  pages = {104029},
  year = {2009},
  eprint = {0902.3652},
  archivePrefix = {arXiv},
  primaryClass = {gr-qc}
}

@article{Alcubierre:2002kk,
  author = {Alcubierre, Miguel and Br{\"u}gmann, Bernd and Diener, Peter and Koppitz, Michael and Pollney, Denis and Seidel, Edward and Takahashi, Ryoji},
  title = {Gauge conditions for long-term numerical black hole evolutions without excision},
  journal = {Physical Review D},
  volume = {67},
  pages = {084023},
  year = {2003},
  eprint = {gr-qc/0206072},
  archivePrefix = {arXiv}
}

@article{Campanelli:2005dd,
  author = {Campanelli, Manuela and Lousto, C. O. and Marronetti, Pedro and Zlochower, Yosef},
  title = {Accurate evolutions of orbiting black-hole binaries without excision},
  journal = {Physical Review Letters},
  volume = {96},
  pages = {111101},
  year = {2006},
  eprint = {gr-qc/0511048},
  archivePrefix = {arXiv}
}

@article{Baker:2005vv,
  author = {Baker, John G. and Centrella, Joan and Choi, Dae-Il and Koppitz, Michael and van Meter, James},
  title = {Gravitational-wave extraction from an inspiraling configuration of merging black holes},
  journal = {Physical Review Letters},
  volume = {96},
  pages = {111102},
  year = {2006},
  eprint = {gr-qc/0511103},
  archivePrefix = {arXiv}
}

@article{vanMeter:2006vi,
  author = {van Meter, James R. and Baker, John G. and Koppitz, Michael and Choi, Dae-Il},
  title = {How to move a black hole without excision: Gauge conditions for the numerical evolution of a moving puncture},
  journal = {Physical Review D},
  volume = {73},
  pages = {124011},
  year = {2006},
  eprint = {gr-qc/0605030},
  archivePrefix = {arXiv}
}

@article{Farris:2012ux,
  author = {Farris, Brian D. and Gold, Roman and Paschalidis, Vasileios and Etienne, Zachariah B. and Shapiro, Stuart L.},
  title = {Binary black-hole mergers in magnetized disks: Simulations in full general relativity},
  journal = {Physical Review Letters},
  volume = {109},
  pages = {221102},
  year = {2012},
  eprint = {1207.3354},
  archivePrefix = {arXiv},
  primaryClass = {astro-ph.HE}
}

@article{Etienne:2012te,
  author = {Etienne, Zachariah B. and Paschalidis, Vasileios and Liu, Yuk Tung and Shapiro, Stuart L.},
  title = {Relativistic magnetohydrodynamics in dynamical spacetimes: Improved electromagnetic gauge condition for adaptive mesh refinement grids},
  journal = {Physical Review D},
  volume = {85},
  pages = {024013},
  year = {2012},
  eprint = {1110.4633},
  archivePrefix = {arXiv},
  primaryClass = {astro-ph.HE}
}

@article{Etienne:2015cea,
  author = {Etienne, Zachariah B. and Paschalidis, Vasileios and Haas, Roland and M{\"o}sta, Philipp and Shapiro, Stuart L.},
  title = {{IllinoisGRMHD}: An Open-Source, User-Friendly GRMHD Code for Dynamical Spacetimes},
  journal = {Classical and Quantum Gravity},
  volume = {32},
  pages = {175009},
  year = {2015},
  eprint = {1501.07276},
  archivePrefix = {arXiv},
  primaryClass = {astro-ph.HE}
}

@article{Uryu:2014tda,
    author = "Ury{\={u}}, K{\={o}}ji and Gourgoulhon, Eric and Markakis, Charalampos M. and Fujisawa, Kotaro and Tsokaros, Antonios and Eriguchi, Yoshiharu",
    title = "{Equilibrium solutions of relativistic rotating stars with mixed poloidal and toroidal magnetic fields}",
    eprint = "1410.3913",
    archivePrefix = "arXiv",
    primaryClass = "astro-ph.HE",
    doi = "10.1103/PhysRevD.90.101501",
    journal = "Phys. Rev. D",
    volume = "90",
    number = "10",
    pages = "101501",
    year = "2014"
}

@article{Uryu:2019ckz,
    author = "Ury{\={u}}, K{\={o}}ji and Yoshida, Shijun and Gourgoulhon, Eric and Markakis, Charalampos and Fujisawa, Kotaro and Tsokaros, Antonios and Taniguchi, Keisuke and Eriguchi, Yoshiharu",
    title = "{New code for equilibriums and quasiequilibrium initial data of compact objects. IV. Rotating relativistic stars with mixed poloidal and toroidal magnetic fields}",
    eprint = "1906.10393",
    archivePrefix = "arXiv",
    primaryClass = "gr-qc",
    doi = "10.1103/PhysRevD.100.123019",
    journal = "Phys. Rev. D",
    volume = "100",
    number = "12",
    pages = "123019",
    year = "2019"
}

@article{Paschalidis:2014qra, 
    author = {Paschalidis, Vasileios and Ruiz, Milton and Shapiro, Stuart L.}, 
    title = {Relativistic Simulations of Black Hole-Neutron Star Coalescence: The Jet Emerges}, 
    journal = {The Astrophysical Journal Letters}, 
    volume = {806}, 
    pages = {L14}, 
    year = {2015}, 
    doi = {10.1088/2041-8205/806/1/L14}, 
    eprint = {1410.7392}, 
    archivePrefix = {arXiv}, 
    primaryClass = {astro-ph.HE}}

@article{Joshi:2026Interior,
  author        = {Joshi, Raj Kishor and Haskell, Brynmor and Cook, William and Bernuzzi, Sebastiano},
  title         = {Interior Magnetic Fields in Magnetars and Radio Pulsars},
  eprint        = {2609.07647},
  archivePrefix = {arXiv},
  primaryClass  = {astro-ph.HE},
  journal       = {arXiv e-prints},
  year          = {2026}
}
\bibliographystyle{apsrev4-2}

\end{document}